\documentclass[11pt,a4paper]{article}
\usepackage[T1]{fontenc}
\usepackage[round]{natbib}
\usepackage{authblk}
\usepackage{mathptmx}
\usepackage{stackrel,amssymb,amsmath}
\usepackage{courier}
\usepackage[cm]{fullpage}
\usepackage[pdftex]{graphicx}
\usepackage{subfigure}
\usepackage{url}
\usepackage{xcolor}
\usepackage[hidelinks,breaklinks]{hyperref}
\usepackage{caption} 
\usepackage{xcolor} 
\usepackage{dcolumn} 
\usepackage{floatrow} 
\newcommand{\ignore}[1]{}
\newcommand{\unix}[1]{\texttt{#1}} 
\newcommand{\email}[1]{\texttt{#1}}
\providecommand{\keywords}[1]{{\small\textbf{\textit{Keywords---}} #1}}

\makeatletter
\let\@fnsymbol\@arabic
\makeatother

\newcommand{\astfootnote}[1]{%
\let\oldthefootnote=\thefootnote%
\setcounter{footnote}{0}%
\renewcommand{\thefootnote}{\fnsymbol{footnote}}%
\footnote{#1}%
\let\thefootnote=\oldthefootnote%
}

\begin{document}

\title{New network models facilitate analysis of biological networks}

\author{Alex Stivala}

\affil{Universit\`a della  Svizzera italiana, Via Giuseppe Buffi 13, 6900 Lugano, Switzerland \\
Email: \email{alexander.stivala@usi.ch}}

\maketitle

\begin{abstract}
  Exponential-family random graph models (ERGMs) are a family of
  network models originating in social network analysis, which have
  also been applied to biological networks. Advances in estimation
  algorithms have increased the practical scope of these models to
  larger networks, however it is still not always possible to estimate
  a model without encountering problems of model near-degeneracy,
  particularly if it is desired to use only simple model parameters,
  rather than more complex parameters designed to overcome the problem
  of near-degeneracy. Two new network models related to the ERGM, the
  Tapered ERGM, and the latent order logistic (LOLOG) model, have
  recently been proposed to overcome this problem. In this work I
  illustrate the application of the Tapered ERGM and the LOLOG to a
  set of biological networks, including protein-protein interaction
  (PPI) networks, gene regulatory networks, and neural networks. I
  find that the Tapered ERGM and the LOLOG are able to estimate models
  for networks for which it was not possible to estimate a
  conventional ERGM, and are able to do so using only simple model
  parameters. In the case of two neural networks where data on the
  spatial position of neurons is available, this allows the estimation
  of models including terms for spatial distance and triangle
  structures, allowing triangle motif statistical significance to be
  estimated while accounting for the effect of spatial proximity on
  connection probability. For some larger networks, however, Tapered
  ERGM and LOLOG estimation was not possible in practical time, while
  conventional ERGM models were able to be estimated only by using the
  Equilibrium Expectation (EE) algorithm.
\end{abstract}


\section{Introduction}
\label{sec:intro}

Networks are of great interest in biology in a variety of contexts,
such as protein-protein interaction (PPI) networks
\citep{delasrivas10}, gene regulatory networks
\citep{emmert-streib14}, and neural networks \citep{allard20}.  It is
therefore important to have appropriate methods for analyzing such
networks \citep{desilva05}, and as discussed in \citet{stivala21}, the
exponential-family random graph model (ERGM) is one such method,
developed in the context of social networks \citep{lusher13}, and
first applied to biological networks by \citet{saul07}. Network models
can be generative, seeking mechanisms and explanations, or null
models, used for hypothesis testing \citep{betzel17}. A generative
model can, of course, also be used as a null model, by generating
simulated networks and comparing the observed statistics of interest to
their distributions in the simulated networks, as done with ERGM in, for
example, \citet{felmlee21,stivala21}.

Null models are often used for testing the statistical significance of
``motifs'', small (usually connected and induced) subgraphs, which are
thought to be over-represented \citep{milo02,shen02}. The conventional
null model for assessing this significance is generated by randomizing
the observed network, preserving the degree of each node
\citep{milo02}, although more sophisticated randomization techniques,
preserving additional properties, can also be used
\citep{mahadevan06,orsini15}. The conventional procedure has intrinsic
limitations, as described in \citet{fodor20}, which can be overcome,
at least in part, by using a more sophisticated model, such as the
ERGM, as a generative model, a null model, or both
\citep{stivala21}. These problems can also be solved using 
an information theory approach, such as that
recently described in \citet{benichou23}.

ERGMs are widely used for modeling social networks
\citep{lusher13,koskinen20,koskinen23}, as well as other application
fields \citep{ghafouri20}, and development of ERGM modelling
techniques is an active research area
\citep{cimini19,lusher20,schweinberger20,schmid21,stivala22_slides,krivitsky22,ergm4,caimo23,giacomarra23,schmid24}.
A brief survey of ERGM applications to biological networks is given in
\citet{stivala21}, but some notable more recent work on the
application of ERGMs to neural networks includes the application of
temporal ERGMs \citep{leifeld18} to model brain network reorganization
after stroke \citep{obando22}, a detailed review of ERGMs in brain
connectivity networks \citep{dichio23b}, and the use of ERGMs as part
of a novel method to explore development of the \textit{C.~elegans}
connectome \citep{dichio23}. The \textit{C.~elegans} neural network
has also been the subject of a related modeling technique, stochastic
blockmodelling \citep{pavlovic14,gross23}.

As discussed in \citet{stivala21}, for some of the biological networks
considered it was not possible to estimate an ERGM, and in particular
neural networks proved particularly problematic. In this work, I use
two new, related, models, the Tapered ERGM
\citep{fellows17,blackburn23} and the latent order logistic model
(LOLOG) \citep{fellows18,clark22} to re-analyze the biological
networks considered in \citet{stivala21} and compare the results to
the ERGMs presented there, as well as to analyze some neural networks
for which ERGM models could not be estimated.

\section{Network models}

\subsection{Exponential-family random graph model (ERGM) and Tapered ERGM}

An ERGM is a probability distribution with the form
\begin{equation}
  \label{eqn:ergm}
  p_{\mathrm{ergm}}(x \mid \theta) = \frac{1}{\kappa(\theta)}\exp\left(\sum_A \theta_A g_A(x)\right)
\end{equation}
where
\begin{itemize}
\item $X = [X_{ij}]$ is square binary matrix of random tie variables,
\item $x$ is a realization of $X$,
\item $A$ is a ``configuration'', a (small) set of nodes and a subset of ties between them,
\item $g_A(x)$ is the network statistic for configuration $A$,
\item $\theta_A$ is a model parameter corresponding to configuration $A$,
\item $\kappa(\theta) = \sum_{x \in G_N} \exp\left(\sum_A \theta_A
  g_A(x)\right)$, where $G_N$ is the set of all square binary matrices
  of order $N$ (graphs with $N$ nodes), is the normalizing constant to
  ensure a proper distribution.
\end{itemize}

Configurations can include nodal attributes, including dyadic
attributes such as, for example, the distance between each pair of
nodes. These are assumed to be fixed (exogenous to the model).

By estimating the parameter vector $\theta$ which maximizes the
probability of an observed graph $x$ under the model, that is, the
maximum likelihood estimate (MLE), together with the estimated
standard errors of the parameter estimates, inferences can be made
about the over-representation (a positive and statistically significant
parameter estimate) or under-representation (a negative and
statistically significant parameter estimate) of the corresponding
configurations. These inferences about over-representation (or
under-representation) are, for each configuration, conditional on all
the other configurations included in the model, which need not be
independent (and which, indeed are usually not independent). For
example, a model might include parameters for density, degree
distribution, triangles, two-paths (note that a triangle is formed by
adding a third edge to a two-path), and spatial distance between
nodes. In this model, a statistically significant positive estimate
for triangles would indicate that the triangle configuration occurs
more frequently than expected by chance, even accounting for density,
degree distribution, number of two-paths, and the effect of spatial
distance on edge probability.

ERGM parameter estimation is, due to the intractable normalizing
constant $\kappa(\theta)$ in (\ref{eqn:ergm}), a computationally
intractable problem, and hence in practice it is necessary (except for
extremely small networks, \citep{vegayon21}) to use Markov chain Monte
Carlo (MCMC) methods \citep{hunter12} such as Markov Chain Monte Carlo
MLE (MCMCMLE) \citep{geyer92} or stochastic approximation
\citep{snijders02}. More recently, the ``Equilibrium Expectation''
(EE) algorithm \citep{byshkin16,byshkin18,borisenko20} has allowed
ERGM models for networks, including biological networks
\citep{byshkin18,stivala21}, far larger than previously possible, to
be estimated in practical time \citep{stivala20b}. Note that
\citet{stivala20b} states that, ``An important next step is the
strengthening of the theoretical basis for the EE algorithm \ldots
there are no theoretical guarantees behind the EE algorithm''
\citep[p.~17]{stivala20b}.  However that referred to the original EE
algorithm \citep{byshkin18}, rather than the simplified EE algorithm
\citep{borisenko20} now (also) implemented in
EstimNetDirected.\footnote{\url{https://github.com/stivalaa/EstimNetDirected}}\textsuperscript{,}\footnote{The
simplified EE algorithm of \citet{borisenko20} is used if
\unix{useBorisenkoUpdate = true} is set in the EstimNetDirected
configuration file, otherwise the original EE algorithm described in
\citet{byshkin18,stivala20b} is used. I recommend always setting
\unix{useBorisenkoUpdate = true}, as was done, for example, for
estimating the ERGM models described in
\citet{stivala21,stivala22_slides,stivala23_slides}.} This algorithm
has been shown to converge to the MLE, if it exists, when the learning
rate is small enough, by \citet{giacomarra23}, using the uncertain
energies framework of \citet{ceperley99}, a proof first outlined
briefly (but not for the specific case of connected networks
considered in \citet{giacomarra23}) by \citet{borisenko20} using the
results of \citet{ceperley99,frenkel17}.

It is a well-known problem with ERGMs that simple model specifications
(such as a model containing only terms for edges and triangles) can
result in ``near-degeneracy'', where the MLE does not exist, or the
probability mass is concentrated in a small subset of graphs, often
(nearly) empty or (nearly) complete graphs
\citep{handcock03,snijders06,hunter07,schweinberger11,chatterjee13,schweinberger20,blackburn23}. The
usual solution to this problem is to use ``alternating'' or
``geometrically weighted'' configurations
\citep{snijders06,robins07,hunter07,lusher13,stivala23}, however this
can result in the model no longer directly testing the hypotheses of
interest \citep{stivala21,blackburn23}, or researchers omitting model
terms due only to their tendency to near-degeneracy, or omitting
models entirely, due to an inability to obtain converged estimates
\citep{martin17,martin20,clark22}.

Tapered ERGMs \citep{fellows17,blackburn23} are a variant of the ERGM
that solves the problem of near-degeneracy by imposing upper bounds on
the variance of the sufficient statistics. The Tapered ERGM works by
assigning lower probability to graphs with statistics far from their
central location, using a new tapering parameter vector $\tau$. The
parameter vector $\tau$ can itself by estimated by optimizing a double
penalized likelihood, penalizing kurtosis values too far from the
target values and values of $\tau$ that are too large
\citep{blackburn23}. The Tapered ERGM is implemented in the
ergm.tapered R package
\citep{ergm.tapered,ergm,krivitsky22,ergm4}.

\subsection{Latent order logistic (LOLOG) model}

The latent order logistic (LOLOG) model \citep{fellows18,clark22} is
related to the ERGM, but is based on the principle of network growth,
and in particular, a (latent) node ordering process. Each edge
variable is sequentially considered for creation, and edges are not
deleted.

In the following, $X$ is a random graph with $N$ nodes ($[X_{ij}]$ is
square binary matrix of random tie variables), $x$ is a
realization of $X$, and  $X_t$, $t = 1, \ldots, \lvert x \rvert$ are latent
random variables representing the sequential formation of $X$. $X_t$
has exactly $t$ edges and is formed from $X_{t-1}$ by the addition
of a single edge \citep{fellows18,clark22}. Here
$\lvert x \rvert$ is the number of dyads in the graph, and hence
for directed graphs $\lvert x \rvert = N(N-1)$ and for undirected graphs
$\lvert x \rvert = N(N-1)/2$.

A LOLOG model is specified by two components \citep{fellows18,clark22}. First,
the probability of observing a graph given a specified edge formation
order $s$, which is a product of logistic likelihoods:
\begin{equation}
  \label{eqn:lolog_full}
  p(x \mid s, \theta) = \prod_{t=1}^{\lvert x \rvert} \frac{1}{Z_t(s)} \exp\left(\theta \cdot C_{s,t}\right)
\end{equation}
where
\begin{itemize}
\item $s = \{s_1, s_2, \ldots, s_{\lvert x \rvert}\}$ is the set of all possible edge formation orders with $\lvert x \rvert$ dyads,
\item $C_{s,t} = g(x_t, s_{\leq t}) - g(x_{t-1}, s_{\leq t-1})$ where $s_{\leq t}$ denotes the first $t$ elements of $s$ and $g(\cdot)$ are the sufficient statistics; $C_{s,t}$ are the change statistics,
\item $Z_t(s) = \exp\left(g(x_t^{+}, s_{\leq t}) - g(x_{t-1},s_{\leq t-1}) \right) + 1$,
  where $x_t^{+}$ is the graph
  $x_{t-1}$ with edge $s_t$ added, are the normalizing constants.
\end{itemize}

The second component is the marginal likelihood of an observed graph, a sum
over all possible edge permutations, where $p(s)$ is the probability 
that the ordering $s$ occurs:

\begin{equation}
  p_{\textrm{lolog}}(x \mid \theta) = \sum_s p(x \mid s, \theta) p(s)
  \label{eqn:lolog_marginal}
\end{equation}

LOLOG model estimation by Monte Carlo method of moments is implemented
in the lolog R package \citep{lolog}. If a partial edge ordering is
observed (based on the order of nodes being added to the
network, for example), that the edge ordering can be constrained
according to this data, otherwise the space of all possible edge
permutations is randomly sampled.

It is suggested in \citet[p.~571]{clark22} that the reason that the
LOLOG does not suffer from the near-degeneracy problem that ERGM
models do, is that, unlike ERGM, the LOLOG simulation procedure
considers each dyad exactly once, limiting the scope for ``explosive''
edge formation (or dissolution, which does not occur at all in the
LOLOG simulation procedure) that can occur in ERGM.

\citet{clark22}, by re-analyzing a substantial set of networks first
analyzed with the ERGM, find that the LOLOG is generally in
qualitative agreement with the ERGM, and is often able to fit with
simpler terms that would result in near-degeneracy in ERGM. Further,
they find that the LOLOG tends to be easier (and faster) to fit.

As a recently developed model, there are far fewer published uses of
the LOLOG than there are of the well-established ERGM. However, as
well as the re-analyses of networks originally modelled with ERGMs in
\citet{clark22}, some applications to date include analysis of YouTube
video recommendations \citep{abul-fottouh20,gruzd23} and boilerplate
language in international trade agreements \citep{peacock19} and
environmental impact statements \citep{scott22}.

\section{Methods}

\subsection{Network data}

Seven biological networks are examined in this work. Four of them are
exactly those used in \citet{stivala21}: two undirected PPI networks,
and two directed gene regulatory networks. The remaining three
are neural networks.

The \textit{Saccharomyces cerevisiae} (yeast) PPI network
\citep{vonmering02}, HIPPIE human PPI network
\citep{schaefer12,schaefer13,suratanee14,alanislobato17}, Alon
\textit{E.~coli} regulatory network \citep{salgado01,shen02}, and
yeast regulatory network \citep{milo02,costanzo01} are exactly those
used in \citet{stivala21}, and their data processing is described in detail
there.

The first neural network is the whole-animal chemical connectome of
the adult male \textit{C.~elegans} worm \citep{cook19}.  This was
mentioned in \citet{stivala21} as a network for which a converged ERGM
could not be found.  The network was obtained from the
\textit{C.~elegans} connectome tables in MATLAB-ready format
(corrected July 2020) by Kamal Premaratne (University of Miami)
downloaded from
\url{https://wormwiring.org/matlab%20scripts/Premaratne%20MATLAB-ready%20files%20.zip}
(accessed 6 September, 2021).  The data was converted using the
R.matlab \citep{Rmatlab} and igraph \citep{csardi06} packages in R.

The second neural network is the hermaphrodite \textit{C.~elegans}
neural network of 277 neurons. This data, including the spatial
two-dimensional positions in the lateral plane (in mm) of the
neurons \citep{choe04,kaiser06}, and the birth times (in minutes) of
the neurons \citep{sulston77,sulston83,varier11} was downloaded from
\url{https://www.dynamic-connectome.org/?page_id=25} (accessed 31 May
2019).\footnote{This data is now available from
\url{https://sites.google.com/view/dynamicconnectomelab/resources}.}
There was no birth time data for the hermaphrodite-specific ventral
chord motor neuron VC6, which was excluded from the analysis in
\citet{varier11}, so this was imputed as the mean birth time of the
other VCn neurons.

The third neural network is the \textit{Drosophila} optic medulla
connectome \citep{takemura13}, obtaineed via the Open Connectome
database \citep{vogelstein18} at
\url{http://openconnecto.me/graph-services/download/} (accessed 6
November 2016).\footnote{This data is now available from
  \url{https://neurodata.io/data/takemura13/}.} Following
\citet{agarwala23}, three-dimensional geometric locations were
assigned to the neurons using the centroids of the synaptic
coordinates associated with them, and the subgraph induced by the
``named'' nodes, consisting of those neurons that have
alphanumeric labels, was extracted. As described in
\citet{takemura13,agarwala23}, the neighbourhoods of the 379 neurons
in the central neural column of the \textit{Drosophila} medulla were
traced, but not necessarily the connections between the other neurons,
and the 358 ``named'' neurons ``correspond roughly to the central 379
nodes'' \citep{agarwala23}. Unlike \citet{agarwala23}, however, I do
not treat the network as undirected, and do not remove the highest degree
node (this is necessary in \citet{agarwala23} as it violates a
condition of their model). I also do not consider the full network,
but only the network induced by the named nodes, since only these have
their connections fully traced.

For all networks, multiple edges and loops (an edge or arc from a node
to itself) are removed. In the case of the \textit{Drosophila}
medulla, this is another difference from \citet{agarwala23}, where
loops are retained.  In the \textit{E.~coli} network, following
\citet{hummel12,stivala21}, where the self-loops indicate
self-regulation, this is represented instead by a binary node
attribute ``self'', which is true when a self-loop was present, and
false otherwise.

Summary statistics of the networks, computed using the igraph
\citep{csardi06} R package, are shown in
Table~\ref{tab:summary_stats}.

\begin{table}[htb]
  \centering
  \caption{Network summary statistics.}
  \label{tab:summary_stats}
  {\small
    \begin{tabular}{llrrrrrr}
      \hline
      Network & Directed & N  &   Components &  Size of largest &Mean degree     &    Density & Clustering   \\
      &          &    &              &  component       &                &            &  coefficient  \\
      \hline
      Yeast PPI & N & 2617 & 92 & 2375 & 9.06 & 0.00346 & 0.46862\\
      Human PPI (HIPPIE) & N & 11517 & 93 & 11322 & 8.19 & 0.00071 & 0.03773\\
      Alon \textit{E.~coli} regulatory & Y & 423 & 34 & 328 & 2.45 & 0.00291 & 0.02382\\
      Alon yeast regulatory & Y & 688 & 11 & 662 & 3.14 & 0.00228 & 0.01625\\
      Cook \textit{C.~elegans} connectome & Y & 575 & 17 & 559 & 18.25 & 0.01589 & 0.25776\\
      Kaiser \textit{C.~elegans} neural & Y & 277 & 1 & 277 & 15.20 & 0.02753 & 0.19809\\
      \textit{Drosophila} medulla (named) & Y & 358 & 8 & 351 & 20.07 & 0.02811 & 0.17798\\
      \hline
    \end{tabular}
  }
  \floatfoot{``Clustering coefficient'' is the global clustering coefficient (transitivity).}
\end{table}

\begin{figure}
  \centering
  \includegraphics[width=\textwidth]{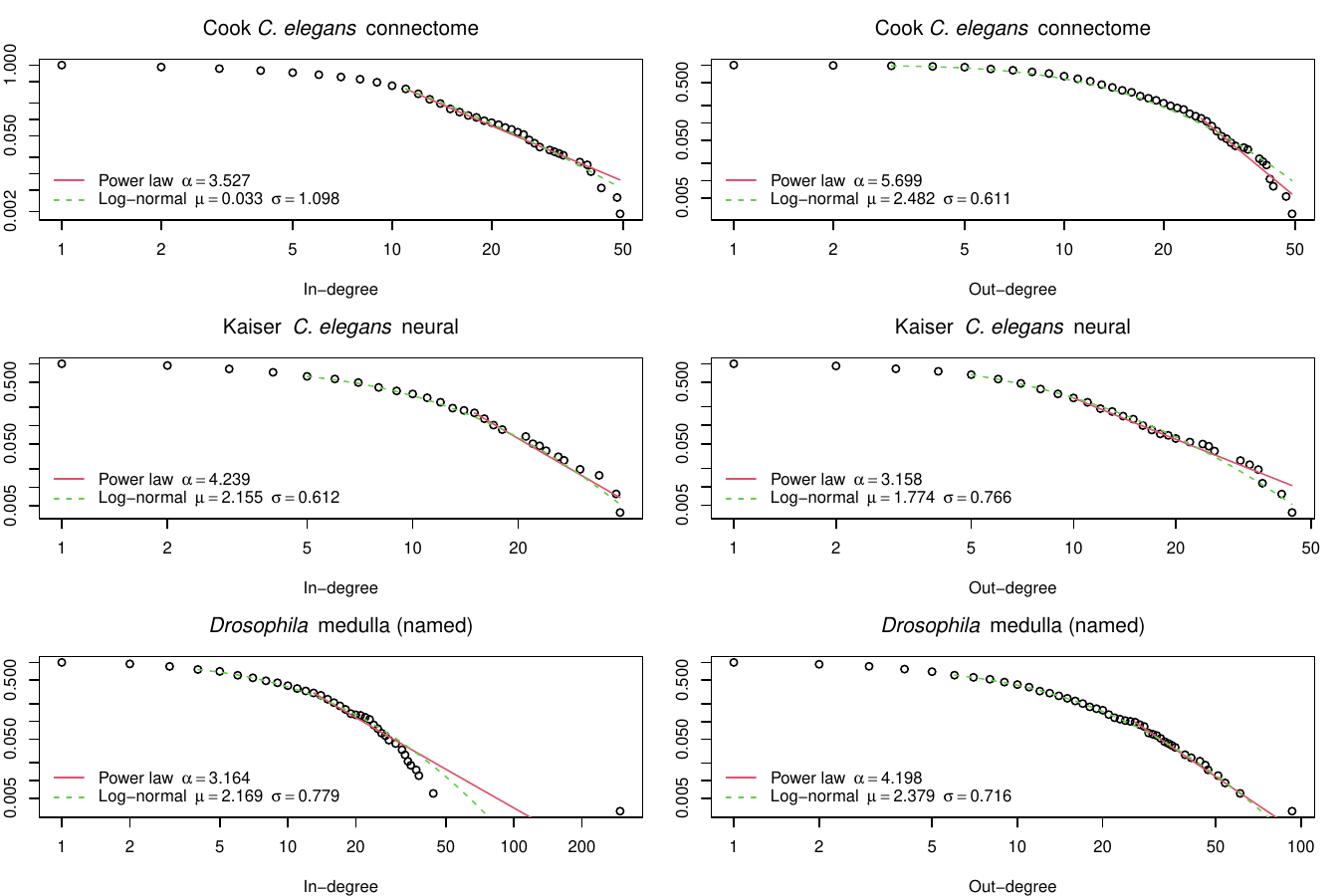}
  \caption{Discrete power law and log-normal distributions fitted to
    the empirical CDFs for in-degree (left) and out-degree (right)
    distributions of the neural networks.}
  \label{fig:degree_distributions}
\end{figure}

The in-degree and out-degree distributions of the neural networks
are shown in Figure~\ref{fig:degree_distributions} as plots of their
empirical cumulative distribution function (CDF). In this figure,
$\alpha$ is the exponent in the discrete power law distribution
$\Pr(X=x) = Cx^{-\alpha}$ (where $C$ is a normalization constant), and
$\mu$ and $\sigma$ are the parameters (respectively, mean and standard
deviation of $\log(x)$) of the discrete log-normal
distribution. Discrete power law and log-normal distributions were
fitted using the methods of \citet{clauset09} implemented in the
poweRlaw package \citep{gillespie15}. Degree distributions of the
other networks are shown in \citet[Fig.3]{stivala21}.

For the Cook \textit{C.~elegans} connectome, for both log-normal and
power law, and for both in-degree and out-degree distributions, the
null hypothesis that the tail of the empirical distribution ($x_{\min}
= 11$ for in-degree for both power law and log-normal, and for
out-degree, $x_{\min} = 26$ for power law and $x_{\min} = 3$ for
log-normal) is consistent with the log-normal or power law
distribution cannot be rejected at the conventional $p < 0.05$ level.

For the Kaiser \textit{C.~elegans} neural network, the null hypotheses
that the tails of the in-degree and out-degree distributions are
consistent with power law ($x_{\min} = 15$ for in-degree, $x_{\min} =
10$ for out-degree) or log-normal distributions ($x_{\min} = 5$ for
both in-degree and out-degree distributions) cannot be rejected at the
conventional $p < 0.05$ level.

For the \textit{Drosophila} optic medulla network, for the in-degree
distribution, the null hypothesis that the tail of the empirical CDF
($x_{\min} = 13$) is consistent with a power law distribution is
rejected ($p < 0.05$), while for the log-normal distribution
($x_{\min} = 4$) the null hypothesis is not rejected at the
conventional $p < 0.05$ level. For the out-degree distribution,
neither of the null hypotheses that the tail of the empirical CDF is
consistent with a power law distribution ($x_{\min} = 26$) or with a
log-normal distribution ($x_{\min} = 6$) can be rejected at the
conventional $p < 0.05$ level.

\subsection{Model estimation}

All estimations, unless otherwise noted, were run using R
\citep[version 4.0.2]{R-manual} packages on an Intel Xeon E5-2650 v3
2.30 GHz CPU on a Linux compute cluster node. Facilitating direct
comparisons of estimation time, this is the same cluster that was used
for ERGM estimations using the Estimnet \citep{byshkin16,byshkin18},
EstimNetDirected \citep{stivala20b}, and statnet
\citep{handcock08,statnet} ergm software in \citet{stivala21}. Note
that these programs implement stochastic (Monte Carlo) algorithms and
terminate when certain convergence criteria are met, and so can
demonstrate a large variance in runtime. Hence the reported times,
for a single run for each of the estimations shown here, are merely
indicative of the approximate computational requirements, and are not
necessarily representative.

For LOLOG estimations, the lolog R package \citep{lolog} was used, and
for Tapered ERGM estimations, the ergm.tapered R package
\citep{ergm.tapered} was used. The latter package builds on the
statnet \citep{handcock08,statnet} ergm package
\citep{ergm,krivitsky22,ergm4}. No parallel computing was used (each
estimation was run on a single core), and a 48 hour maximum time limit
was imposed by the cluster job management system.

For estimating Tapered ERGMs with the ergm.tapered package, tapering
for dependent parameters was estimated using kurtosis-penalized
likelihood, as described in \citet{blackburn23}. For geometrically
weighted LOLOG or ERGM terms, the decay parameter, denoted $\alpha$ in
models presented here, was fixed at values based on those used in
\citet{stivala21}, and adjusted if necessary by trial and error to
improve model convergence and fit, as estimating the decay parameter
is not implemented in LOLOG, and curved ERGM model estimation
\citep{hunter06,hunter07} resulted in estimation not converging within
the time limit. LOLOG models were estimated with default estimation
parameters, and model convergence was confirmed by inspecting the model
diagnostic plots (shown in \ref{sec:si_figures}).  Tapered ERGM
models were estimated with ergm.tapered with default estimation parameters,
using the ergm version 4 package \citep{ergm}  adaptive MCMC via effective sample
size feature to automatically adjust the required number of iterations
\citep{krivitsky22}. Model convergence was confirmed by inspecting the
MCMC diagnostic plots, and only converged non-degenerate models were
used.

\section{Results and discussion}

\subsection{Protein-protein interaction (PPI) networks}

Yeast PPI and human PPI (HIPPIE) LOLOG estimations, and Tapered ERGM
estimations of the HIPPIE network, did not converge within the 48 hour
maximum time limit. Hence the only PPI network model I was able to
estimate was a Tapered ERGM model for the yeast PPI network
(Table~\ref{tab:yeast_ppi_tapered_estimations}). This estimation took
approximately 9 hours.

\begin{table}[htb]
  \centering
  \caption{Tapered ERGM \citep{fellows17,blackburn23} parameter estimates
    (with estimated standard errors in parentheses)
    for the yeast PPI network. Estimated using the
    ergm.tapered package \citep{ergm.tapered}.}
  \label{tab:yeast_ppi_tapered_estimations}
  \begin{tabular}{l*{1}{D{)}{)}{11)3}}}
    \hline
    Effect  & \multicolumn{1}{c}{Model 1}\\
    \hline
    Edges  & -6.215 \; (0.013) ^{***}\\
    GW degree ($\alpha = 0.1$)  & -2.478 \; (0.125) ^{***}\\
    Two-paths  & 0.015 \; (0.002) ^{***}\\
    Triangles  & 0.100 \; (0.005) ^{***}\\
    \hline
    AIC  & 154767.00\\
    BIC  & 154832.00\\
    \hline
  \end{tabular}
  \floatfoot{*** $p < 0.001$; ** $p < 0.01$; * $p < 0.05$; $^{\boldsymbol{\cdot}} p < 0.1$.}
\end{table}

The model in Table~\ref{tab:yeast_ppi_tapered_estimations} shows that
triangles are over-represented in this network (the Triangles
parameter is positive and statistically significant). The
``geometrically weighted degree'' parameter is negative and
significant, indicating centralization of edges
\citep{hunter07,levy16,levy16poster}. These results are consistent
with the ERGM model of \citet[Table~4]{stivala21}, in which there are
a positive and significant estimated parameters for both the
``alternating $k$-triangle`` and ``alternating $k$-star'' effects.
Note that, confusingly \citep{levy16poster,martin20,stivala20d}, the
interpretation of the geometrically weighted degree (GW degree)
parameter defined in \citet{hunter07} and used in statnet (and hence
the tapered ERGM in Table~\ref{tab:yeast_ppi_tapered_estimations}) has
a different interpretation regarding the sign to that of the
``alternating $k$-star'' effect \citep{snijders06,robins07} used in
\citet{stivala21}.  A positive value of the alternating $k$-star
effect indicates centralization based on high-degree nodes, a
situation which is indicated by a negative value of the
geometrically-weighted degree parameter.

Goodness-of-fit plots for this model are shown in
Figure~\ref{fig:yeast_ppi_tapered_gof}, showing that the fit to the
triad census (and hence number of triangles) is excellent. However the
fit to the other network statistics, and in particular the geodesic
distance distribution, is poor.

Using the tapered ERGM, therefore, allowed estimation of a model more
directly testing motif significance (using a parameter for triangles
directly, rather than requiring alternating $k$-triangles or
``geometrically weighted edgewise shared partners''). However I was
not able to estimate a Tapered ERGM model for the larger
human PPI network, nor a LOLOG model for either of the PPI networks.
ERGM models for these networks could also not be estimated with
``standard'' ERGM estimation software, specifically PNet \citep{pnet},
MPNet \citep{mpnet14,mpnet22}, or the statnet ergm package
\citep{ergm,ergm4}. In contrast, ERGM models for both networks could
be estimated using the ``Equilibrium Expectation'' (EE) algorithm
\citep{byshkin18,borisenko20} as described in \citet{stivala21}.

It is also worth noting that, while the ``alternating $k$-two-paths''
parameter, when included, was found to be negative and significant in
\citet[Table~4]{stivala21}, in the different Tapered ERGM model (in
which simple Triangles and Two-paths are used, rather than
``alternating'' versions) shown in
Table~\ref{tab:yeast_ppi_tapered_estimations}, the Two-paths parameter is
found to be positive and significant. No interpretation of these
parameters is made, but they are included to ``control for'' the
triangle (or alternating $k$-triangle) parameters, since, as discussed
in \citet{stivala21}, these ERGM configurations are not \emph{induced}
subgraphs. That is, a triangle contains two-paths within it as a
subgraph (but not an induced subgraph), and so it is usual ERGM
modeling practice to include a term for two-paths when a term for
triangles is included \citep{koskinen13}.

One further point to note, is that in \citet{stivala21}, a nodal
attribute for the functional category (class) of the protein was used,
and a parameter to test for matching class included
\citep[Table~4]{stivala21}. However a significant number of proteins
are ``uncharacterized'' (have an NA value for class)
\cite[S.~I.]{vonmering02}, and NA values for nodal
attributes are not currently allowed for Tapered ERGM models in statnet;
some form of imputation or other technique for handling missing
data \citep{koskinen13a} would have to be used.

\subsection{Gene regulatory networks}

\begin{table}[htb]
  \caption{Latent order logistic (LOLOG) model \citep{fellows18} parameter estimates for the Alon \textit{E. coli}
    regulatory network. Estimated using the lolog package
    \citep{lolog}.}
  \label{tab:ecoli_lolog}
  \begin{tabular}{lrrr}
    \hline
    Effect & Estimate & Std. error & p-value\\
    \hline
    Edges & $-5.9596$ & $0.5955$ & $< 0.001$ \\
    GW in-degree ($\alpha = 0.2$) & $-2.7560$ & $0.1564$ & $< 0.001$ \\
    GW out-degree ($\alpha = 0.2$) & $1.5308$ & $0.2016$ & $< 0.001$ \\
    Two-paths & $-0.8845$ & $0.1564$ & $< 0.001$ \\
    Triangles & $2.9323$ & $0.2463$ & $< 0.001$ \\
    Nodecov self & $0.7223$ & $0.2084$ & $< 0.001$ \\
    Nodematch self & $-0.8923$ & $0.1420$ & $< 0.001$ \\
    \hline
  \end{tabular}
\end{table}

\begin{table}[htb]
  \centering
  \caption{Tapered ERGM \citep{fellows17,blackburn23} parameter estimates
    (with estimated standard errors in parentheses)
    for the Alon \textit{E. coli} regulatory network. Estimated using the
    ergm.tapered package \citep{ergm.tapered}.}
  \label{tab:ecoli_tapered_estimations}
  \begin{tabular}{l*{2}{D{)}{)}{11)3}}}
    \hline
    Effect  & \multicolumn{1}{c}{Model 1} & \multicolumn{1}{c}{Model 2}\\
    \hline
    Edges  & -3.062 \; (0.077) ^{***} & -2.716 \; (0.156) ^{***}\\
    GW in-degree ($\alpha = 2$)  & -3.979 \; (0.134) ^{***} & -3.831 \; (0.134) ^{***}\\
    GW out-degree ($\alpha = 0$)  & 1.570 \; (0.190) ^{***} & 1.479 \; (0.188) ^{***}\\
    Two-paths  & -0.495 \; (0.102) ^{***} & -0.501 \; (0.101) ^{***}\\
    Transitive triples  & 2.171 \; (0.234) ^{***} & 2.156 \; (0.177) ^{***}\\
    Nodecov self  &   & -0.230 \; (0.122) ^{\boldsymbol{\cdot}}\\
    Nodematch self  &   & -0.524 \; (0.139) ^{***}\\
    \hline
    AIC  & 5656.00 & 5635.00\\
    BIC  & 5717.00 & 5715.00\\
    \hline
  \end{tabular}
  \floatfoot{*** $p < 0.001$; ** $p < 0.01$; * $p < 0.05$; $^{\boldsymbol{\cdot}} p < 0.1$.}
\end{table}

A LOLOG model for the \text{E.~coli} gene regulatory network is shown
in Table~\ref{tab:ecoli_lolog}. This estimation took
approximately 20 minutes. Tapered ERGM models for the network are
shown in Table~\ref{tab:ecoli_tapered_estimations}. Estimation of
Model 1 and Model 2 took approximately 1 minute and 33 minutes,
respectively.  Consistent with the ERGM models for this network in
\citet{stivala21}, both the LOLOG and Tapered ERGM models indicate
centralization based on high-in-degree nodes, but not (indeed, a
tendency against such centralization) for out-degree, an
over-representation of transitive triangles, and a tendency against
self-regulating operons regulating other self-regulating operons
(negative and significant estimate for the ``Nodematch self''
parameter).

This is the only network examined here in which an ERGM was able to be
estimated with ``standard'' statnet ergm package
\citep[Table~S1]{stivala21}, rather than requiring the use of the EE
algorithm in EstimNetDirected. The LOLOG and Tapered ERGM models shown
here were able to be estimated with the Triangles or Transitive
triples, respectively, parameter, rather than requiring the
alternating $k$-triangles or geometrically weighted edgewise shared
partners (GWESP) parameter used in \cite{stivala21}, however. The
LOLOG and statnet (and hence the Tapered ERGM), packages, however, do
not allowing the modelling of loops (self-edges), and hence the nodal
covariate ``self'' is used here, as in \citet{hummel12} to model
self-regulation in a simplistic way. In contrast, EstimNetDirected can
model networks containing loops, and it was found that loops are
over-represented in this network \citep{stivala21}.

Goodness-of-fit plots for the LOLOG (Table~\ref{tab:ecoli_lolog}) and
Tapered ERGM (Table~\ref{tab:ecoli_tapered_estimations}) are shown
Figure~\ref{fig:ecoli_lolog_gof} and
Figure~\ref{fig:ecoli_tapered_gof}, respectively, showing all models
fit the data well.

\begin{table}[htb]
  \caption{Latent order logistic (LOLOG) model \citep{fellows18} parameter estimates for the Alon yeast
    regulatory network. Estimated using the lolog package
    \citep{lolog}.}
  \label{tab:alon_yeast_transcription_lolog}
  \begin{tabular}{lrrr}
    \hline
    Effect & Estimate & Std. error & p-value\\
    \hline
    Edges & $-4.1067$ & $0.0777$ & $< 0.001$ \\
    GW in-degree ($\alpha = 0.2$) & $1.0386$ & $0.1048$ & $< 0.001$ \\
    GW out-degree ($\alpha = 0.2$) & $-3.9870$ & $0.1178$ & $< 0.001$ \\
    Two-paths & $-0.7062$ & $0.0656$ & $< 0.001$ \\
    Triangles & $2.6633$ & $0.1659$ & $< 0.001$ \\
    \hline
  \end{tabular}
\end{table}

\begin{table}[htb]
  \centering
  \caption{Tapered ERGM \citep{fellows17,blackburn23} parameter estimates
    (with estimated standard errors in parentheses)
    for the Alon yeast regulatory network. Estimated using the
    ergm.tapered package \citep{ergm.tapered}.}
  \label{tab:yeast_transcription_tapered_estimations}
  \begin{tabular}{l*{1}{D{)}{)}{11)3}}}
    \hline
    Effect  & \multicolumn{1}{c}{Model 1}\\
    \hline
    Edges  & -3.858 \; (0.067) ^{***}\\
    Reciprocity  & -4.707 \; (2.747) ^{\boldsymbol{\cdot}}\\
    GW in-degree ($\alpha = 0.5$)  & 1.112 \; (0.170) ^{***}\\
    GW out-degree ($\alpha = 1.5$)  & -4.540 \; (0.108) ^{***}\\
    Two-paths  & -0.291 \; (0.065) ^{***}\\
    Transitive triples  & 1.815 \; (0.145) ^{***}\\
    \hline
    AIC  & 11964.00\\
    BIC  & 12041.00\\
    \hline
  \end{tabular}
  \floatfoot{*** $p < 0.001$; ** $p < 0.01$; * $p < 0.05$; $^{\boldsymbol{\cdot}} p < 0.1$.}
\end{table}

A LOLOG model for the yeast regulatory network is shown in
Table~\ref{tab:alon_yeast_transcription_lolog}. This estimation took
approximately 36 minutes. The Tapered ERGM estimation of this network
shown in Table~\ref{tab:yeast_transcription_tapered_estimations} took
approximately 5 minutes. Consistent with the results in
\citet[Table~7]{stivala21}, these models show centralization on
out-degree, but not in-degree, and an over-representation of (transitive)
triangles.

Goodness-of-fit plots for the LOLOG model
(Table~\ref{tab:alon_yeast_transcription_lolog}) and Tapered ERGM
model (Table~\ref{tab:yeast_transcription_tapered_estimations}) are
shown in Figure~\ref{fig:yeast_transcription_lolog_gof} and
Figure~\ref{fig:yeast_transcription_tapered_gof}, respectively.  These
figures show a good fit to the data, with the exception of the
reciprocity (mutual) statistic for the tapered ERGM model
(Table~\ref{tab:yeast_transcription_tapered_estimations},
Fig.~\ref{fig:yeast_transcription_tapered_gof}).  As discussed in
\citet{stivala21}, there is only a single reciprocated arc in this
network, and hence ERGM models that do not include the Reciprocity
parameter can show poor goodness-of-fit on statistics involving
reciprocated arcs, while models that do include this parameter can
exhibit poor convergence.  I also found this to be the case with the
Tapered ERGM, with the model shown in
Table~\ref{tab:yeast_transcription_tapered_estimations} showing good
convergence diagnostics, while a model (not shown) without the
Reciprocity (mutual) parameter does not. This issue does not affect
the LOLOG model with no Reciprocity (mutual) parameter
(Table~\ref{tab:alon_yeast_transcription_lolog}), however, where the
goodness-of-fit for the mutual statistic is good
(Fig.~\ref{fig:yeast_transcription_lolog_gof}).

\subsection{Neural networks}

\begin{table}[htb]
  \caption{Latent order logistic (LOLOG) model \citep{fellows18}
    parameter estimates for the Cook \textit{C.~elegans} connectome.
    Estimated using the lolog package \citep{lolog}.}
  \label{tab:celegans_neural_lolog}
  \begin{tabular}{lrrr}
    \hline
    Effect & Estimate & Std. error & p-value\\
    \hline
    Edges & $-2.8910$ & $0.1179$ & $< 0.001$ \\
    GW in-degree ($\alpha = 0.1$) & $0.3020$ & $0.1874$ & $0.1071$ \\
    GW out-degree ($\alpha = 0.1$) & $-3.9835$ & $0.2112$ & $< 0.001$ \\
    Two-paths & $-0.4339$ & $0.0317$ & $< 0.001$ \\
    Triangles & $3.6632$ & $0.2081$ & $< 0.001$ \\
    \hline
  \end{tabular}
\end{table}

\begin{table}[htb]
  \caption{Tapered ERGM \citep{fellows17,blackburn23} parameter estimates
    (with estimated standard errors in parentheses)
    for the Cook \textit{C.~elegans} connectome. Estimated using the
    ergm.tapered package \citep{ergm.tapered}.}
  \label{tab:celegans_tapered_estimations}
  \begin{tabular}{l*{2}{D{)}{)}{11)3}}}
    \hline
    Effect  & \multicolumn{1}{c}{Model 1} & \multicolumn{1}{c}{Model 2}\\
    \hline
    Edges  & -3.860 \; (0.027) ^{***} & -3.747 \; (0.029) ^{***}\\
    GW in-degree ($\alpha = 0.2$)  & -4.458 \; (0.184) ^{***} & -4.499 \; (0.189) ^{***}\\
    GW out-degree ($\alpha = 0.2$)  & -9.269 \; (0.164) ^{***} & -9.381 \; (0.164) ^{***}\\
    Two-paths  & 0.001 \; (0.003) ^{} &  \\
    GWDSP ($\alpha = 1$)  &   & -0.006 \; (0.005) ^{}\\
    Transitive triples  & 0.101 \; (0.005) ^{***} & 0.102 \; (0.008) ^{***}\\
    \hline
    AIC  & 50449.00 & 50515.00\\
    BIC  & 50514.00 & 50580.00\\
    \hline
  \end{tabular}
  \floatfoot{*** $p < 0.001$; ** $p < 0.01$; * $p < 0.05$; $^{\boldsymbol{\cdot}} p < 0.1$.}
\end{table}

A LOLOG model for the Cook \textit{C.~elegans} connectome is shown in
Table~\ref{tab:celegans_neural_lolog}. This estimation
took approximately 3 hours.
Tapered ERGM models for this network are shown in
Table~\ref{tab:celegans_tapered_estimations}.  Estimation of Model 1
and Model 2 took approximately 3 hours, and 7 hours, respectively.

These models both show centralization based on high out-degree nodes,
and an over-representation of (transitive) triangles. The Tapered ERGM
models also have a negative and statistically significant GW in-degree
parameter, indicating centralization also for the in-degree
distribution, but this parameter is not statistically significant in the
LOLOG model. Conversely, the LOLOG model has a negative and
statistically significant Two-paths parameter, but neither Two-paths
(Model 1) nor geometrically-weighted dyadwise shared partners (GWDSP,
Model 2) are significant in the Tapered ERGM model
(Table~\ref{tab:celegans_tapered_estimations}).

Goodness-of-fit plots for the LOLOG model
(Table~\ref{tab:celegans_neural_lolog}) and Tapered ERGM models
(Table~\ref{tab:celegans_tapered_estimations}) are shown in
Figure~\ref{fig:celegans_neural_lolog_gof} and
Figure~\ref{fig:celegans_tapered_gof}, respectively. The LOLOG model
has a poor fit on reciprocity (mutual) and the in-degree distribution,
but a better fit on the out-degree distribution and edgewise shared
partners. I also estimated a model (not shown) including the mutual
(Reciprocity) parameter, however this model showed poor convergence on
the diagnostic plots. The Tapered ERGM models show not very good fit
on the degree distributions, and bad fit on the dyadwise and edgewise
shared partner and geodesic distance distributions, but reasonable fit
on the triad census.

As discussed in \citet{stivala21}, a converged ERGM could not be found
for this network using statnet (ergm) or EstimNetDirected, and hence
this is an example where the LOLOG and Tapered ERGM can both estimate
models that could not be estimated with ``standard'' ERGMs. This
network, as well as the Kaiser \textit{C.~elegans} and
\textit{Drosophila} medulla neural networks, for which I also could
not find good ERGM models with statnet (ergm) or EstimNetDirected,
although not larger than the other, non-neural, networks, have far
higher mean degree and density (Table~\ref{tab:summary_stats}), which
might be one reason contributing to my inability to find a
converged model for such networks with other ERGM estimation
algorithms.

\begin{table}[htb]
  \centering
  \caption{Latent order logistic (LOLOG) model \citep{fellows18}
    parameter estimates for the Kaiser \textit{C.~elegans} neural
    network. Model 1 has node inclusion order specified by birth time,
    while Model 2 does not. Estimated using the lolog package
    \citep{lolog}.}
  \label{tab:celegans277_neural_lolog}
  \begin{tabular}{l*{2}{D{)}{)}{14)3}}}
    \hline
    Effect  & \multicolumn{1}{c}{Model 1} & \multicolumn{1}{c}{Model 2}\\
    \hline
    Edges  & -2.8248 \; (0.1103) ^{***} & -2.7212 \; (0.1276) ^{***}\\
    Reciprocity  & -0.4218 \; (0.6775) ^{} & -0.6095 \; (0.7846) ^{}\\
    GW in-degree ($\alpha = 0.2$)  & -0.9906 \; (0.1547) ^{***} & -0.9640 \; (0.1576) ^{***}\\
    GW out-degree ($\alpha = 0.2$)  & -0.7592 \; (0.1628) ^{***} & -0.7105 \; (0.1408) ^{***}\\
    Two-paths  & -0.0798 \; (0.0198) ^{***} & -0.0916 \; (0.0154) ^{***}\\
    Triangles  & 1.1689 \; (0.2579) ^{***} & 1.2790 \; (0.2533) ^{***}\\
    Euclidean distance  & -1.0682 \; (0.1152) ^{***} & -0.8780 \; (0.1040) ^{***}\\
    Nodecov birthtime  &   & -0.0001 \; (0.0000) ^{**}\\
    \hline
  \end{tabular}
  \floatfoot{*** $p < 0.001$; ** $p < 0.01$; * $p < 0.05$.}  
\end{table}

Table~\ref{tab:celegans277_neural_lolog} shows LOLOG models for the
Kaiser \textit{C.~elegans} neural network. These estimations took
approximately 54 minutes and 11 minutes for Model 1 and Model 2
respectively, on a Lenovo PC with an Intel Core i5-10400 2.90 GHz CPU
under Windows with cygwin (R version 4.2.1, lolog version
1.3).\footnote{This estimation was run on a different system from the
others due to the compute cluster becoming unexpectedly, and
regrettably, unavailable during the course of the work described
here.}  Model diagnostic and goodness-of-fit plots for Model 1 and
Model 2 (Table~\ref{tab:celegans277_neural_lolog}) are shown in
Figure~\ref{fig:celegans277_neural_lolog_gof1} and
Figure~\ref{fig:celegans277_neural_lolog_gof2}, respectively. These
show the models fit will on the model parameters, as well as in-degree
and out-degree distribution and edgewise shared partners distribution
(although the fit for small values on the latter is not very good).
I also estimated models without the Reciprocity (mutual) parameter,
and the results were consistent, including a good fit on the
distribution of reciprocal edges --- note that the Reciprocity
parameter in the models in Table~\ref{tab:celegans277_neural_lolog} is
not statistically significant.

These models show centralization on both in-degree and out-degree, and
an over-representation of triangles, as well as a statistically
significant negative estimate for the Euclidean distance parameter,
meaning that spatially closer neurons are more likely to be connected.
The centralization on degree and over-representation of triangles is
consistent with the findings reported by \citet{varshney11}, for a
different version of the data, by fitting power law
distributions to the degree distributions (just as in
Figure~\ref{fig:degree_distributions} for the data used here), and
using a conventional motif-finding method based on randomizing the
network while preserving the degree of each node \citep{varshney11}.

The finding that triangles are over-represented in the
\textit{C.~elegans} neural network was also described in the
well-known ``Network motifs'' paper \citep{milo02}, and, as noted
by\citet{varshney11}, was already reported by \citet{white86}. No
spatial data was used in \citet{varshney11}, and they state that such
over-representation of the triangle motif ``would arise naturally if
proximity was a limiting factor for connectivity, however there is no
evidence for this limitation'' \citep[p.~12]{varshney11}. However, as
noted by \citet{artzy04}, it was already shown by \citet{white86} that
neighbouring neurons are more likely than distant neurons to form a
connection --- and this spatial feature of clustering is not accounted
for in conventional motif finding methods, as used in
\citet{milo02,varshney11}, where only degree distribution is preserved
in the randomized networks constituting the null model. So, although
spatial proximity is not necessarily a ``limiting factor'' for
connectivity, since long distance connections do exist, and indeed in
most cases can do so because the connections were formed early on in
development when the neurons were spatially closer \citep{varier11},
it is certainly a relevant factor that should be accounted for in
determining the statistical significance of the triangle motif
\citep{artzy04}. The LOLOG model (as does ERGM) allows for the
inclusion of both the triangle structure and spatial proximity (as
well as other relevant and non-independent features, including terms to
model degree distribution), and as shown by the models in
Table~\ref{tab:celegans277_neural_lolog}, both the triangle structure
and spatial proximity of connected neurons are statistically
significant features of this network. That is, even accounting
for the spatial proximity effect (and all other effects in the model),
there is significant over-representation of the triangle structure.

Although this model shows that increasing Euclidean distance between
two neurons makes them less likely to be directly connected, this does
not necessarily imply that the network has grown in such a way as to
minimize the length of connections. \citet{kaiser06} find that neural
networks (in examples including \textit{C.~elegnas}) seem to minimize
geodesic distance of processing paths (processing steps), rather than
``wiring length''. The non-optimal wiring length arises from
long-distance connections, suggesting a trade-off between wiring
length and average number of processing steps \citep{kaiser06}.

Model 1 (Table~\ref{tab:celegans277_neural_lolog}) was estimated with
the node inclusion order specified in the LOLOG model as the neuronal
birth times, while Model 2 does not specify the inclusion order, but
instead includes the birth times as a nodal covariate. The results for
both models are consistent, but Model 2 shows a negative and
statistically significant estimate of the birth time covariate,
indicating that older neurons tend to have more connections.  This is
consistent with the finding in \citet{varier11} that early-born
neurons are more highly connected than later-born neurons.

I was unable to find a converged Tapered ERGM estimate, using simple
model terms such as those used for the LOLOG model
(Table~\ref{tab:celegans277_neural_lolog}), for the Kaiser
\textit{C.~elegans} neural network.

\begin{table}[htb]
  \centering
  \caption{Latent order logistic (LOLOG) model \citep{fellows18}
    parameter estimates for the \textit{Drosophila} medulla
    network. Estimated using the lolog package \citep{lolog}.}
  \label{tab:drosophila_medulla_lolog}
  \begin{tabular}{lrrr}
    \hline
    Effect & Estimate & Std. error & p-value\\
    \hline
    Edges & $-1.4418$ & $0.1173$ & $< 0.001$ \\
    Reciprocity & $-1.2554$ & $0.8765$ & $0.1521$ \\
    GW in-degree ($\alpha = 0.2$) & $-0.8541$ & $0.1597$ & $< 0.001$ \\
    GW out-degree ($\alpha = 0.2$) & $-0.9857$ & $0.1320$ & $< 0.001$ \\
    Two-paths & $-0.1128$ & $0.0164$ & $< 0.001$ \\
    Triangles & $1.2676$ & $0.2523$ & $< 0.001$ \\
    Euclidean distance & $-0.0012$ & $0.0001$ & $< 0.001$ \\
    \hline
  \end{tabular}
\end{table}

\begin{table}[htb]
  \centering
  \caption{Tapered ERGM \citep{fellows17,blackburn23} parameter estimates
    for the \textit{Drosophila} medulla network. Estimated using the
    ergm.tapered package \citep{ergm.tapered}.}
  \label{tab:drosophila_medulla_tapered_estimations}
  \begin{tabular}{l*{1}{D{)}{)}{11)3}}}
    \hline
    Effect  & \multicolumn{1}{c}{Model 1}\\
    \hline
    Edges  & -2.579 \; (0.062) ^{***}\\
    Reciprocity  & 1.406 \; (0.095) ^{***}\\
    GW in-degree ($\alpha = 0.2$)  & -3.384 \; (0.025) ^{***}\\
    GW out-degree ($\alpha = 0.2$)  & -3.733 \; (0.027) ^{***}\\
    Two-paths  & -0.000 \; (0.005) ^{}\\
    Transitive triples  & 0.073 \; (0.011) ^{***}\\
    Euclidean distance  & -0.001 \; (0.000) ^{***}\\
    \hline
    AIC  & 29150.00\\
    BIC  & 29229.00\\
    \hline
  \end{tabular}
  \floatfoot{*** $p < 0.001$; ** $p < 0.01$; * $p < 0.05$; $^{\boldsymbol{\cdot}} p < 0.1$.}
\end{table}

Table~\ref{tab:drosophila_medulla_lolog} shows a LOLOG model for the
\textit{Drosophila} optic medulla network. This estimation took
approximately 1.3 hours on a Lenovo PC with an Intel Core i5-10400
2.90 GHz CPU under Windows with cygwin (R version 4.2.1, lolog version
1.3). A Tapered ERGM model for the \textit{Drosophila} optic medulla
network is shown in
Table~\ref{tab:drosophila_medulla_tapered_estimations}.  This
estimation took approximately 6 minutes (on the same cluster system as
the other estimations). The diagnostic and goodness-of-fit plots in
Figure~\ref{fig:drosophila_medulla_lolog_gof} show that the LOLOG
model fits well, and the goodness-of-fit plots in
Figure~\ref{fig:drosophila_medulla_tapered_gof} show that the Tapered
ERGM fits well, aside from the geodesic distance distribution, and
some of the triad census (for example 300, three mutually connected
nodes).

Both the LOLOG model (Table~\ref{tab:drosophila_medulla_lolog}) and
the Tapered ERGM model
(Table~\ref{tab:drosophila_medulla_tapered_estimations}) show
centralization on both in-degree and out-degree (as we might expect
from the power law and log-normal distributions fit to the degree
distributions in Figure~\ref{fig:degree_distributions}), and an
over-representation of (transitive) triangles, as well as a negative
and statistically significant estimate of the Euclidean distance
parameter, indicating that spatially closer neurons are more likely to
be connected.

That spatial distance between nodes is a contributor to the structure
of neural networks (closer neurons are more likely to be connected) is
expected. As well as the previous findings in the context of the
\textit{C.~elegans} neural network discussed above, it was found by
\citet{stillman17} using generalized ERGM \citep{desmarais12} models
of the human Default Mode Network, and previously noted by
\citet{vertes12,roberts16} for the human brain, and
\citet{betzel17,allard20} more generally, including for
\textit{Drosophila} connectomes, and is noted in general in a review
of trade-offs in brain network structure \citep{bullmore12}. As
discussed previously for the \textit{C.~elegans} neural network, that
the (transitive) triangles parameter is positive and significant in a
model that also includes Euclidean distance between nodes is evidence that the
transitive triangle structure is over-represented, even when
accounting for the spatial proximity effect.

\begin{table}[htb]
  \centering
  \caption{Summary of model fits}
  \label{tab:model_summary}
  \begin{tabular}{llp{0.25\linewidth}p{0.25\linewidth}}
    \hline
    Network                             & ERGM        & Tapered ERGM                                  &  LOLOG \\
    \hline
    Yeast PPI                           & EE only     & Only fits triad census well                   & No     \\ 
    Human PPI (HIPPIE)                  & EE only     & No                                            & No     \\
    Alon \textit{E.~coli} regulatory    & EE, statnet & Good                                          & Good   \\
    Alon yeast regulatory               & EE only     & Good, except reciprocity                      & Good   \\
    Cook \textit{C.~elegans} connectome & No          & Only fits triad census well                   & Poor fit on reciprocity and in-degree distribution \\
    Kaiser \textit{C.~elegans} neural   & No          & No (with simple triangle terms)               & Good \\
    \textit{Drosophila} medulla (named) & No          & Poor on geodesic distance distribution and some triads & Good \\
    \hline
  \end{tabular}
  \floatfoot{``No'' means a non-degenerate model was unable to be
    estimated. ``EE'' is the Equilibrium Expectation algorithm
    implemented in the Estimnet or EstimNetDirected software; models
    estimated with the EE algorithm or statnet are described in
    \citet{stivala21}, and use the ``alternating'' or ``geometrically
    weighted'' model terms for triangles to avoid degeneracy. Note
    that the LOLOG goodness-of-fit does not include geodesic distance
    distribution or triad census.}
\end{table}

Table~\ref{tab:model_summary} summarizes results described here for
the ability to find non-degenerate models, and their goodness-of-fit,
for ``standard'' ERGMs, Tapered ERGMs, and LOLOGs. These results show
that, with the exception of the HIPPIE network (and the Yeast PPI
network for LOLOG and \textit{C.~elegans} neural network for Tapered
ERGM), the Tapered ERGM and LOLOG are able to estimate models for
biological networks that could not be estimated using standard ERGMs
with either statnet or the EstimNetDirected software. Further, the
Tapered ERGM and LOLOG are able to estimate models with simple terms,
such as triangle, while ``alternating'' \citep{snijders06,robins07}
or ``geometrically weighted'' \citep{hunter07} terms are required to
avoid near-degeneracy in estimating these models using standard ERGMs.

\section{Conclusions and future work}

I was able to estimate both Tapered ERGM and LOLOG models for
biological networks for which I was unable to estimate ``standard''
ERGM models.  Furthermore, these models can be estimated with simple
model terms such as the triangle, without having to consider the
problem of near-degeneracy which frequently occurs using such
parameters in standard ERGMs, necessitating the use of ``geometrically
weighted'' or ``alternating'' parameters instead. As discussed in
\citet{clark22} for LOLOGs, and \citet{blackburn23} for Tapered ERGMs,
this simplifies both the model and the modeling procedure, and allows
the use of a wider set of statistics, without being constrained by
problems with near-degeneracy. In the context of biological networks
as considered in this work, it overcomes the limitation discussed in
\citet{stivala21}, that the use of more complex statistics such as
``alternating $k$-stars'' or ``geometrically weighted edgewise shared
partners'' (GWESP) is a further step away from directly testing for
over-representation of the motif. (Note that even using the triangle
term is not quite the same as a motif as usually defined, since motifs
are usually considered to be induced subgraphs, and ERGM
configurations are usually subgraphs, not induced subgraphs, although
they can also be defined to be induced subgraphs, as for example in
\citet{yaveroglu15}).

For the networks considered here, the LOLOG and Tapered ERGMs have
consistent interpretations, and, where ``standard'' ERGMs were able to
be estimated, their interpretations are also consistent with the LOLOG
and Tapered ERGM models estimated in this work.

For networks where spatial information is available --- the
\textit{C.~elegans} neural network and the \textit{Drosophxila}
medulla network --- the ability to estimate a model
with both spatial information and triangle terms overcomes the problem
with standard motif significance testing methods that spatial
aggregation is not accounted for in determining the
over-representation of triangles \citep{artzy04}.  Further, using the
ERGM (including Tapered ERGM) or LOLOG, we do not have to make the
assumption of the geometric Chung-Lu model in \citet{agarwala23} that
the ``intensities'' (expected degrees) of nodes and the distances
between them are independent --- an assumption that has biological
evidence contradicting it, as noted by \citet{agarwala23}.

A further advantage of the LOLOG is its ability to account for edge
orderings \citep{fellows18,clark22}. This was used in
\citet{fellows18} to model the order in which nodes join a social
network (as they have a joining date), and although only used in the
one example for which relevant information was available in
\citet{clark22}, it was suggested that citation networks are another
natural use. In this work, I was able to use the birth times of
neurons in the \textit{C.~elegans} neural network as the ordering
variable in a LOLOG model, although it did not result in any clear
improvement in model fit relative to using the birth time simply as a
nodal covariate.

Currently, the LOLOG package has a rather limited number of model
terms implemented, although there is a facility to define new model
terms \citep{lolog}. In contrast, an advantage of the Tapered ERGM is
that its implementation builds on the statnet ergm package, and hence
can use any of the (dauntingly) many model terms available in
the ergm package \citep{ergm}, as well as allowing user-defined terms
\citep{hunter13}, and automatically takes advantage of computational
improvements made in that software \citep{krivitsky22}.

For larger networks, standard estimation algorithms such as the
MCMCMLE algorithm used in the statnet ergm package or the Monte Carlo
method of moments approach used in the LOLOG package may still not be
able to estimate models in practical time.  Specifically, I was only
able to estimate a Tapered ERGM model for the smaller of the two PPI
networks (both of which are considerably larger than the other
networks considered in this work), and a LOLOG for neither of them. In
contrast, ERGM models were able to be estimated for both, using the
Equilibrium Expectation (EE) algorithm \citep{stivala21}.  This
suggests that a fruitful line of future work could be to investigate
the applicability of this algorithm to the LOLOG and the Tapered ERGM
models. Another future application could be to implement the
``tapering'' algorithm \citep{fellows17,clark22} for the autologistic
actor attribute model (ALAAM), a variant of the ERGM for modeling
social influence \citep{robins01b,daraganova13,koskinen22,parker22},
which, like the ERGM, can suffer from problems of near-degeneracy,
which currently can only be overcome by using ``geometrically
weighted'' model terms \citep{stivala23}, as implemented in the
ALAAMEE software \citep{ALAAMEE}.

The development of more scalable estimation algorithms (such as the EE
algorithm) for LOLOG seems a particularly useful line of
investigation, since very large networks can have ERGM models
estimated using the EE algorithm, however such networks can be too
large for the simulation-based goodness-of-fit procedure usually used
for ERGMs, due to the computational difficulty of the MCMC procedure
for ERGM simulation \citep{stivala20b}, a limitation which is yet to
be overcome.  In contrast, a ``key advantage of the LOLOG model is the
ease of simulation from the model'' \citep[p.~570]{clark22}; it
requires only a draw from the distribution of edge orderings and
sequential logistic regression on the change statistics
\citep{fellows18}. This suggests that, unlike ERGMs, this simulation
procedure would be practical even on very large networks, which are
too large for LOLOG models to be estimated in practical time with the
algorithms currently implemented.

\section*{Funding}

This work was funded by the Swiss National Science Foundation (SNSF) project number 200778.

\section*{Acknowledgements}

I used the high performance computing cluster at the Institute of
Computing, Universit\`a della Svizzera italiana, for data
processing and computations.

\section*{Data availability statement}

Source code, scripts, configuration files, and datasets are freely
available from
\url{https://github.com/stivalaa/bionetworks_estimations}.

\newpage


\begin{thebibliography}{113}
\providecommand{\natexlab}[1]{#1}
\providecommand{\url}[1]{\texttt{#1}}
\expandafter\ifx\csname urlstyle\endcsname\relax
  \providecommand{\doi}[1]{doi: #1}\else
  \providecommand{\doi}{doi: \begingroup \urlstyle{rm}\Url}\fi

\bibitem[Abul-Fottouh et~al.(2020)Abul-Fottouh, Song, and
  Gruzd]{abul-fottouh20}
D.~Abul-Fottouh, M.~Y. Song, and A.~Gruzd.
\newblock Examining algorithmic biases in {YouTube}’s recommendations of
  vaccine videos.
\newblock \emph{International Journal of Medical Informatics}, 140:\penalty0
  104175, 2020.
\newblock \doi{https://doi.org/10.1016/j.ijmedinf.2020.104175}.

\bibitem[Agarwala and Kenter(2023)]{agarwala23}
S.~Agarwala and F.~Kenter.
\newblock {A geometric Chung–Lu model and the \textit{Drosophila} medulla
  connectome}.
\newblock \emph{Journal of Complex Networks}, 11\penalty0 (3):\penalty0
  cnad010, 2023.
\newblock \doi{10.1093/comnet/cnad010}.

\bibitem[Alanis-Lobato et~al.(2016)Alanis-Lobato, Andrade-Navarro, and
  Schaefer]{alanislobato17}
G.~Alanis-Lobato, M.~A. Andrade-Navarro, and M.~H. Schaefer.
\newblock {{HIPPIE v2.0}: enhancing meaningfulness and reliability of
  protein–protein interaction networks}.
\newblock \emph{Nucleic Acids Research}, 45\penalty0 (D1):\penalty0 D408--D414,
  10 2016.

\bibitem[Allard and Serrano(2020)]{allard20}
A.~Allard and M.~{\'A}. Serrano.
\newblock Navigable maps of structural brain networks across species.
\newblock \emph{PLoS Computational Biology}, 16\penalty0 (2):\penalty0
  e1007584, 2020.

\bibitem[Artzy-Randrup et~al.(2004)Artzy-Randrup, Fleishman, Ben-Tal, and
  Stone]{artzy04}
Y.~Artzy-Randrup, S.~J. Fleishman, N.~Ben-Tal, and L.~Stone.
\newblock Comment on ``network motifs: simple building blocks of complex
  networks'' and ``superfamilies of evolved and designed networks''.
\newblock \emph{Science}, 305\penalty0 (5687):\penalty0 1107--1107, 2004.

\bibitem[Bengtsson(2018)]{Rmatlab}
H.~Bengtsson.
\newblock \emph{R.matlab: Read and Write MAT Files and Call MATLAB from Within
  R}, 2018.
\newblock URL \url{https://CRAN.R-project.org/package=R.matlab}.
\newblock R package version 3.6.2.

\bibitem[B{\'e}nichou et~al.(2023)B{\'e}nichou, Masson, and
  Vestergaard]{benichou23}
A.~B{\'e}nichou, J.-B. Masson, and C.~L. Vestergaard.
\newblock Compression-based inference of network motif sets.
\newblock \emph{arXiv preprint arXiv:2311.16308}, 2023.

\bibitem[Betzel and Bassett(2017)]{betzel17}
R.~F. Betzel and D.~S. Bassett.
\newblock Generative models for network neuroscience: prospects and promise.
\newblock \emph{Journal of The Royal Society Interface}, 14\penalty0
  (136):\penalty0 20170623, 2017.

\bibitem[Blackburn and Handcock(2023)]{blackburn23}
B.~Blackburn and M.~S. Handcock.
\newblock Practical network modeling via tapered exponential-family random
  graph models.
\newblock \emph{Journal of Computational and Graphical Statistics}, 32\penalty0
  (2):\penalty0 388--401, 2023.

\bibitem[Borisenko et~al.(2020)Borisenko, Byshkin, and Lomi]{borisenko20}
A.~Borisenko, M.~Byshkin, and A.~Lomi.
\newblock A simple algorithm for scalable {Monte Carlo} inference.
\newblock \emph{arXiv preprint arXiv:1901.00533v4}, 2020.

\bibitem[Bullmore and Sporns(2012)]{bullmore12}
E.~Bullmore and O.~Sporns.
\newblock The economy of brain network organization.
\newblock \emph{Nature Reviews Neuroscience}, 13\penalty0 (5):\penalty0
  336--349, 2012.

\bibitem[Byshkin et~al.(2016)Byshkin, Stivala, Mira, Krause, Robins, and
  Lomi]{byshkin16}
M.~Byshkin, A.~Stivala, A.~Mira, R.~Krause, G.~Robins, and A.~Lomi.
\newblock Auxiliary parameter {MCMC} for exponential random graph models.
\newblock \emph{Journal of Statistical Physics}, 165\penalty0 (4):\penalty0
  740--754, 2016.

\bibitem[Byshkin et~al.(2018)Byshkin, Stivala, Mira, Robins, and
  Lomi]{byshkin18}
M.~Byshkin, A.~Stivala, A.~Mira, G.~Robins, and A.~Lomi.
\newblock Fast maximum likelihood estimation via equilibrium expectation for
  large network data.
\newblock \emph{Scientific Reports}, 8:\penalty0 11509, 2018.

\bibitem[Caimo and Gollini(2023)]{caimo23}
A.~Caimo and I.~Gollini.
\newblock Recent advances in exponential random graph modelling.
\newblock In \emph{Mathematical Proceedings of the Royal Irish Academy}, volume
  123, pages 1--12. Royal Irish Academy, 2023.
\newblock \doi{10.1353/mpr.2023.0000}.

\bibitem[Ceperley and Dewing(1999)]{ceperley99}
D.~Ceperley and M.~Dewing.
\newblock The penalty method for random walks with uncertain energies.
\newblock \emph{Journal of Chemical Physics}, 110\penalty0 (20):\penalty0
  9812--9820, 1999.

\bibitem[Chatterjee and Diaconis(2013)]{chatterjee13}
S.~Chatterjee and P.~Diaconis.
\newblock Estimating and understanding exponential random graph models.
\newblock \emph{The Annals of Statistics}, 41\penalty0 (5):\penalty0
  2428--2461, 2013.

\bibitem[Choe et~al.(2004)Choe, McCormick, and Koh]{choe04}
Y.~Choe, B.~McCormick, and W.~Koh.
\newblock Network connectivity analysis on the temporally augmented
  \textit{{C}.~elegans} web: A pilot study.
\newblock \emph{Soc Neurosci Abstr}, 30\penalty0 (921.9), 2004.

\bibitem[Cimini et~al.(2019)Cimini, Squartini, Saracco, Garlaschelli,
  Gabrielli, and Caldarelli]{cimini19}
G.~Cimini, T.~Squartini, F.~Saracco, D.~Garlaschelli, A.~Gabrielli, and
  G.~Caldarelli.
\newblock The statistical physics of real-world networks.
\newblock \emph{Nature Reviews Physics}, 1:\penalty0 58--71, 2019.

\bibitem[Clark and Handcock(2022)]{clark22}
D.~A. Clark and M.~S. Handcock.
\newblock Comparing the real-world performance of exponential-family random
  graph models and latent order logistic models for social network analysis.
\newblock \emph{Journal of the Royal Statistical Society Series A: Statistics
  in Society}, 185\penalty0 (2):\penalty0 566--587, 2022.

\bibitem[Clauset et~al.(2009)Clauset, Shalizi, and Newman]{clauset09}
A.~Clauset, C.~R. Shalizi, and M.~E. Newman.
\newblock Power-law distributions in empirical data.
\newblock \emph{SIAM Review}, 51\penalty0 (4):\penalty0 661--703, 2009.

\bibitem[Cook et~al.(2019)Cook, Jarrell, Brittin, Wang, Bloniarz, Yakovlev,
  Nguyen, Tang, Bayer, Duerr, et~al.]{cook19}
S.~J. Cook, T.~A. Jarrell, C.~A. Brittin, Y.~Wang, A.~E. Bloniarz, M.~A.
  Yakovlev, K.~C. Nguyen, L.~T.-H. Tang, E.~A. Bayer, J.~S. Duerr, et~al.
\newblock Whole-animal connectomes of both \textit{Caenorhabditis elegans}
  sexes.
\newblock \emph{Nature}, 571\penalty0 (7763):\penalty0 63--71, 2019.

\bibitem[Costanzo et~al.(2001)Costanzo, Crawford, Hirschman, Kranz, Olsen,
  Robertson, Skrzypek, Braun, Hopkins, Kondu, Lengieza, Lew-Smith, Tillberg,
  and Garrels]{costanzo01}
M.~C. Costanzo, M.~E. Crawford, J.~E. Hirschman, J.~E. Kranz, P.~Olsen, L.~S.
  Robertson, M.~S. Skrzypek, B.~R. Braun, K.~L. Hopkins, P.~Kondu, C.~Lengieza,
  J.~E. Lew-Smith, M.~Tillberg, and J.~I. Garrels.
\newblock {YPD\textsuperscript{TM}, PombePD\textsuperscript{TM} and
  WormPD\textsuperscript{TM}: model organism volumes of the
  BioKnowledge\textsuperscript{TM} Library, an integrated resource for protein
  information}.
\newblock \emph{Nucleic Acids Research}, 29\penalty0 (1):\penalty0 75--79, 01
  2001.
\newblock \doi{10.1093/nar/29.1.75}.

\bibitem[Cs\'ardi and Nepusz(2006)]{csardi06}
G.~Cs\'ardi and T.~Nepusz.
\newblock The igraph software package for complex network research.
\newblock \emph{InterJournal}, Complex Systems:\penalty0 1695, 2006.
\newblock URL \url{https://igraph.org}.

\bibitem[Daraganova and Robins(2013)]{daraganova13}
G.~Daraganova and G.~Robins.
\newblock Autologistic actor attribute models.
\newblock In D.~Lusher, J.~Koskinen, and G.~Robins, editors, \emph{Exponential
  Random Graph Models for Social Networks}, chapter~9, pages 102--114.
  Cambridge University Press, New York, 2013.

\bibitem[De~Las~Rivas and Fontanillo(2010)]{delasrivas10}
J.~De~Las~Rivas and C.~Fontanillo.
\newblock Protein--protein interactions essentials: key concepts to building
  and analyzing interactome networks.
\newblock \emph{PLoS Computational Biology}, 6\penalty0 (6):\penalty0 e1000807,
  2010.

\bibitem[de~Silva and Stumpf(2005)]{desilva05}
E.~de~Silva and M.~P. Stumpf.
\newblock Complex networks and simple models in biology.
\newblock \emph{Journal of The Royal Society Interface}, 2\penalty0
  (5):\penalty0 419--430, 2005.

\bibitem[Desmarais and Cranmer(2012)]{desmarais12}
B.~A. Desmarais and S.~J. Cranmer.
\newblock Statistical inference for valued-edge networks: The generalized
  exponential random graph model.
\newblock \emph{PLoS ONE}, 7\penalty0 (1):\penalty0 e30136, 2012.

\bibitem[Dichio and De~Vico~Fallani(2023{\natexlab{a}})]{dichio23}
V.~Dichio and F.~De~Vico~Fallani.
\newblock The exploration-exploitation paradigm for networked biological
  systems.
\newblock \emph{arXiv preprint arXiv:2306.17300v1}, 2023{\natexlab{a}}.

\bibitem[Dichio and De~Vico~Fallani(2023{\natexlab{b}})]{dichio23b}
V.~Dichio and F.~De~Vico~Fallani.
\newblock Statistical models of complex brain networks: a maximum entropy
  approach.
\newblock \emph{Reports on Progress in Physics}, 86\penalty0 (10):\penalty0
  102601, 2023{\natexlab{b}}.
\newblock \doi{10.1088/1361-6633/ace6bc}.

\bibitem[Emmert-Streib et~al.(2014)Emmert-Streib, Dehmer, and
  Haibe-Kains]{emmert-streib14}
F.~Emmert-Streib, M.~Dehmer, and B.~Haibe-Kains.
\newblock Gene regulatory networks and their applications: understanding
  biological and medical problems in terms of networks.
\newblock \emph{Frontiers in Cell and Developmental Biology}, 2, 2014.
\newblock \doi{10.3389/fcell.2014.00038}.

\bibitem[Fellows and Handcock(2017)]{fellows17}
I.~Fellows and M.~Handcock.
\newblock Removing phase transitions from {Gibbs} measures.
\newblock In A.~Singh and J.~Zhu, editors, \emph{Proceedings of the 20th
  International Conference on Artificial Intelligence and Statistics},
  volume~54 of \emph{Proceedings of Machine Learning Research}, pages 289--297,
  20--22 Apr 2017.

\bibitem[Fellows(2018)]{fellows18}
I.~E. Fellows.
\newblock A new generative statistical model for graphs: The latent order
  logistic ({LOLOG}) model.
\newblock \emph{arXiv preprint arXiv:1804.04583v1}, 2018.

\bibitem[Fellows(2019)]{lolog}
I.~E. Fellows.
\newblock \emph{lolog: Latent Order Logistic Graph Models}, 2019.
\newblock URL \url{https://CRAN.R-project.org/package=lolog}.
\newblock R package version 1.2.

\bibitem[Felmlee et~al.(2021)Felmlee, McMillan, and Whitaker]{felmlee21}
D.~Felmlee, C.~McMillan, and R.~Whitaker.
\newblock Dyads, triads, and tetrads: a multivariate simulation approach to
  uncovering network motifs in social graphs.
\newblock \emph{Applied Network Science}, 6\penalty0 (1):\penalty0 63, 2021.

\bibitem[Fodor et~al.(2020)Fodor, Brand, Stones, and Buckle]{fodor20}
J.~Fodor, M.~Brand, R.~J. Stones, and A.~M. Buckle.
\newblock Intrinsic limitations in mainstream methods of identifying network
  motifs in biology.
\newblock \emph{BMC Bioinformatics}, 21:\penalty0 165, 2020.

\bibitem[Frenkel et~al.(2017)Frenkel, Schrenk, and Martiniani]{frenkel17}
D.~Frenkel, K.~J. Schrenk, and S.~Martiniani.
\newblock {Monte Carlo} sampling for stochastic weight functions.
\newblock \emph{Proceedings of the National Academy of Sciences of the USA},
  114\penalty0 (27):\penalty0 6924--6929, 2017.

\bibitem[Geyer and Thompson(1992)]{geyer92}
C.~J. Geyer and E.~A. Thompson.
\newblock Constrained {Monte Carlo} maximum likelihood for dependent data.
\newblock \emph{Journal of the Royal Statistical Society Series B: Statistical
  Methdology}, 54\penalty0 (3):\penalty0 657--699, 1992.

\bibitem[Ghafouri and Khasteh(2020)]{ghafouri20}
S.~Ghafouri and S.~H. Khasteh.
\newblock A survey on exponential random graph models: an application
  perspective.
\newblock \emph{PeerJ Computer Science}, 6:\penalty0 e269, 2020.

\bibitem[Giacomarra et~al.(2023)Giacomarra, Bet, and Zocca]{giacomarra23}
F.~Giacomarra, G.~Bet, and A.~Zocca.
\newblock Generating synthetic power grids using exponential random graphs
  models.
\newblock \emph{arXiv preprint arXiv:2310.19662v1}, 2023.

\bibitem[Gillespie(2015)]{gillespie15}
C.~S. Gillespie.
\newblock Fitting heavy tailed distributions: The {poweRlaw} package.
\newblock \emph{Journal of Statistical Software}, 64\penalty0 (2), 2015.

\bibitem[Gross et~al.(2023)Gross, Petrovi{\'c}, and Stasi]{gross23}
E.~Gross, S.~Petrovi{\'c}, and D.~Stasi.
\newblock Goodness of fit for log-linear {ERGM}s.
\newblock \emph{arXiv preprint arXiv:2104.03167v5}, 2023.

\bibitem[Gruzd et~al.(2023)Gruzd, Abul-Fottouh, Song, and Saiphoo]{gruzd23}
A.~Gruzd, D.~Abul-Fottouh, M.~Y. Song, and A.~Saiphoo.
\newblock From {Facebook} to {YouTube}: The potential exposure to {COVID-19}
  anti-vaccine videos on social media.
\newblock \emph{Social Media + Society}, 9\penalty0 (1):\penalty0
  20563051221150403, 2023.
\newblock \doi{10.1177/20563051221150403}.

\bibitem[Handcock(2003)]{handcock03}
M.~S. Handcock.
\newblock Assessing degeneracy in statistical models of social networks.
\newblock Technical Report~39, Center for Statistics and the Social Sciences,
  University of Washington, 2003.
\newblock URL \url{https://csss.uw.edu/Papers/wp39.pdf}.

\bibitem[Handcock et~al.(2008)Handcock, Hunter, Butts, Goodreau, {Morris}, and
  {Martina}]{handcock08}
M.~S. Handcock, D.~R. Hunter, C.~T. Butts, S.~M. Goodreau, {Morris}, and
  {Martina}.
\newblock statnet: Software tools for the representation, visualization,
  analysis and simulation of network data.
\newblock \emph{Journal of Statistical Software}, 24\penalty0 (1):\penalty0
  1--11, 2008.
\newblock URL \url{http://www.jstatsoft.org/v24/i01}.

\bibitem[Handcock et~al.(2016)Handcock, Hunter, Butts, Goodreau, Krivitsky,
  Bender-deMoll, and Morris]{statnet}
M.~S. Handcock, D.~R. Hunter, C.~T. Butts, S.~M. Goodreau, P.~N. Krivitsky,
  S.~Bender-deMoll, and M.~Morris.
\newblock \emph{statnet: Software Tools for the Statistical Analysis of Network
  Data}.
\newblock The Statnet Project (\url{http://www.statnet.org}), 2016.
\newblock URL \url{http://CRAN.R-project.org/package=statnet}.
\newblock R package version 2016.9.

\bibitem[Handcock et~al.(2022{\natexlab{a}})Handcock, Hunter, Butts, Goodreau,
  Krivitsky, and Morris]{ergm}
M.~S. Handcock, D.~R. Hunter, C.~T. Butts, S.~M. Goodreau, P.~N. Krivitsky, and
  M.~Morris.
\newblock \emph{ergm: Fit, Simulate and Diagnose Exponential-Family Models for
  Networks}.
\newblock The Statnet Project (\url{http://www.statnet.org}),
  2022{\natexlab{a}}.
\newblock URL \url{http://CRAN.R-project.org/package=ergm}.
\newblock R package version 4.3.1.

\bibitem[Handcock et~al.(2022{\natexlab{b}})Handcock, Krivitsky, and
  Fellows]{ergm.tapered}
M.~S. Handcock, P.~N. Krivitsky, and I.~Fellows.
\newblock \emph{ergm.tapered: Tapered Exponential-Family Models for Networks},
  2022{\natexlab{b}}.
\newblock URL \url{https://github.com/statnet/ergm.tapered}.
\newblock R package version 1.1-0.

\bibitem[Hummel et~al.(2012)Hummel, Hunter, and Handcock]{hummel12}
R.~M. Hummel, D.~R. Hunter, and M.~S. Handcock.
\newblock Improving simulation-based algorithms for fitting {ERGMs}.
\newblock \emph{Journal of Computational and Graphical Statistics}, 21\penalty0
  (4):\penalty0 920--939, 2012.

\bibitem[Hunter(2007)]{hunter07}
D.~R. Hunter.
\newblock Curved exponential family models for social networks.
\newblock \emph{Social Networks}, 29\penalty0 (2):\penalty0 216--230, 2007.

\bibitem[Hunter and Handcock(2006)]{hunter06}
D.~R. Hunter and M.~S. Handcock.
\newblock Inference in curved exponential family models for networks.
\newblock \emph{Journal of Computational and Graphical Statistics}, 15\penalty0
  (3):\penalty0 565--583, 2006.

\bibitem[Hunter et~al.(2012)Hunter, Krivitsky, and Schweinberger]{hunter12}
D.~R. Hunter, P.~N. Krivitsky, and M.~Schweinberger.
\newblock Computational statistical methods for social network models.
\newblock \emph{Journal of Computational and Graphical Statistics}, 21\penalty0
  (4):\penalty0 856--882, 2012.

\bibitem[Hunter et~al.(2013)Hunter, Goodreau, and Handcock]{hunter13}
D.~R. Hunter, S.~M. Goodreau, and M.~S. Handcock.
\newblock ergm.userterms: A template package for extending statnet.
\newblock \emph{Journal of Statistical Software}, 52\penalty0 (2):\penalty0
  1–25, 2013.
\newblock \doi{10.18637/jss.v052.i02}.

\bibitem[Kaiser and Hilgetag(2006)]{kaiser06}
M.~Kaiser and C.~C. Hilgetag.
\newblock Nonoptimal component placement, but short processing paths, due to
  long-distance projections in neural systems.
\newblock \emph{PLoS Computational Biology}, 2\penalty0 (7):\penalty0 e95,
  2006.

\bibitem[Koskinen(2020)]{koskinen20}
J.~Koskinen.
\newblock Exponential random graph modelling.
\newblock In P.~Atkinson, S.~Delamont, A.~Cernat, J.~Sakshaug, and R.~Williams,
  editors, \emph{SAGE Research Methods Foundations}. SAGE, London, 2020.
\newblock \doi{10.4135/9781526421036888175}.

\bibitem[Koskinen(2023)]{koskinen23}
J.~Koskinen.
\newblock Exponential random graph models.
\newblock In J.~McLevey, J.~Scott, and P.~J. Carrington, editors, \emph{The
  Sage Handbook of Social Network Analysis}, chapter~33. Sage, second edition,
  2023.

\bibitem[Koskinen and Daraganova(2013)]{koskinen13}
J.~Koskinen and G.~Daraganova.
\newblock Exponential random graph model fundamentals.
\newblock In D.~Lusher, J.~Koskinen, and G.~Robins, editors, \emph{Exponential
  Random Graph Models for Social Networks}, chapter~6, pages 49--76. Cambridge
  University Press, New York, 2013.

\bibitem[Koskinen and Daraganova(2022)]{koskinen22}
J.~Koskinen and G.~Daraganova.
\newblock Bayesian analysis of social influence.
\newblock \emph{Journal of the Royal Statistical Society Series A: Statistics
  in Society}, 185\penalty0 (4):\penalty0 1855--1881, 2022.

\bibitem[Koskinen et~al.(2013)Koskinen, Robins, Wang, and
  Pattison]{koskinen13a}
J.~H. Koskinen, G.~L. Robins, P.~Wang, and P.~E. Pattison.
\newblock Bayesian analysis for partially observed network data, missing ties,
  attributes and actors.
\newblock \emph{Social Networks}, 35\penalty0 (4):\penalty0 514--527, 2013.

\bibitem[Krivitsky et~al.(2022)Krivitsky, Hunter, Morris, and
  Klumb]{krivitsky22}
P.~N. Krivitsky, D.~R. Hunter, M.~Morris, and C.~Klumb.
\newblock ergm 4: Computational improvements.
\newblock \emph{arXiv preprint arXiv:2203.08198v1}, 2022.

\bibitem[Krivitsky et~al.(2023)Krivitsky, Hunter, Morris, and Klumb]{ergm4}
P.~N. Krivitsky, D.~R. Hunter, M.~Morris, and C.~Klumb.
\newblock ergm 4: New features for analyzing exponential-family random graph
  models.
\newblock \emph{Journal of Statistical Software}, 105\penalty0 (6):\penalty0
  1–44, 2023.
\newblock \doi{10.18637/jss.v105.i06}.

\bibitem[Leifeld et~al.(2018)Leifeld, Cranmer, and Desmarais]{leifeld18}
P.~Leifeld, S.~J. Cranmer, and B.~A. Desmarais.
\newblock Temporal exponential random graph models with btergm: Estimation and
  bootstrap confidence intervals.
\newblock \emph{Journal of Statistical Software}, 83\penalty0 (6):\penalty0
  1–36, 2018.
\newblock \doi{10.18637/jss.v083.i06}.

\bibitem[Levy(2016)]{levy16}
M.~Levy.
\newblock gwdegree: Improving interpretation of geometrically-weighted degree
  estimates in exponential random graph models.
\newblock \emph{Journal of Open Source Software}, 1\penalty0 (3):\penalty0 36,
  2016.

\bibitem[Levy et~al.(2016)Levy, Lubell, Leifeld, and Cranmer]{levy16poster}
M.~Levy, M.~Lubell, P.~Leifeld, and S.~Cranmer.
\newblock Interpretation of gw-degree estimates in {ERGM}s, June 2016.
\newblock URL \url{https://doi.org/10.6084/m9.figshare.3465020.v1}.

\bibitem[Lusher et~al.(2013)Lusher, Koskinen, and Robins]{lusher13}
D.~Lusher, J.~Koskinen, and G.~Robins, editors.
\newblock \emph{Exponential Random Graph Models for Social Networks}.
\newblock Structural Analysis in the Social Sciences. Cambridge University
  Press, New York, 2013.

\bibitem[Lusher et~al.(2020)Lusher, Wang, Brennecke, Brailly, Faye, and
  Gallagher]{lusher20}
D.~Lusher, P.~Wang, J.~Brennecke, J.~Brailly, M.~Faye, and C.~Gallagher.
\newblock Advances in exponential random graph models.
\newblock In R.~Light and J.~Moody, editors, \emph{The Oxford Handbook of
  Social Networks}, chapter~13, pages 234--253. Oxford University Press, 2020.
\newblock \doi{10.1093/oxfordhb/9780190251765.013.18}.

\bibitem[Mahadevan et~al.(2006)Mahadevan, Krioukov, Fall, and
  Vahdat]{mahadevan06}
P.~Mahadevan, D.~Krioukov, K.~Fall, and A.~Vahdat.
\newblock Systematic topology analysis and generation using degree
  correlations.
\newblock \emph{ACM SIGCOMM Computer Communication Review}, 36\penalty0
  (4):\penalty0 135--146, 2006.

\bibitem[Martin(2017)]{martin17}
J.~L. Martin.
\newblock The structure of node and edge generation in a delusional social
  network.
\newblock \emph{Journal of Social Structure}, 18\penalty0 (1):\penalty0 1--22,
  2017.
\newblock \doi{10.21307/joss-2018-005}.

\bibitem[Martin(2020)]{martin20}
J.~L. Martin.
\newblock Comment on geodesic cycle length distributions in delusional and
  other social networks.
\newblock \emph{Journal of Social Structure}, 21\penalty0 (1):\penalty0 77--93,
  2020.
\newblock \doi{10.21307/joss-2020-003}.

\bibitem[Milo et~al.(2002)Milo, Shen-Orr, Itzkovitz, Kashtan, Chklovskii, and
  Alon]{milo02}
R.~Milo, S.~Shen-Orr, S.~Itzkovitz, N.~Kashtan, D.~Chklovskii, and U.~Alon.
\newblock Network motifs: simple building blocks of complex networks.
\newblock \emph{Science}, 298\penalty0 (5594):\penalty0 824--827, 2002.

\bibitem[Obando et~al.(2022)Obando, Rosso, Siegel, Corbetta, and
  De~Vico~Fallani]{obando22}
C.~Obando, C.~Rosso, J.~Siegel, M.~Corbetta, and F.~De~Vico~Fallani.
\newblock Temporal exponential random graph models of longitudinal brain
  networks after stroke.
\newblock \emph{Journal of The Royal Society Interface}, 19\penalty0
  (188):\penalty0 20210850, 2022.

\bibitem[Orsini et~al.(2015)Orsini, Dankulov, Colomer-de Sim{\'o}n, Jamakovic,
  Mahadevan, Vahdat, Bassler, Toroczkai, Bogun{\'a}, Caldarelli,
  et~al.]{orsini15}
C.~Orsini, M.~M. Dankulov, P.~Colomer-de Sim{\'o}n, A.~Jamakovic, P.~Mahadevan,
  A.~Vahdat, K.~E. Bassler, Z.~Toroczkai, M.~Bogun{\'a}, G.~Caldarelli, et~al.
\newblock Quantifying randomness in real networks.
\newblock \emph{Nature Communications}, 6:\penalty0 8627, 2015.

\bibitem[Parker et~al.(2022)Parker, Pallotti, and Lomi]{parker22}
A.~Parker, F.~Pallotti, and A.~Lomi.
\newblock New network models for the analysis of social contagion in
  organizations: an introduction to autologistic actor attribute models.
\newblock \emph{Organizational Research Methods}, 25\penalty0 (3):\penalty0
  513--540, 2022.

\bibitem[Pavlovic et~al.(2014)Pavlovic, V{\'e}rtes, Bullmore, Schafer, and
  Nichols]{pavlovic14}
D.~M. Pavlovic, P.~E. V{\'e}rtes, E.~T. Bullmore, W.~R. Schafer, and T.~E.
  Nichols.
\newblock Stochastic blockmodeling of the modules and core of the
  \textit{{Caenorhabditis} elegans} connectome.
\newblock \emph{PLoS ONE}, 9\penalty0 (7):\penalty0 e97584, 2014.

\bibitem[Peacock et~al.(2019)Peacock, Milewicz, and Snidal]{peacock19}
C.~Peacock, K.~Milewicz, and D.~Snidal.
\newblock Boilerplate in international trade agreements.
\newblock \emph{International Studies Quarterly}, 63\penalty0 (4):\penalty0
  923--937, 2019.
\newblock \doi{10.1093/isq/sqz069}.

\bibitem[{R Core Team}(2022)]{R-manual}
{R Core Team}.
\newblock \emph{R: A Language and Environment for Statistical Computing}.
\newblock R Foundation for Statistical Computing, Vienna, Austria, 2022.
\newblock URL \url{http://www.R-project.org}.

\bibitem[Roberts et~al.(2016)Roberts, Perry, Lord, Roberts, Mitchell, Smith,
  Calamante, and Breakspear]{roberts16}
J.~A. Roberts, A.~Perry, A.~R. Lord, G.~Roberts, P.~B. Mitchell, R.~E. Smith,
  F.~Calamante, and M.~Breakspear.
\newblock The contribution of geometry to the human connectome.
\newblock \emph{Neuroimage}, 124:\penalty0 379--393, 2016.

\bibitem[Robins et~al.(2001)Robins, Pattison, and Elliott]{robins01b}
G.~Robins, P.~Pattison, and P.~Elliott.
\newblock Network models for social influence processes.
\newblock \emph{Psychometrika}, 66\penalty0 (2):\penalty0 161--189, 2001.

\bibitem[Robins et~al.(2007)Robins, Snijders, Wang, Handcock, and
  Pattison]{robins07}
G.~Robins, T.~A.~B. Snijders, P.~Wang, M.~Handcock, and P.~Pattison.
\newblock Recent developments in exponential random graph ($p^*$) models for
  social networks.
\newblock \emph{Social Networks}, 29:\penalty0 192--215, 2007.

\bibitem[Salgado et~al.(2001)Salgado, Santos-Zavaleta, Gama-Castro,
  Mill{\'a}n-Z{\'a}rate, D{\'\i}az-Peredo, S{\'a}nchez-Solano, P{\'e}rez-Rueda,
  Bonavides-Mart{\'\i}nez, and Collado-Vides]{salgado01}
H.~Salgado, A.~Santos-Zavaleta, S.~Gama-Castro, D.~Mill{\'a}n-Z{\'a}rate,
  E.~D{\'\i}az-Peredo, F.~S{\'a}nchez-Solano, E.~P{\'e}rez-Rueda,
  C.~Bonavides-Mart{\'\i}nez, and J.~Collado-Vides.
\newblock {RegulonDB} (version 3.2): transcriptional regulation and operon
  organization in \textit{Escherichia coli} {K-12}.
\newblock \emph{Nucleic Acids Research}, 29\penalty0 (1):\penalty0 72--74,
  2001.

\bibitem[Saul and Filkov(2007)]{saul07}
Z.~M. Saul and V.~Filkov.
\newblock Exploring biological network structure using exponential random graph
  models.
\newblock \emph{Bioinformatics}, 23\penalty0 (19):\penalty0 2604--2611, 2007.

\bibitem[Schaefer et~al.(2012)Schaefer, Fontaine, Vinayagam, Porras, Wanker,
  and Andrade-Navarro]{schaefer12}
M.~H. Schaefer, J.-F. Fontaine, A.~Vinayagam, P.~Porras, E.~E. Wanker, and
  M.~A. Andrade-Navarro.
\newblock {HIPPIE}: Integrating protein interaction networks with experiment
  based quality scores.
\newblock \emph{PLoS ONE}, 7\penalty0 (2):\penalty0 e31826, 2012.

\bibitem[Schaefer et~al.(2013)Schaefer, Lopes, Mah, Shoemaker, Matsuoka,
  Fontaine, Louis-Jeune, Eisfeld, Neumann, Perez-Iratxeta, et~al.]{schaefer13}
M.~H. Schaefer, T.~J. Lopes, N.~Mah, J.~E. Shoemaker, Y.~Matsuoka, J.-F.
  Fontaine, C.~Louis-Jeune, A.~J. Eisfeld, G.~Neumann, C.~Perez-Iratxeta,
  et~al.
\newblock Adding protein context to the human protein-protein interaction
  network to reveal meaningful interactions.
\newblock \emph{PLoS Computational Biology}, 9\penalty0 (1):\penalty0 e1002860,
  2013.

\bibitem[Schmid and Hunter(2024)]{schmid24}
C.~S. Schmid and D.~R. Hunter.
\newblock Improving {ERGM} starting values using simulated annealing.
\newblock \emph{Social Networks}, 76:\penalty0 209--214, 2024.
\newblock \doi{10.1016/j.socnet.2023.10.002}.

\bibitem[Schmid et~al.(2022)Schmid, Chen, and Desmarais]{schmid21}
C.~S. Schmid, T.~H.~Y. Chen, and B.~A. Desmarais.
\newblock Generative dynamics of {Supreme Court} citations: Analysis with a new
  statistical network model.
\newblock \emph{Political Analysis}, 30\penalty0 (4):\penalty0 515–534, 2022.
\newblock \doi{10.1017/pan.2021.20}.

\bibitem[Schweinberger(2011)]{schweinberger11}
M.~Schweinberger.
\newblock Instability, sensitivity, and degeneracy of discrete exponential
  families.
\newblock \emph{Journal of the American Statistical Association}, 106\penalty0
  (496):\penalty0 1361--1370, 2011.
\newblock \doi{10.1198/jasa.2011.tm10747}.

\bibitem[Schweinberger et~al.(2020)Schweinberger, Krivitsky, Butts, and
  Stewart]{schweinberger20}
M.~Schweinberger, P.~N. Krivitsky, C.~T. Butts, and J.~R. Stewart.
\newblock Exponential-family models of random graphs: inference in finite,
  super and infinite population scenarios.
\newblock \emph{Statistical Science}, 35\penalty0 (4):\penalty0 627--662, 2020.

\bibitem[Scott et~al.(2021)Scott, Marantz, and Ulibarri]{scott22}
T.~A. Scott, N.~Marantz, and N.~Ulibarri.
\newblock Use of boilerplate language in regulatory documents: Evidence from
  environmental impact statements.
\newblock \emph{Journal of Public Administration Research and Theory},
  32\penalty0 (3):\penalty0 576--590, 2021.
\newblock \doi{10.1093/jopart/muab048}.

\bibitem[Shen-Orr et~al.(2002)Shen-Orr, Milo, Mangan, and Alon]{shen02}
S.~S. Shen-Orr, R.~Milo, S.~Mangan, and U.~Alon.
\newblock Network motifs in the transcriptional regulation network of
  \textit{Escherichia coli}.
\newblock \emph{Nature Genetics}, 31\penalty0 (1):\penalty0 64--68, 2002.

\bibitem[Snijders(2002)]{snijders02}
T.~A.~B. Snijders.
\newblock Markov chain {Monte Carlo} estimation of exponential random graph
  models.
\newblock \emph{Journal of Social Structure}, 3\penalty0 (2):\penalty0 1--40,
  2002.

\bibitem[Snijders et~al.(2006)Snijders, Pattison, Robins, and
  Handcock]{snijders06}
T.~A.~B. Snijders, P.~E. Pattison, G.~L. Robins, and M.~S. Handcock.
\newblock New specifications for exponential random graph models.
\newblock \emph{Sociological Methodology}, 36\penalty0 (1):\penalty0 99--153,
  2006.

\bibitem[Stillman et~al.(2017)Stillman, Wilson, Denny, Desmarais, Bhamidi,
  Cranmer, and Lu]{stillman17}
P.~E. Stillman, J.~D. Wilson, M.~J. Denny, B.~A. Desmarais, S.~Bhamidi, S.~J.
  Cranmer, and Z.-L. Lu.
\newblock Statistical modeling of the default mode brain network reveals a
  segregated highway structure.
\newblock \emph{Scientific Reports}, 7\penalty0 (1):\penalty0 11694, 2017.

\bibitem[Stivala(2020)]{stivala20d}
A.~Stivala.
\newblock Reply to ``{C}omment on geodesic cycle length distributions in
  delusional and other social networks''.
\newblock \emph{Journal of Social Structure}, 21\penalty0 (1):\penalty0
  94--106, 2020.
\newblock \doi{10.21307/joss-2020-004}.

\bibitem[Stivala(2023)]{stivala23}
A.~Stivala.
\newblock Overcoming near-degeneracy in the autologistic actor attribute model.
\newblock \emph{arXiv preprint arXiv:2309.07338v2}, 2023.

\bibitem[Stivala and Lomi(2021)]{stivala21}
A.~Stivala and A.~Lomi.
\newblock Testing biological network motif significance with exponential random
  graph models.
\newblock \emph{Applied Network Science}, 6\penalty0 (1):\penalty0 91, 2021.

\bibitem[Stivala and Lomi(2022)]{stivala22_slides}
A.~Stivala and A.~Lomi.
\newblock {A new scalable implementation of the citation exponential random
  graph model (cERGM) and its application to a large patent citation network}.
\newblock Talk presented at INSNA Sunbelt XLII conference, July 2022.
\newblock URL \url{https://doi.org/10.5281/zenodo.7951927}.

\bibitem[Stivala et~al.(2020)Stivala, Robins, and Lomi]{stivala20b}
A.~Stivala, G.~Robins, and A.~Lomi.
\newblock Exponential random graph model parameter estimation for very large
  directed networks.
\newblock \emph{PLoS ONE}, 15\penalty0 (1):\penalty0 e0227804, 2020.

\bibitem[Stivala et~al.(2023{\natexlab{a}})Stivala, Wang, and Lomi]{ALAAMEE}
A.~Stivala, P.~Wang, and A.~Lomi.
\newblock {ALAAMEE}.
\newblock Computer software, 2023{\natexlab{a}}.
\newblock URL \url{https://github.com/stivalaa/ALAAMEE}.

\bibitem[Stivala et~al.(2023{\natexlab{b}})Stivala, Wang, and
  Lomi]{stivala23_slides}
A.~Stivala, P.~Wang, and A.~Lomi.
\newblock Numbers and structural positions of women in a national director
  interlock network.
\newblock Talk preseented at INSNA Sunbelt XLIII Conference, June
  2023{\natexlab{b}}.
\newblock URL \url{https://doi.org/10.5281/zenodo.8092829}.

\bibitem[Sulston and Horvitz(1977)]{sulston77}
J.~E. Sulston and H.~R. Horvitz.
\newblock Post-embryonic cell lineages of the nematode,
  \textit{{Caenorhabditis} elegans}.
\newblock \emph{Developmental biology}, 56\penalty0 (1):\penalty0 110--156,
  1977.

\bibitem[Sulston et~al.(1983)Sulston, Schierenberg, White, and
  Thomson]{sulston83}
J.~E. Sulston, E.~Schierenberg, J.~G. White, and J.~N. Thomson.
\newblock The embryonic cell lineage of the nematode \textit{{Caenorhabditis}
  elegans}.
\newblock \emph{Developmental biology}, 100\penalty0 (1):\penalty0 64--119,
  1983.

\bibitem[Suratanee et~al.(2014)Suratanee, Schaefer, Betts, Soons, Mannsperger,
  Harder, Oswald, Gipp, Ramminger, Marcus, et~al.]{suratanee14}
A.~Suratanee, M.~H. Schaefer, M.~J. Betts, Z.~Soons, H.~Mannsperger, N.~Harder,
  M.~Oswald, M.~Gipp, E.~Ramminger, G.~Marcus, et~al.
\newblock Characterizing protein interactions employing a genome-wide {siRNA}
  cellular phenotyping screen.
\newblock \emph{PLoS Computational Biology}, 10\penalty0 (9):\penalty0
  e1003814, 2014.

\bibitem[Takemura et~al.(2013)Takemura, Bharioke, Lu, Nern, Vitaladevuni,
  Rivlin, Katz, Olbris, Plaza, Winston, et~al.]{takemura13}
S.-y. Takemura, A.~Bharioke, Z.~Lu, A.~Nern, S.~Vitaladevuni, P.~K. Rivlin,
  W.~T. Katz, D.~J. Olbris, S.~M. Plaza, P.~Winston, et~al.
\newblock A visual motion detection circuit suggested by \textit{Drosophila}
  connectomics.
\newblock \emph{Nature}, 500\penalty0 (7461):\penalty0 175--181, 2013.

\bibitem[Varier and Kaiser(2011)]{varier11}
S.~Varier and M.~Kaiser.
\newblock Neural development features: spatio-temporal development of the
  \textit{{Caenorhabditis} elegans} neuronal network.
\newblock \emph{PLoS Computational Biology}, 7\penalty0 (1):\penalty0 e1001044,
  2011.

\bibitem[Varshney et~al.(2011)Varshney, Chen, Paniagua, Hall, and
  Chklovskii]{varshney11}
L.~R. Varshney, B.~L. Chen, E.~Paniagua, D.~H. Hall, and D.~B. Chklovskii.
\newblock Structural properties of the \textit{{Caenorhabditis} elegans}
  neuronal network.
\newblock \emph{PLoS Computational Biology}, 7\penalty0 (2):\penalty0 e1001066,
  2011.

\bibitem[{Vega Yon} et~al.(2021){Vega Yon}, Slaughter, and {de la
  Haye}]{vegayon21}
G.~G. {Vega Yon}, A.~Slaughter, and K.~{de la Haye}.
\newblock Exponential random graph models for little networks.
\newblock \emph{Social Networks}, 64:\penalty0 225--238, 2021.
\newblock \doi{https://doi.org/10.1016/j.socnet.2020.07.005}.

\bibitem[V\'{e}rtes et~al.(2012)V\'{e}rtes, Alexander-Bloch, Gogtay, Giedd,
  Rapoport, and Bullmore]{vertes12}
P.~E. V\'{e}rtes, A.~F. Alexander-Bloch, N.~Gogtay, J.~N. Giedd, J.~L.
  Rapoport, and E.~T. Bullmore.
\newblock Simple models of human brain functional networks.
\newblock \emph{Proceedings of the National Academy of Sciences of the USA},
  109\penalty0 (15):\penalty0 5868--5873, 2012.
\newblock \doi{10.1073/pnas.1111738109}.

\bibitem[Vogelstein et~al.(2018)Vogelstein, Perlman, Falk, Baden, Gray~Roncal,
  Chandrashekhar, Collman, Seshamani, Patsolic, Lillaney, et~al.]{vogelstein18}
J.~T. Vogelstein, E.~Perlman, B.~Falk, A.~Baden, W.~Gray~Roncal,
  V.~Chandrashekhar, F.~Collman, S.~Seshamani, J.~L. Patsolic, K.~Lillaney,
  et~al.
\newblock A community-developed open-source computational ecosystem for big
  neuro data.
\newblock \emph{Nature methods}, 15\penalty0 (11):\penalty0 846--847, 2018.

\bibitem[von Mering et~al.(2002)von Mering, Krause, Snel, Cornell, Oliver,
  Fields, and Bork]{vonmering02}
C.~von Mering, R.~Krause, B.~Snel, M.~Cornell, S.~G. Oliver, S.~Fields, and
  P.~Bork.
\newblock Comparative assessment of large-scale data sets of protein--protein
  interactions.
\newblock \emph{Nature}, 417\penalty0 (6887):\penalty0 399--403, 2002.

\bibitem[Wang et~al.(2009)Wang, Robins, and Pattison]{pnet}
P.~Wang, G.~Robins, and P.~Pattison.
\newblock \emph{{PNet}: A program for the simulation and estimation of
  exponential random graph models}.
\newblock Melbourne School of Psychological Sciences, The University of
  Melbouxrne, 2009.
\newblock URL \url{http://www.melnet.org.au/s/PNetManual.pdf}.

\bibitem[Wang et~al.(2014)Wang, Robins, Pattison, and Koskinen]{mpnet14}
P.~Wang, G.~Robins, P.~Pattison, and J.~Koskinen.
\newblock \emph{{MPNet}: Program for the simulation and estimation of (p*)
  exponential random graph models for multilevel networks}.
\newblock Melbourne School of Psychological Sciences, The University of
  Melbourne, 2014.
\newblock URL \url{http://www.melnet.org.au/s/MPNetManual.pdf}.

\bibitem[Wang et~al.(2022)Wang, Stivala, Robins, Pattison, Koskinen, and
  Lomi]{mpnet22}
P.~Wang, A.~Stivala, G.~Robins, P.~Pattison, J.~Koskinen, and A.~Lomi.
\newblock \emph{PNet: Program for the simulation and estimation of (p*)
  exponential random graph models for multilevel networks}, 2022.
\newblock URL \url{http://www.melnet.org.au/s/MPNetManual2022.pdf}.

\bibitem[White et~al.(1986)White, Southgate, Thomson, and Brenner]{white86}
J.~G. White, E.~Southgate, J.~N. Thomson, and S.~Brenner.
\newblock The structure of the nervous system of the nematode
  \textit{{Caenorhabditis} elegans}.
\newblock \emph{Philosophical Transactions of the Royal Society B: Biological
  Sciences}, 314\penalty0 (1165):\penalty0 1--340, 1986.

\bibitem[Yavero\v{g}lu et~al.(2015)Yavero\v{g}lu, Fitzhugh, Kurant,
  Markopoulou, Butts, and Pr{\v{z}}ulj]{yaveroglu15}
O.~N. Yavero\v{g}lu, S.~M. Fitzhugh, M.~Kurant, A.~Markopoulou, C.~T. Butts,
  and N.~Pr{\v{z}}ulj.
\newblock ergm.graphlets: A package for {ERG} modeling based on graphlet
  statistics.
\newblock \emph{Journal of Statistical Software}, 65\penalty0 (12):\penalty0
  1--29, 2015.
\newblock URL \url{https://www.jstatsoft.org/v065/i12}.

\end{thebibliography}

\addcontentsline{toc}{section}{\refname}

\newpage
\appendix
\setcounter{section}{0}
\renewcommand{\thesection}{Appendix \Alph{section}}
\setcounter{table}{0}
\renewcommand{\thetable}{\Alph{section}\arabic{table}}
\setcounter{figure}{0}
\renewcommand{\thefigure}{\Alph{section}\arabic{figure}}

\renewcommand{\theHsection}{Appendix \Alph{section}}
\renewcommand{\theHfigure}{\Alph{section}\arabic{figure}}
\renewcommand{\theHtable}{\Alph{section}\arabic{table}}

\section{Supplementary figures}
\label{sec:si_figures}

\begin{figure}[h]
  \centering
  \includegraphics[angle=270,width=\textwidth]{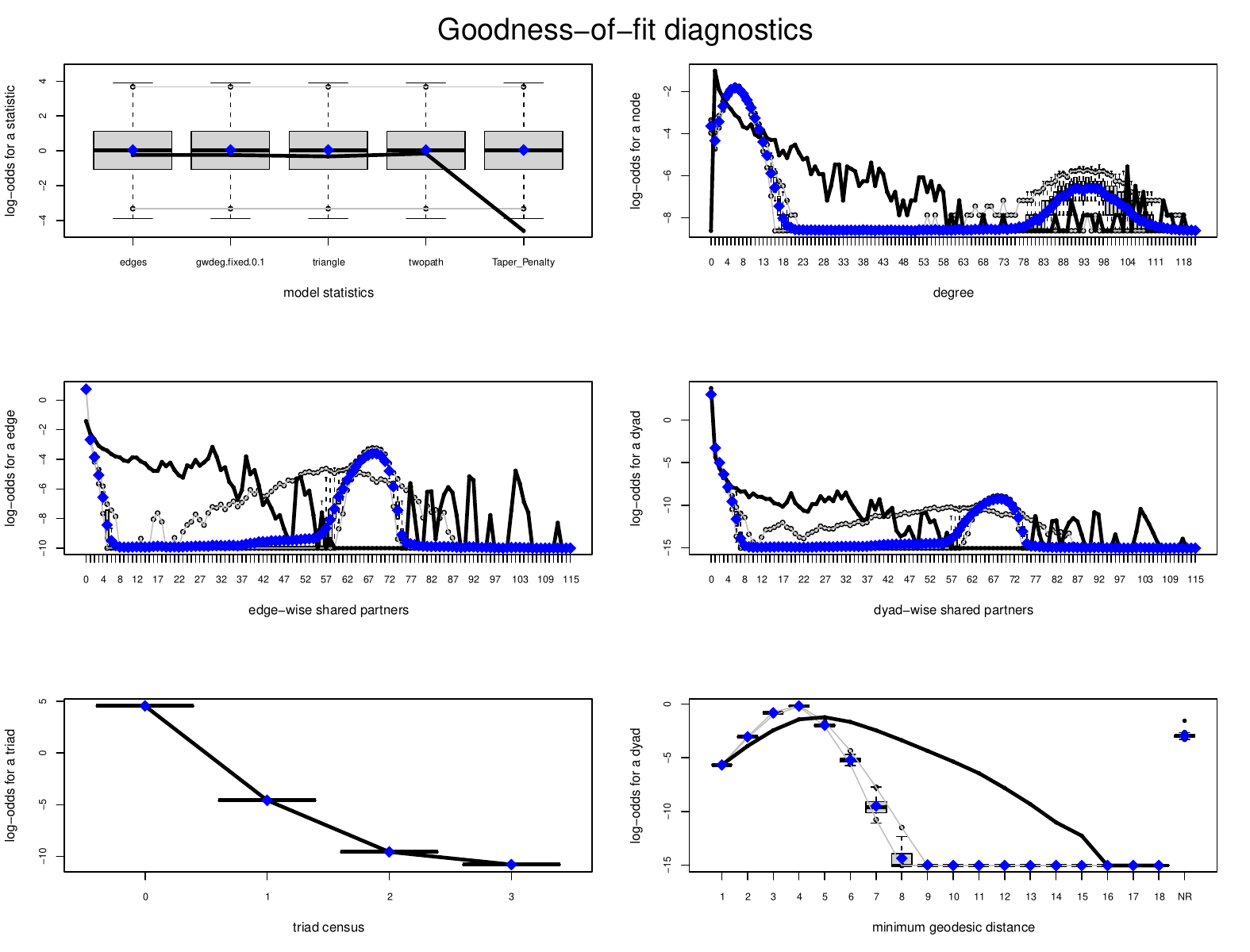}
  \caption{Goodness-of-fit plots for the tapered ERGM model of the
    yeast PPI network
    (Table~\ref{tab:yeast_ppi_tapered_estimations}).}
  \label{fig:yeast_ppi_tapered_gof}
\end{figure}

\begin{figure}
  \centering
  \subfigure[Model diagnostic plots]{\includegraphics[scale=.5]{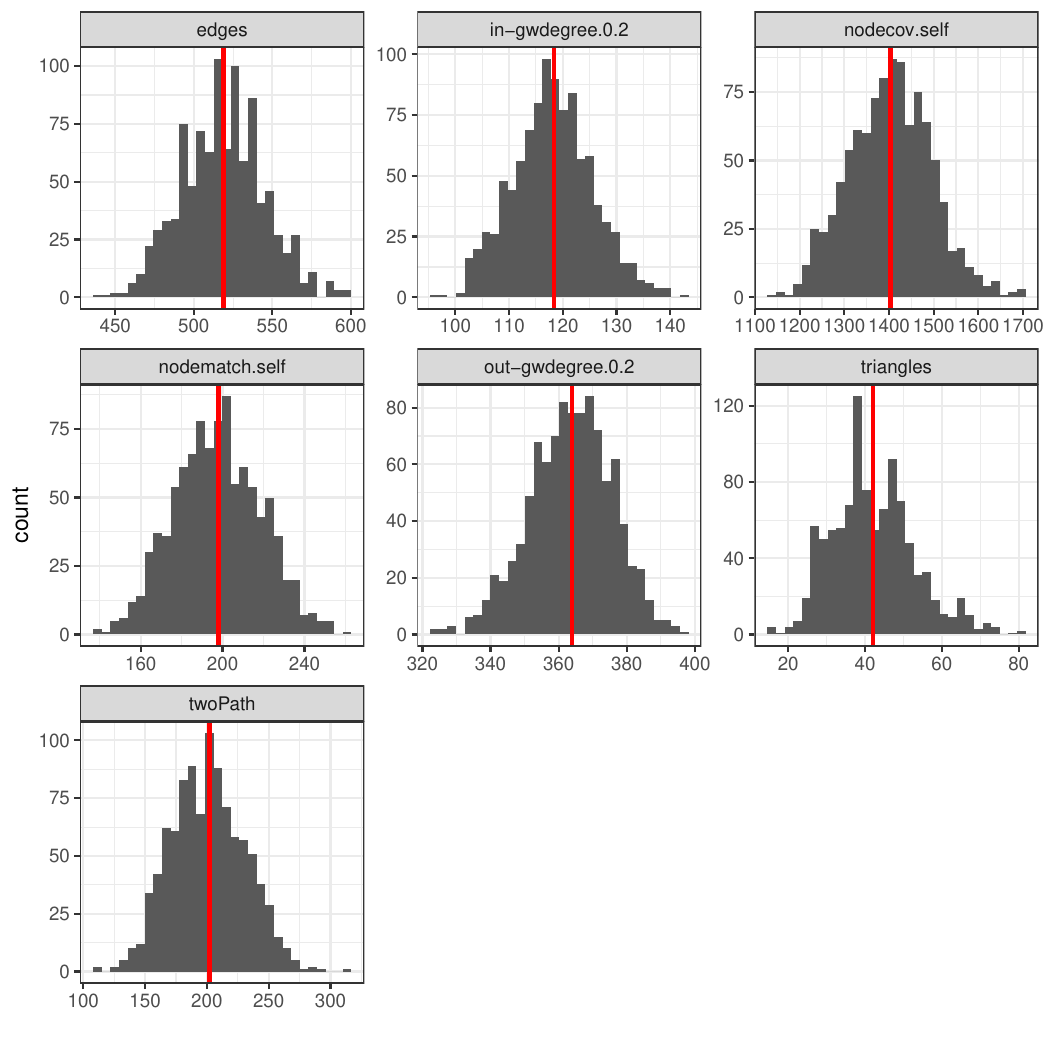}}
  \subfigure[Edges]{\includegraphics[scale=0.4]{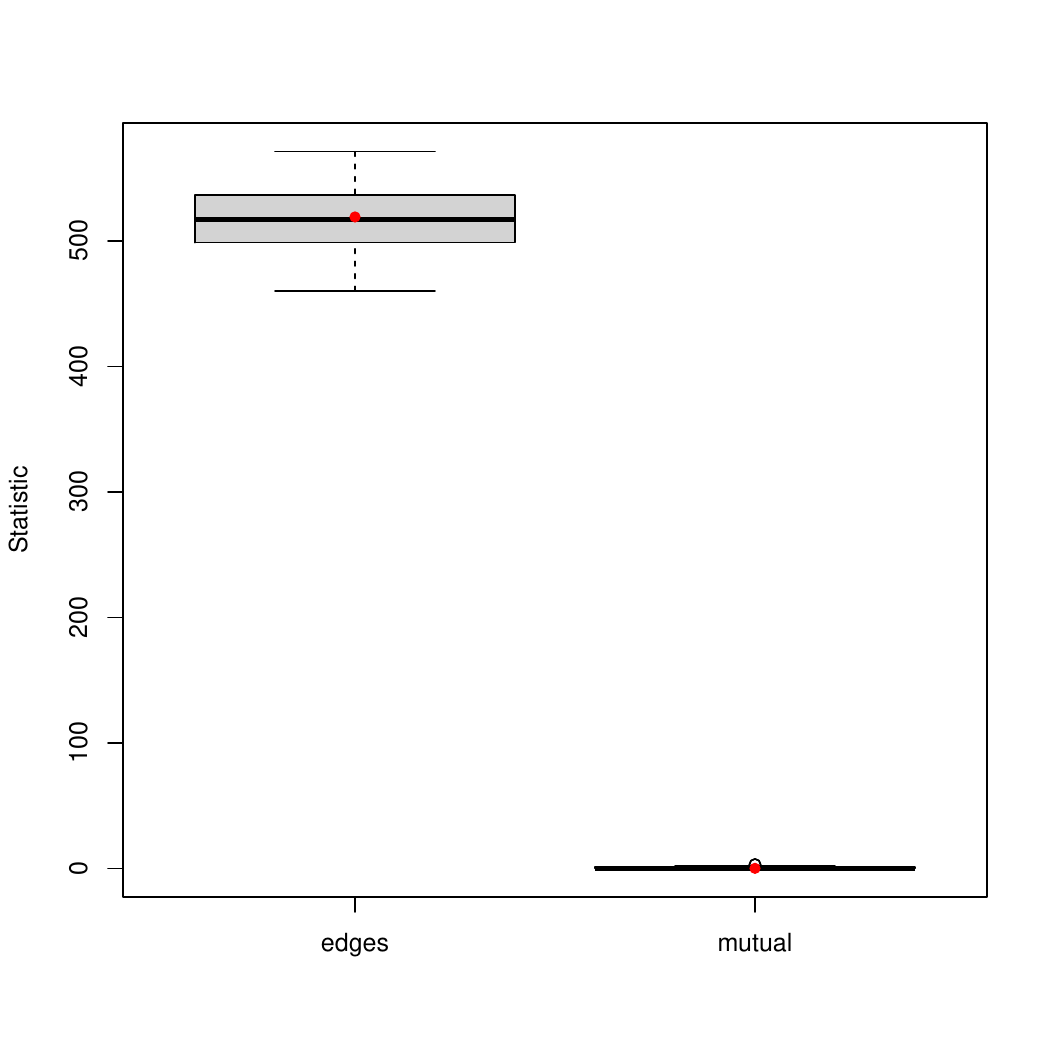}}
  \subfigure[In-degree distribution]{\includegraphics[scale=0.3]{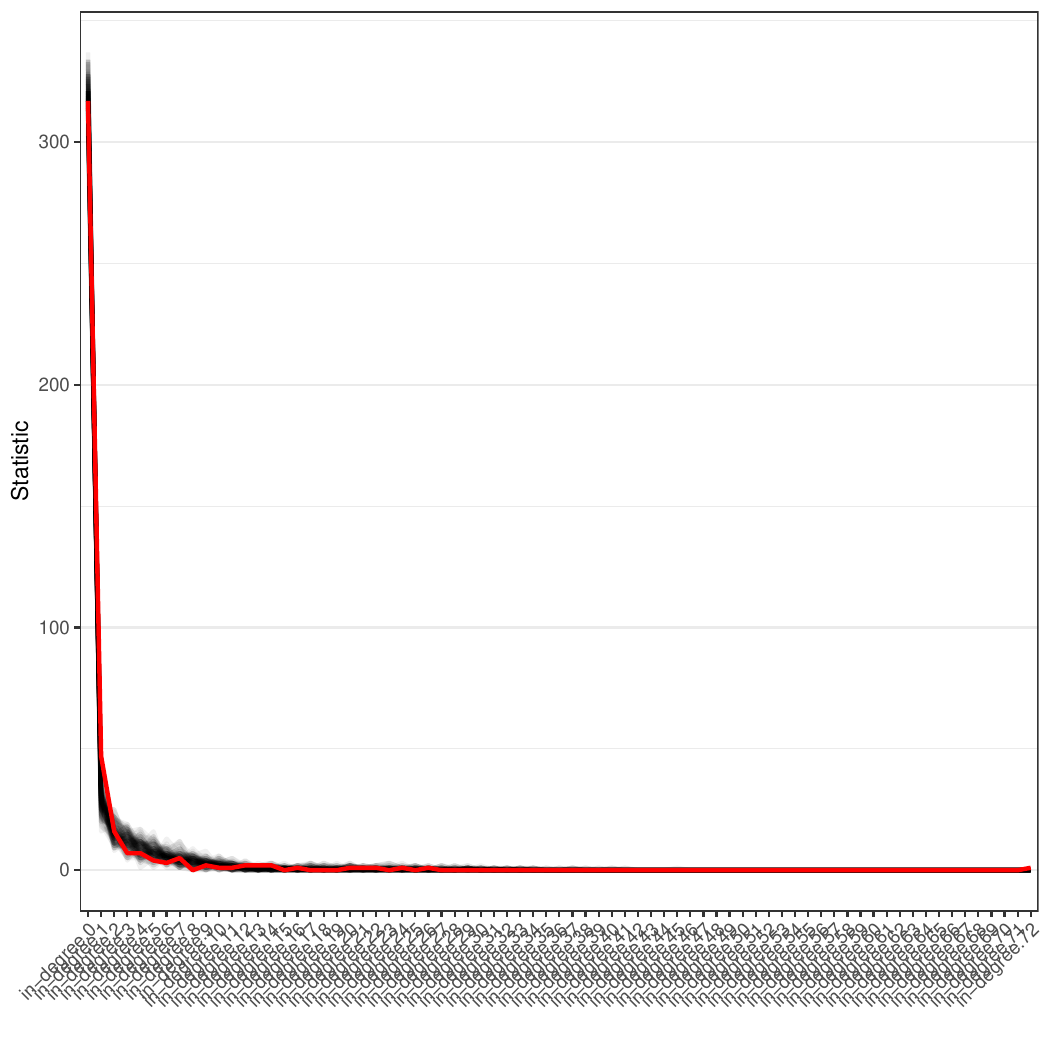}}
  \subfigure[Out-degree distribution]{\includegraphics[scale=0.3]{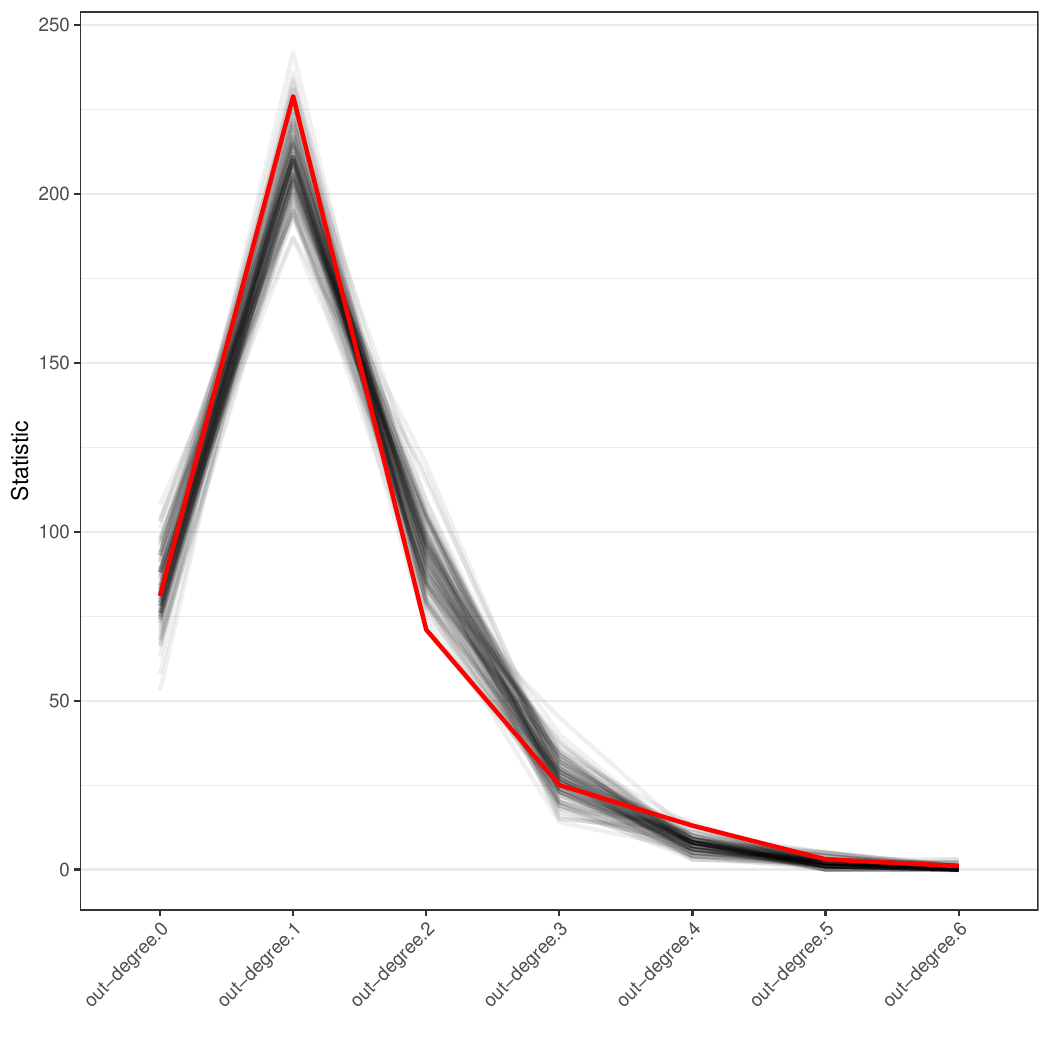}}
  \subfigure[Edgewise shared partners]{\includegraphics[scale=0.3]{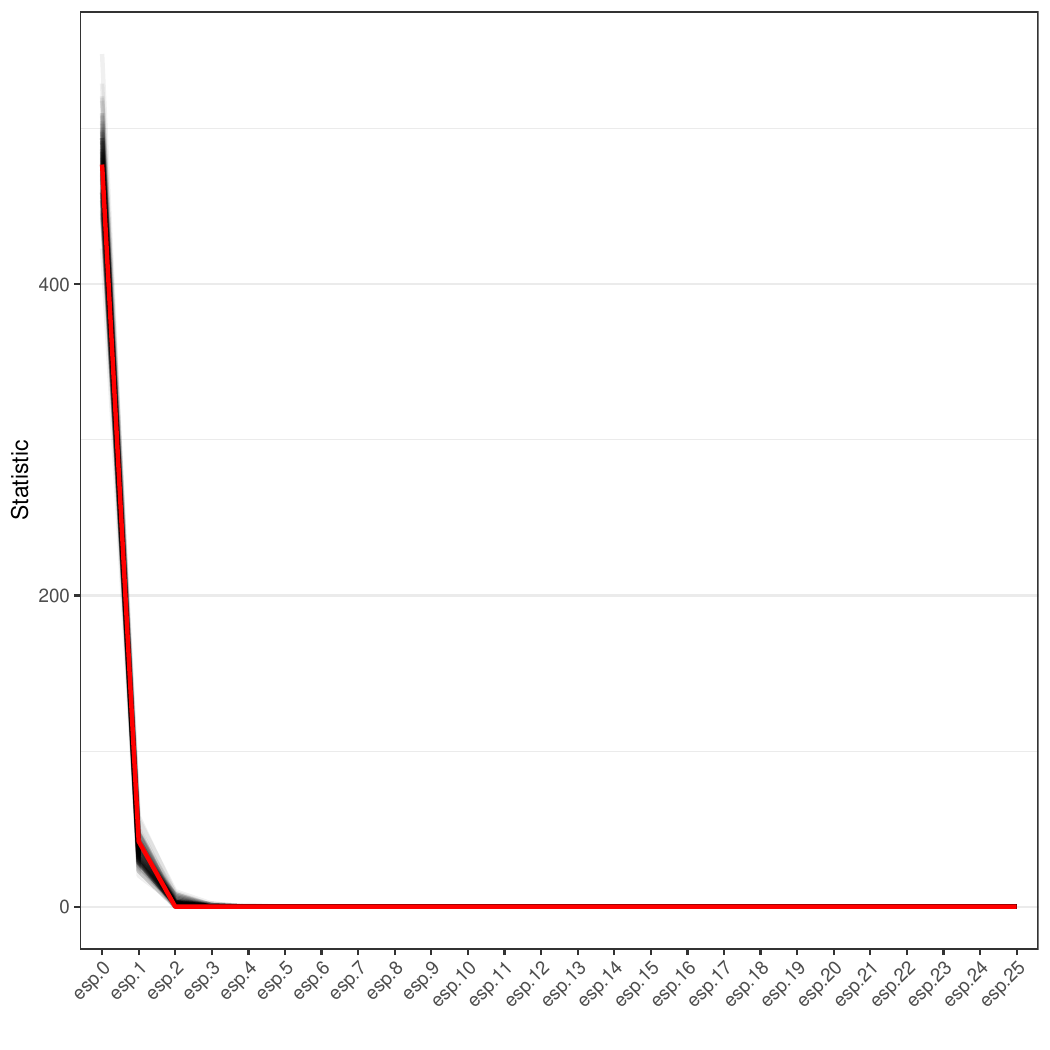}}
  \caption{Model diagnostic and goodness-of-fit plots for the Alon \textit{E. coli} regulatory network LOLOG model, Table~\ref{tab:ecoli_lolog}.}
  \label{fig:ecoli_lolog_gof}
\end{figure}

\begin{figure}
  \centering
  \subfigure[Model 1]{\includegraphics[angle=270,width=.8\textwidth]{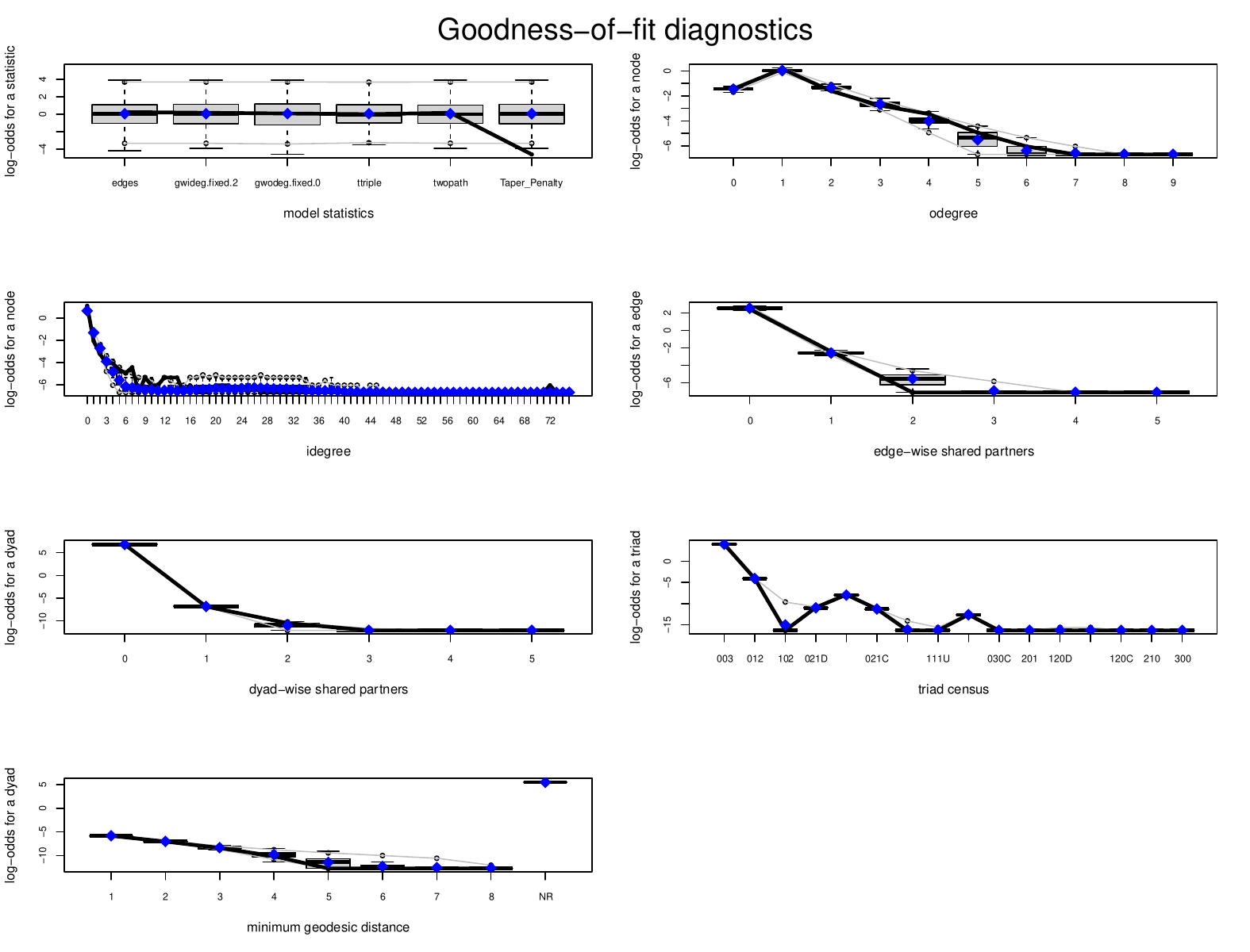}}  
  \subfigure[Model 2]{\includegraphics[angle=270,width=.8\textwidth]{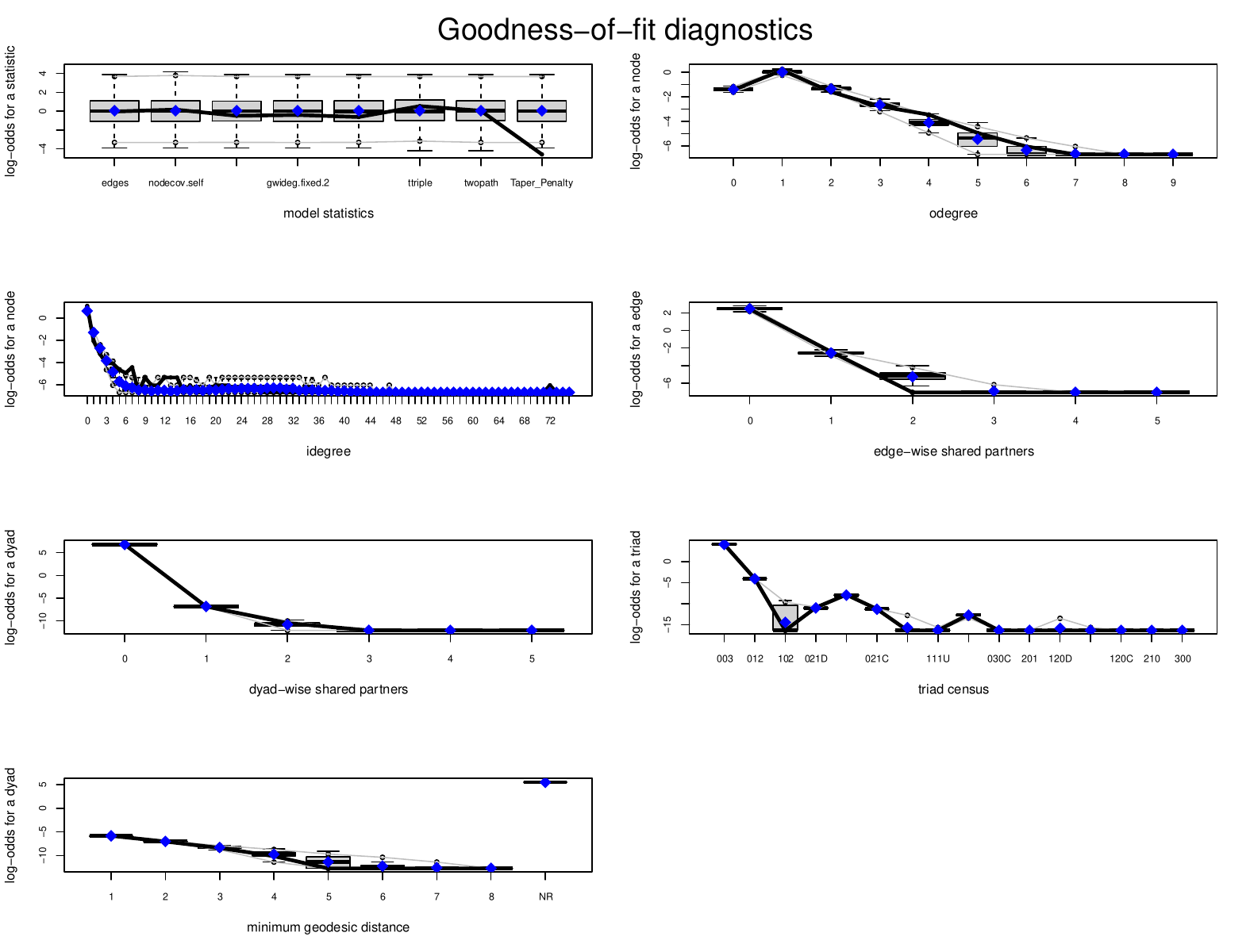}}
  \caption{Goodness-of-fit plots for the tapered ERGM models of the
    Alon \textit{E. coli} regulatory network
    (Table~\ref{tab:ecoli_tapered_estimations}).}
  \label{fig:ecoli_tapered_gof}
\end{figure}

\begin{figure}
  \subfigure[Model diagnostic plots]{\includegraphics[scale=.5]{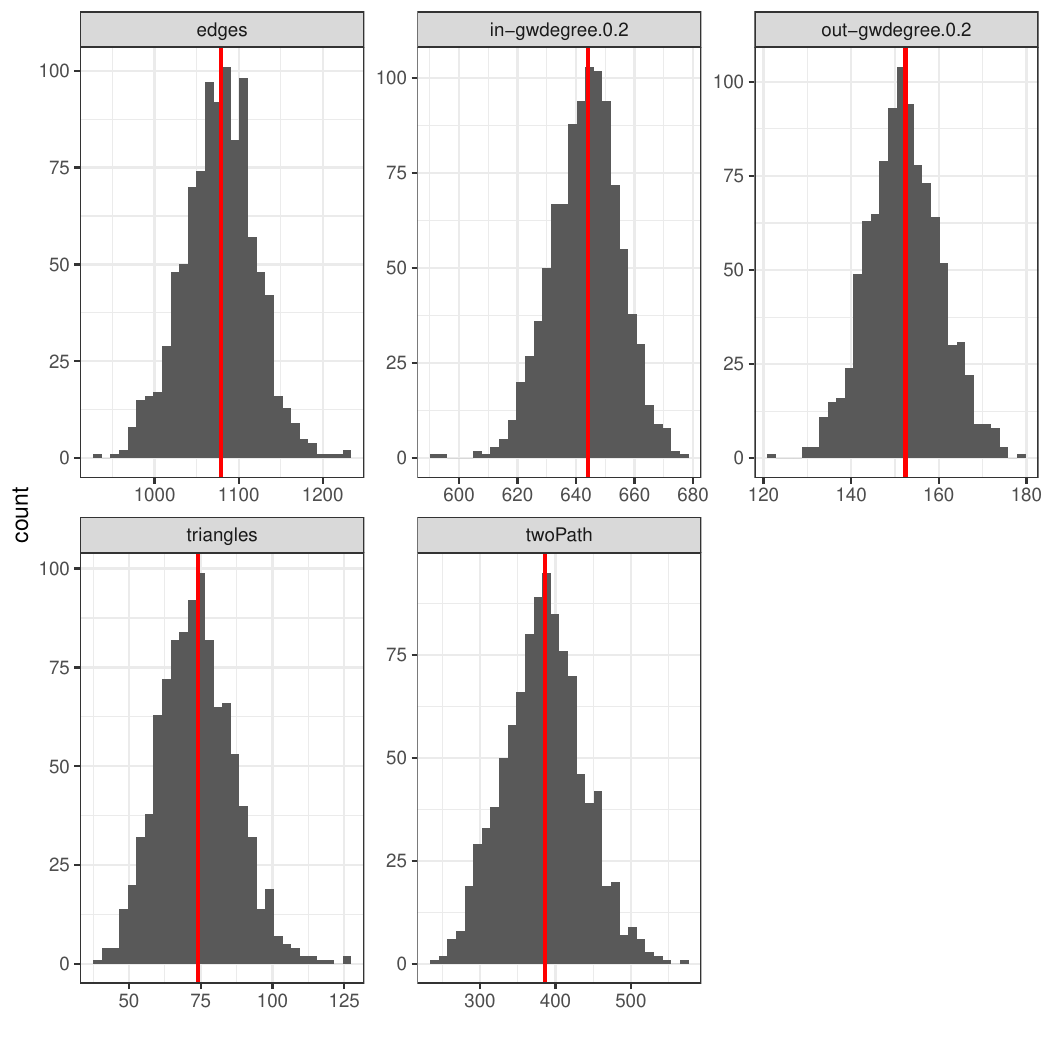}}
  \subfigure[Edges]{\includegraphics[scale=0.4]{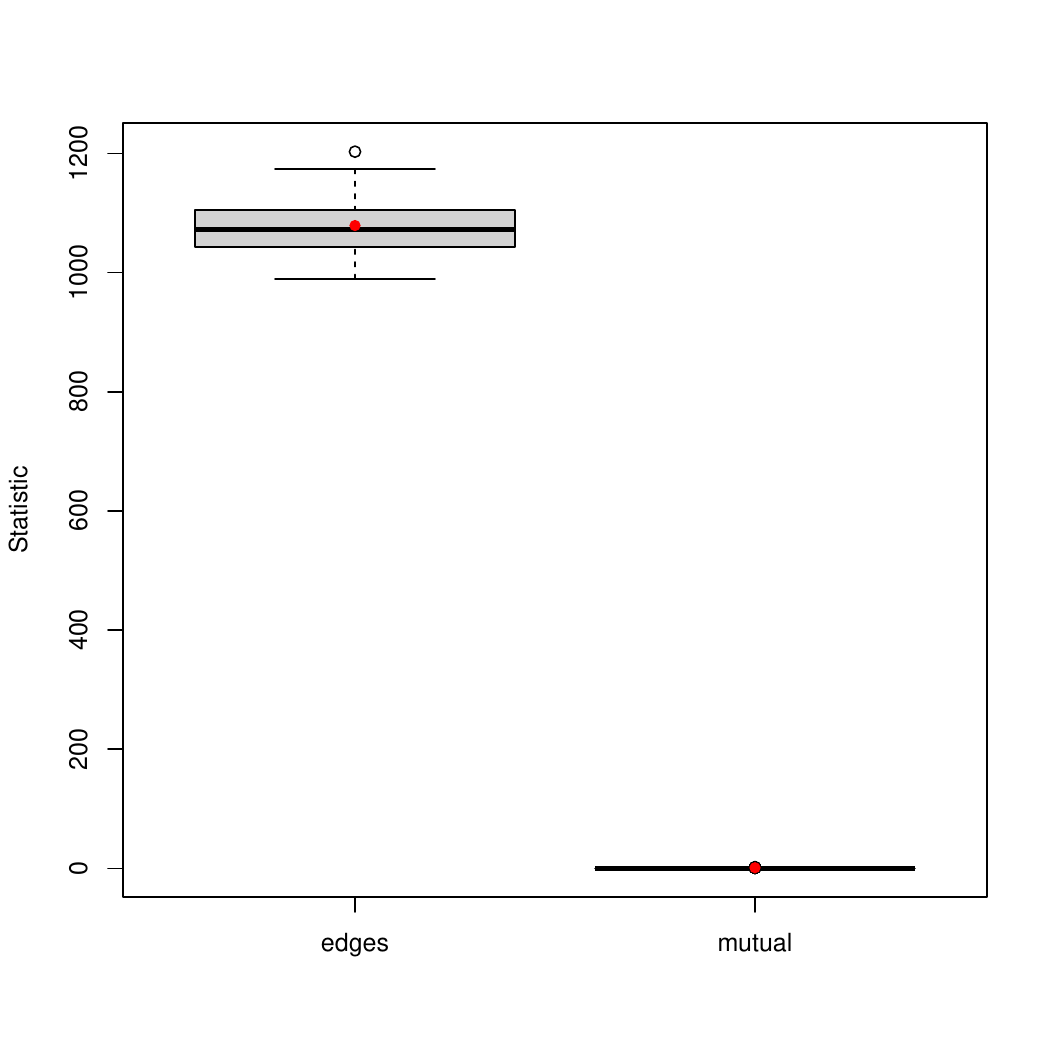}}
  \subfigure[In-degree distribution]{\includegraphics[scale=0.3]{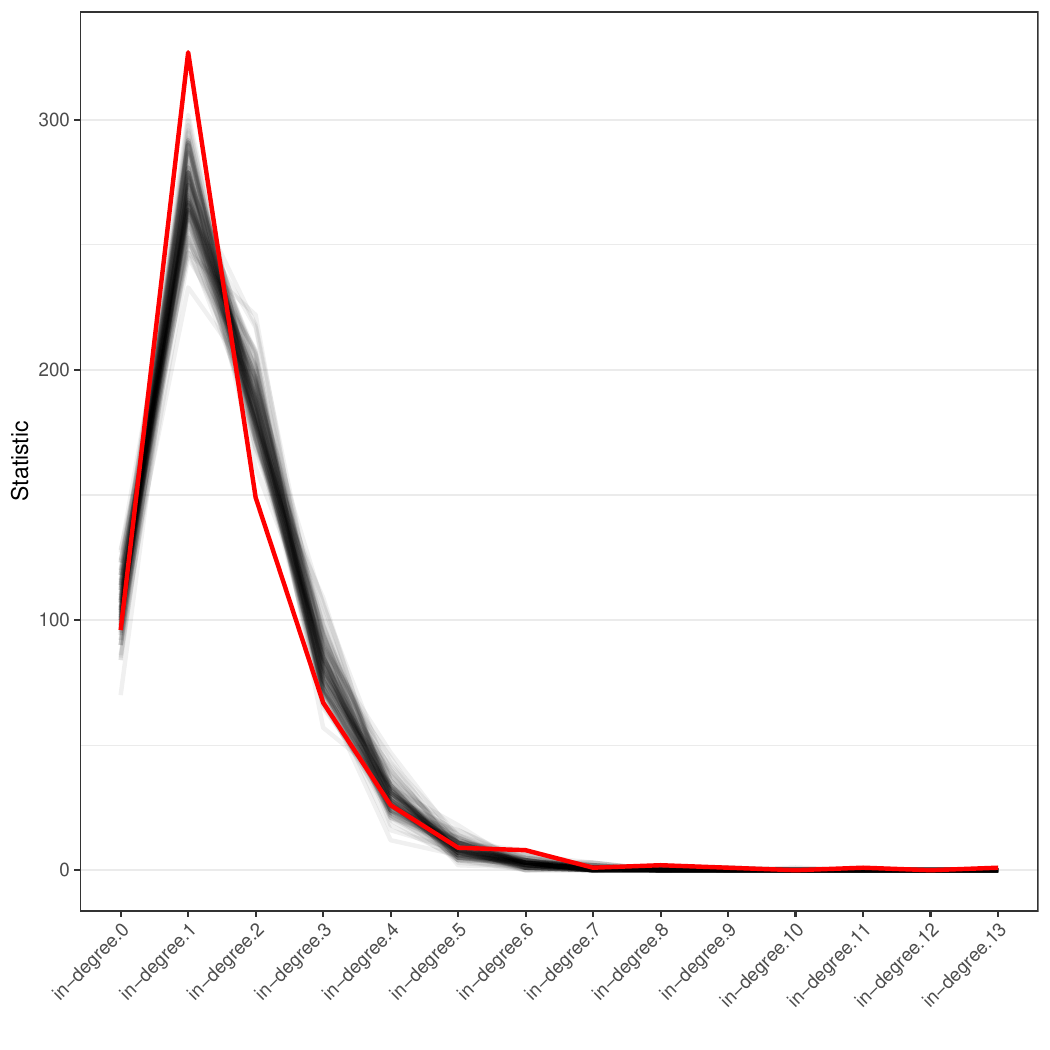}}
  \subfigure[Out-degree distribution]{\includegraphics[scale=0.3]{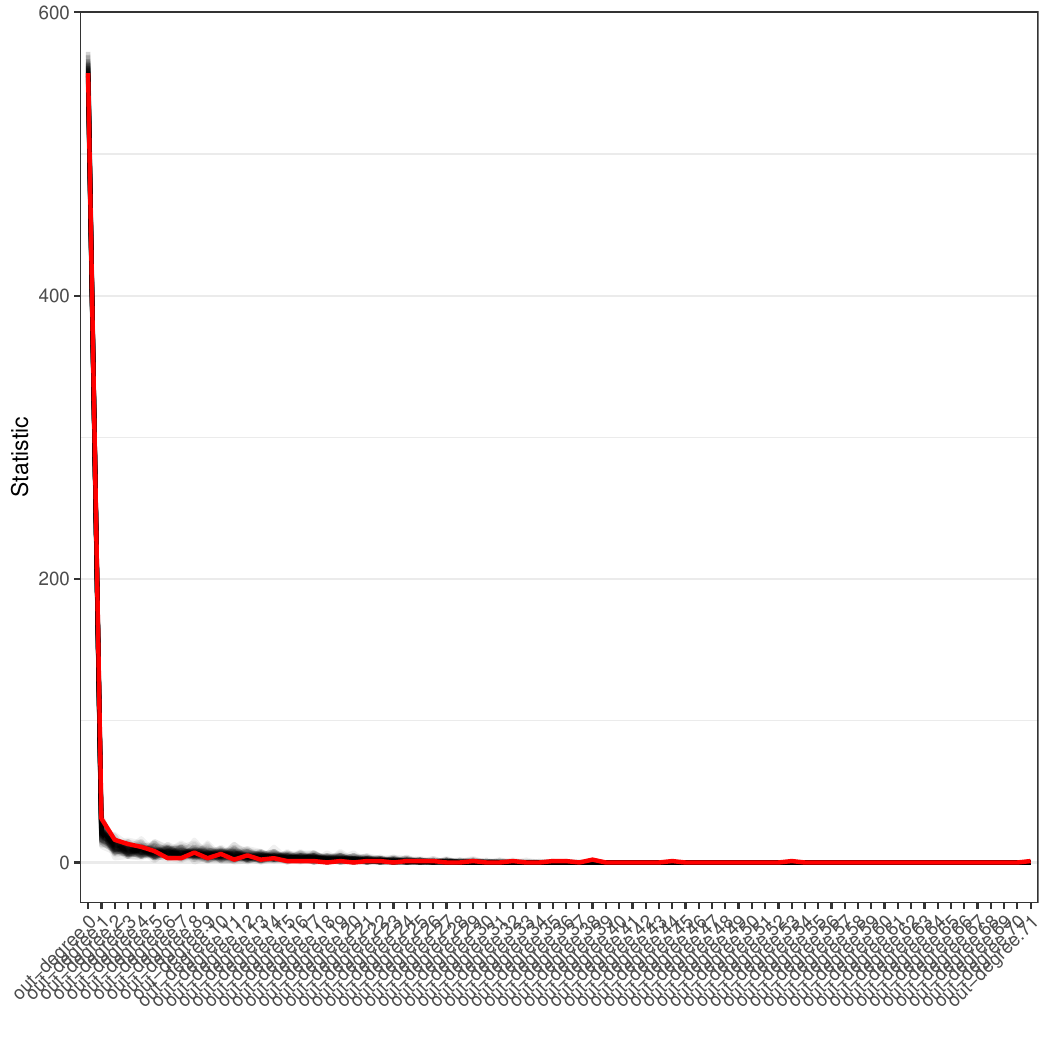}}
  \subfigure[Edgewise shared partners]{\includegraphics[scale=0.3]{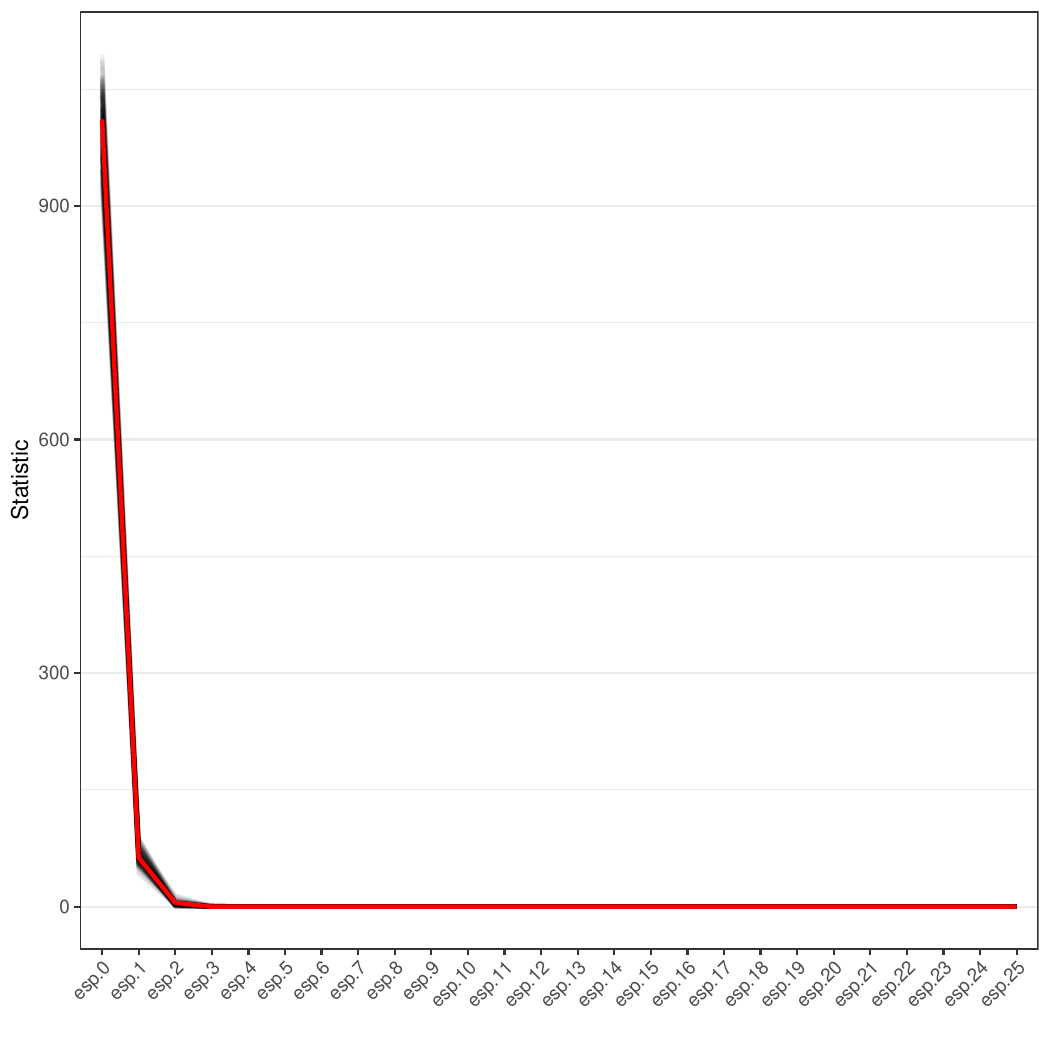}}
  \caption{Model diagnostic and goodness-of-fit plots for the Alon yeast network LOLOG model, Table~\ref{tab:alon_yeast_transcription_lolog}.}
  \label{fig:yeast_transcription_lolog_gof}
\end{figure}

\begin{figure}
  \centering
  \includegraphics[angle=270,width=\textwidth]{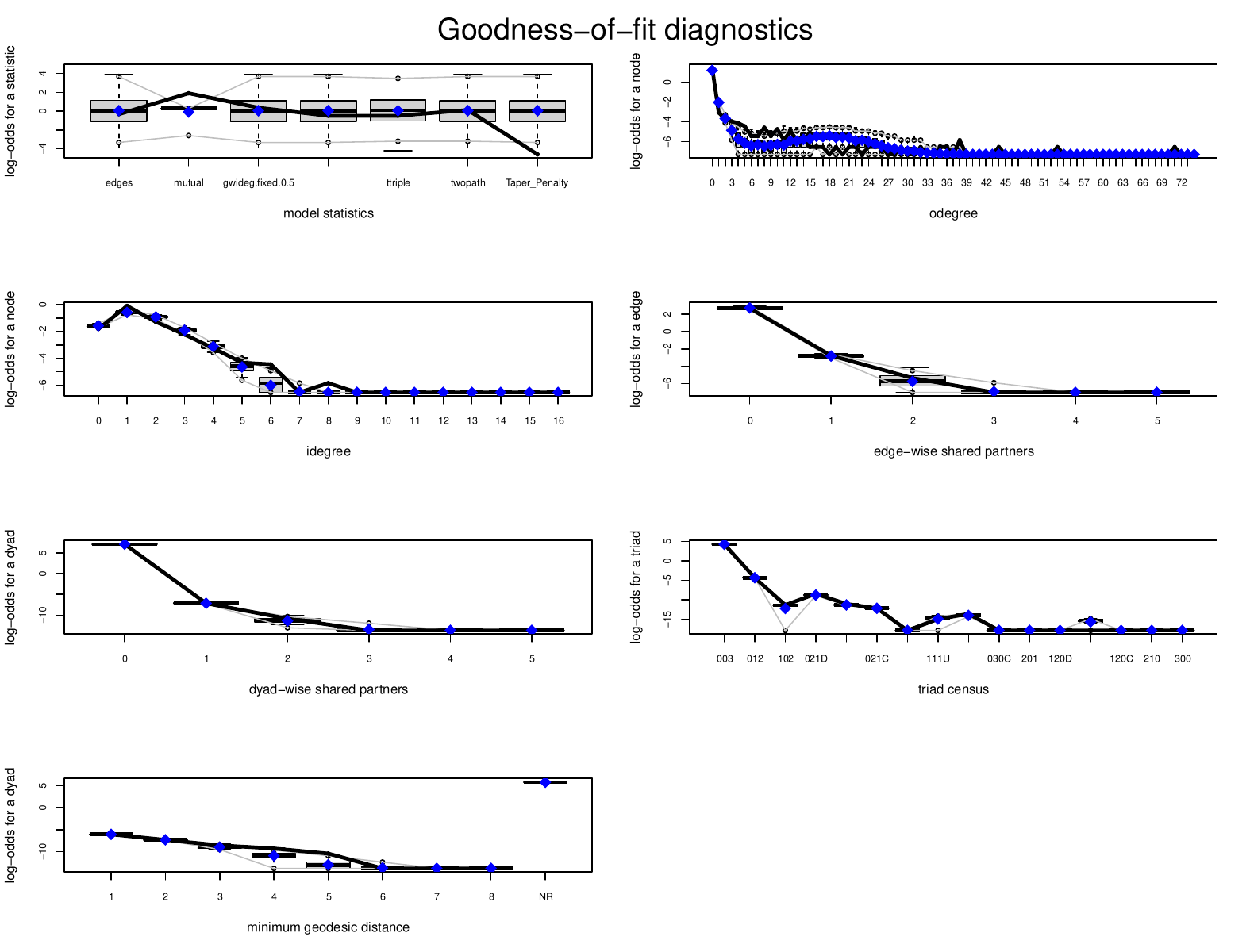}
  \caption{Goodness-of-fit plots for the tapered ERGM model of the
    Alon yeast regulatory network
    (Table~\ref{tab:yeast_transcription_tapered_estimations}).}
  \label{fig:yeast_transcription_tapered_gof}
\end{figure}

\begin{figure}
  \centering
  \subfigure[Model diagnostic plots]{\includegraphics[scale=.5]{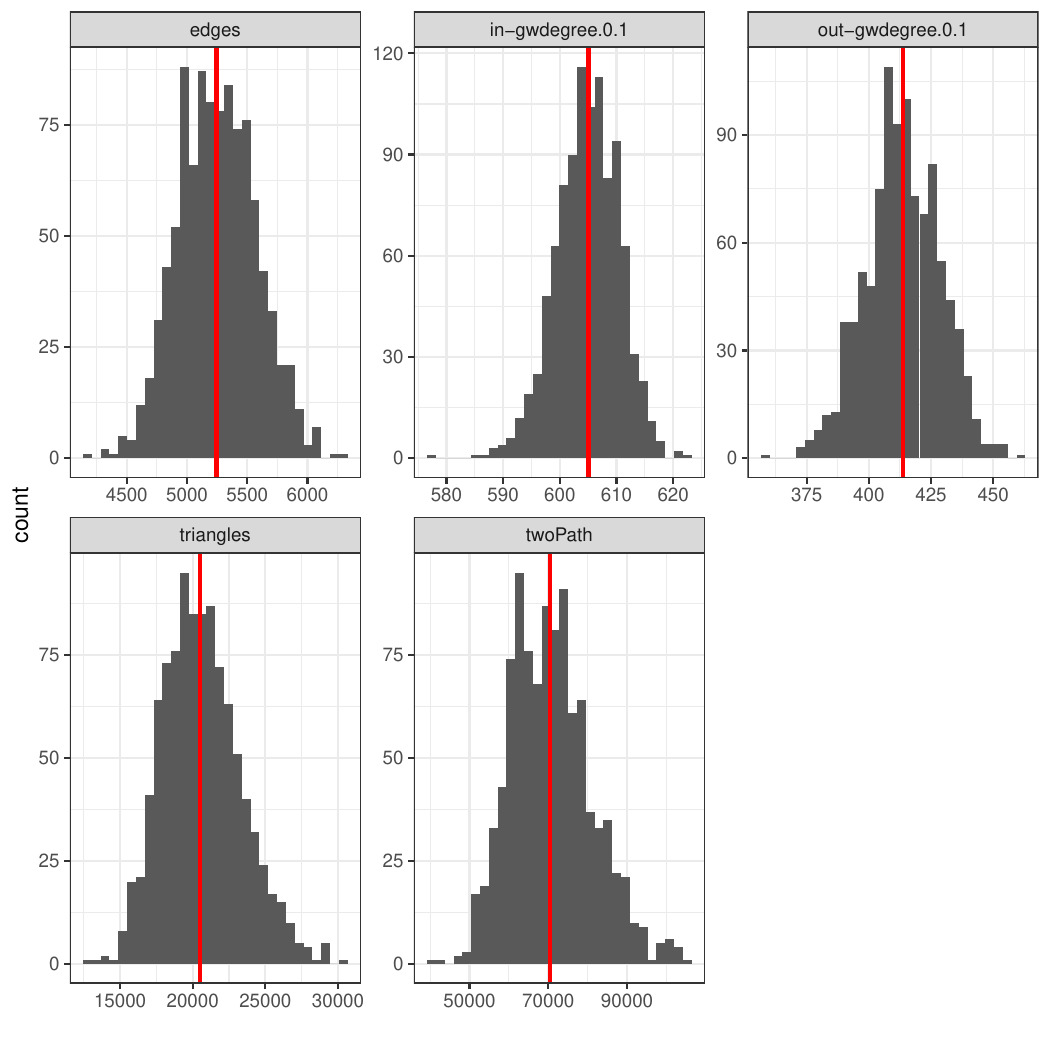}}
  \subfigure[Edges]{\includegraphics[scale=0.4]{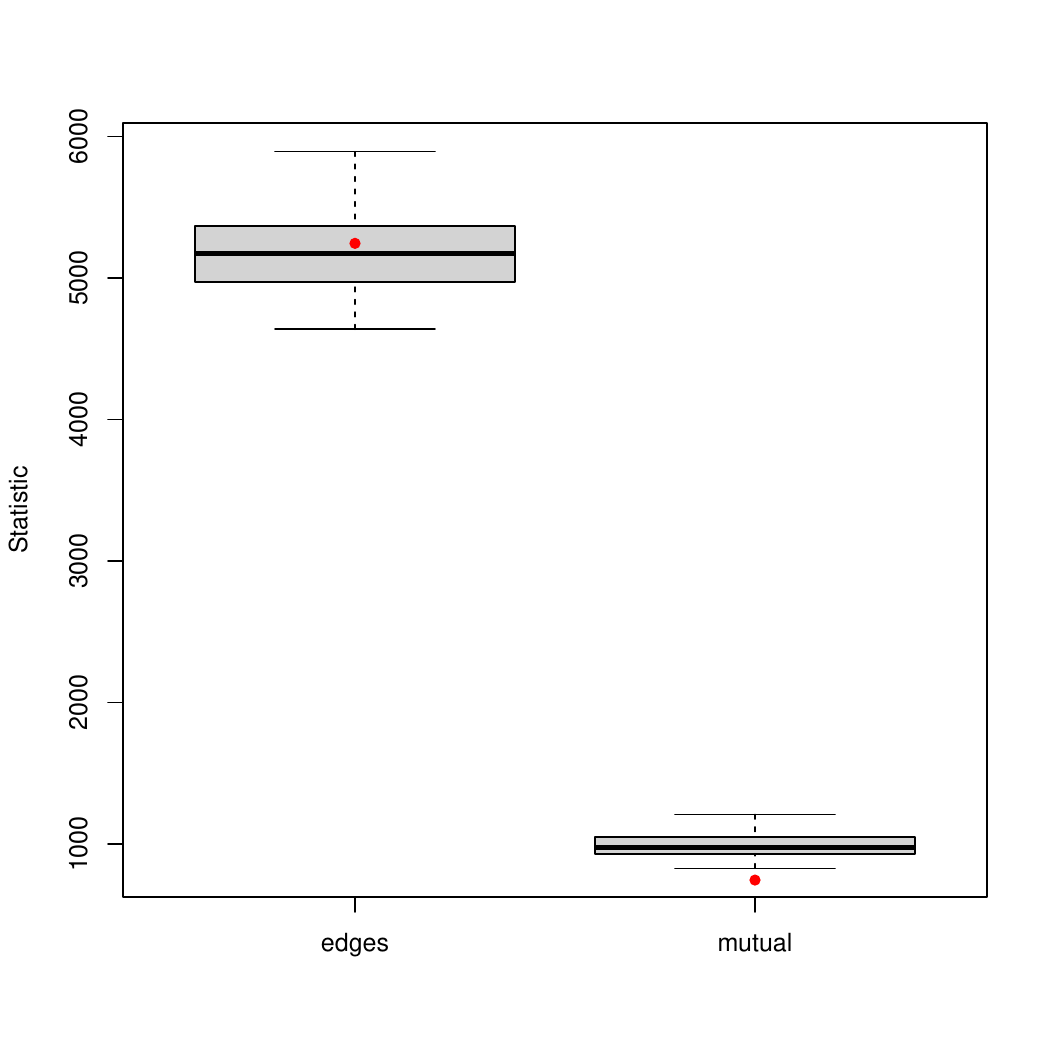}}
  \subfigure[In-degree distribution]{\includegraphics[scale=0.3]{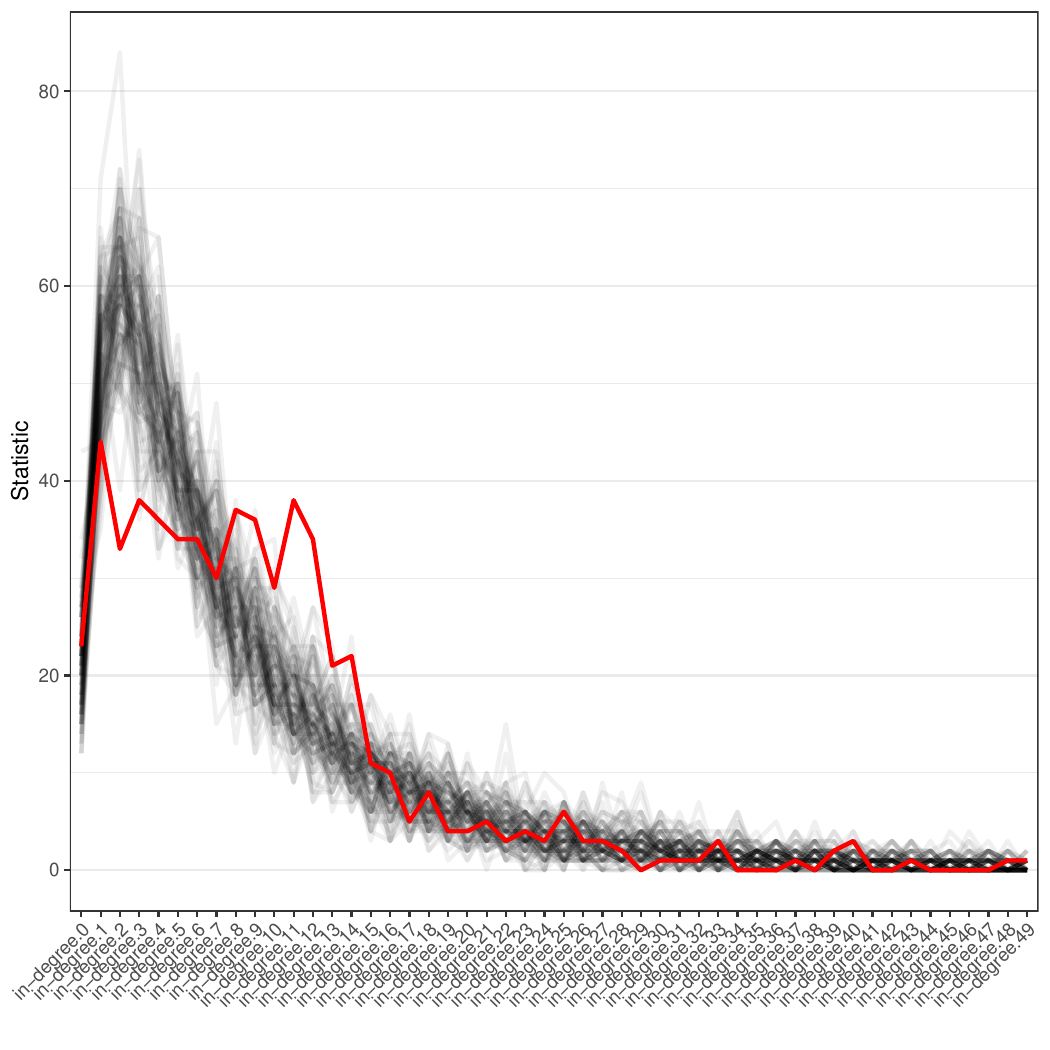}}
  \subfigure[Out-degree distribution]{\includegraphics[scale=0.3]{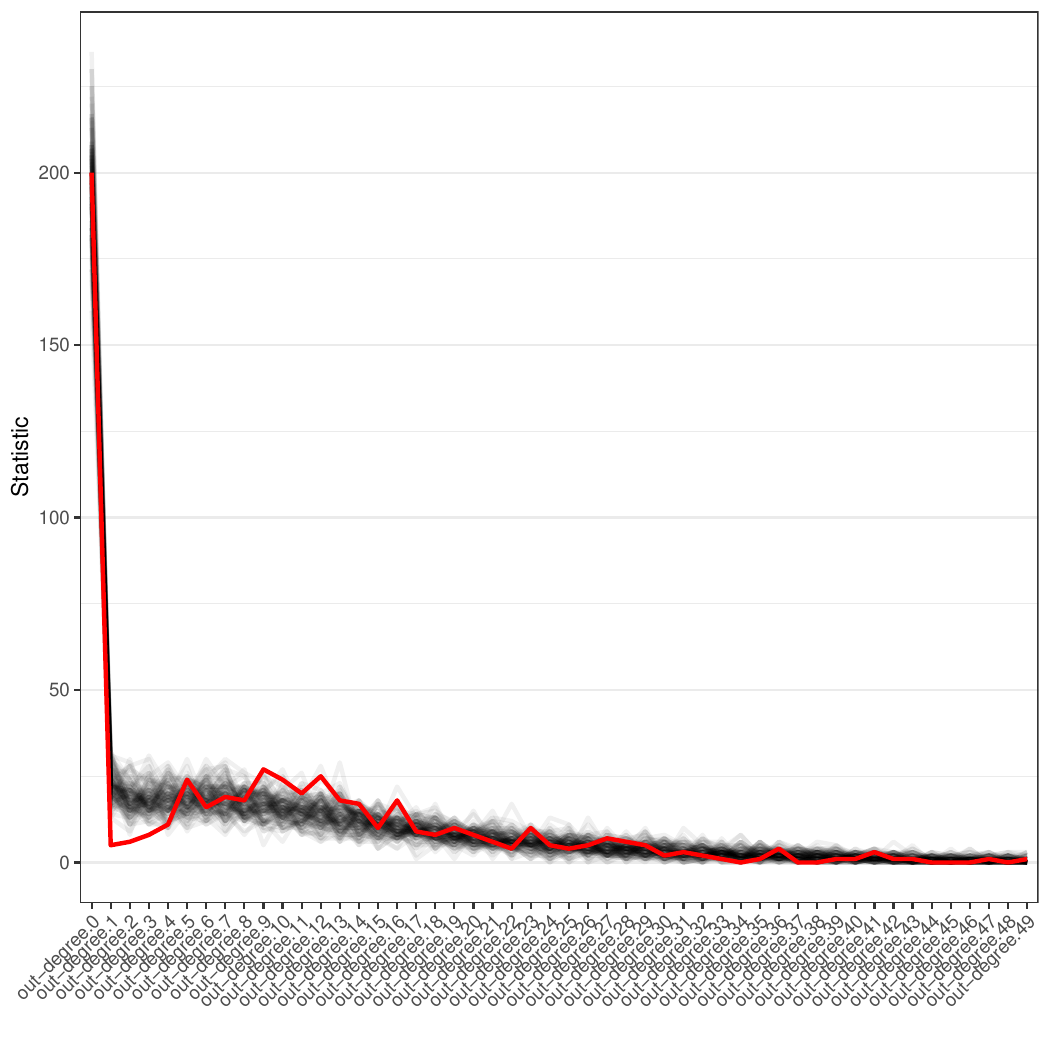}}
  \subfigure[Edgewise shared partners]{\includegraphics[scale=0.3]{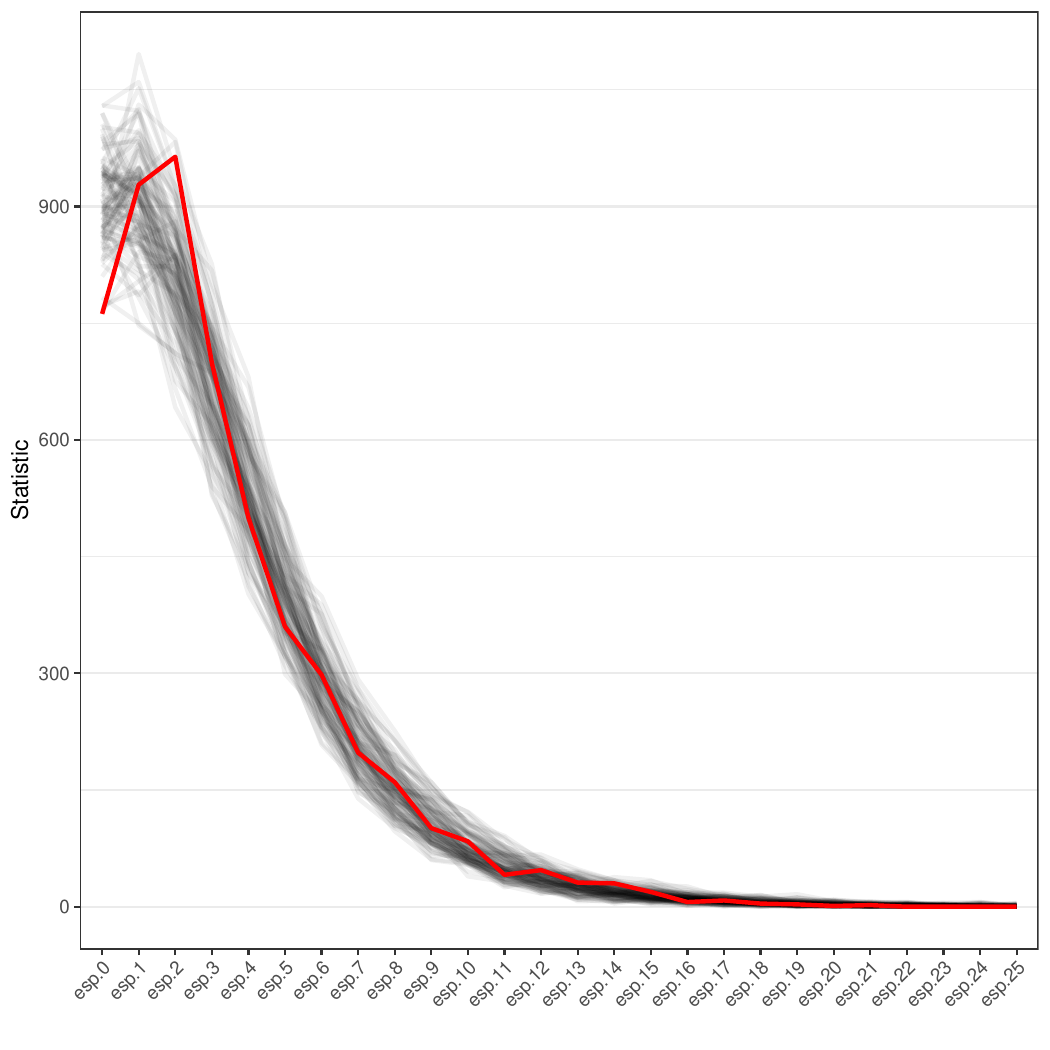}}
  \caption{Model diagnostic and goodness-of-fit plots for the Cook \textit{C.~elegans} connectome LOLOG model, Table~\ref{tab:celegans_neural_lolog}.}
  \label{fig:celegans_neural_lolog_gof}
\end{figure}

\begin{figure}
  \centering
  \subfigure[Model 1]{\includegraphics[angle=270,width=.8\textwidth]{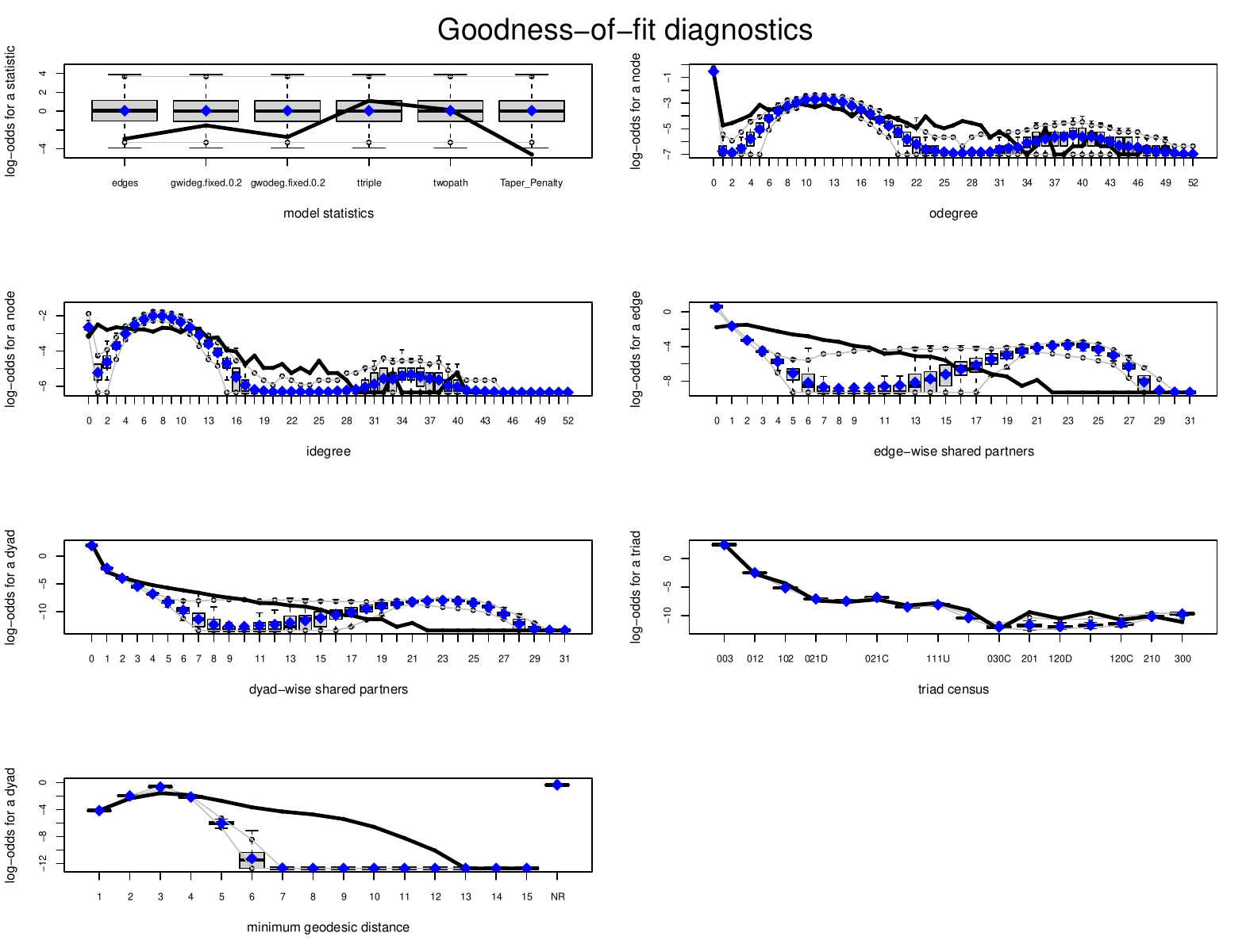}  }
  \subfigure[Model 2]{\includegraphics[angle=270,width=.8\textwidth]{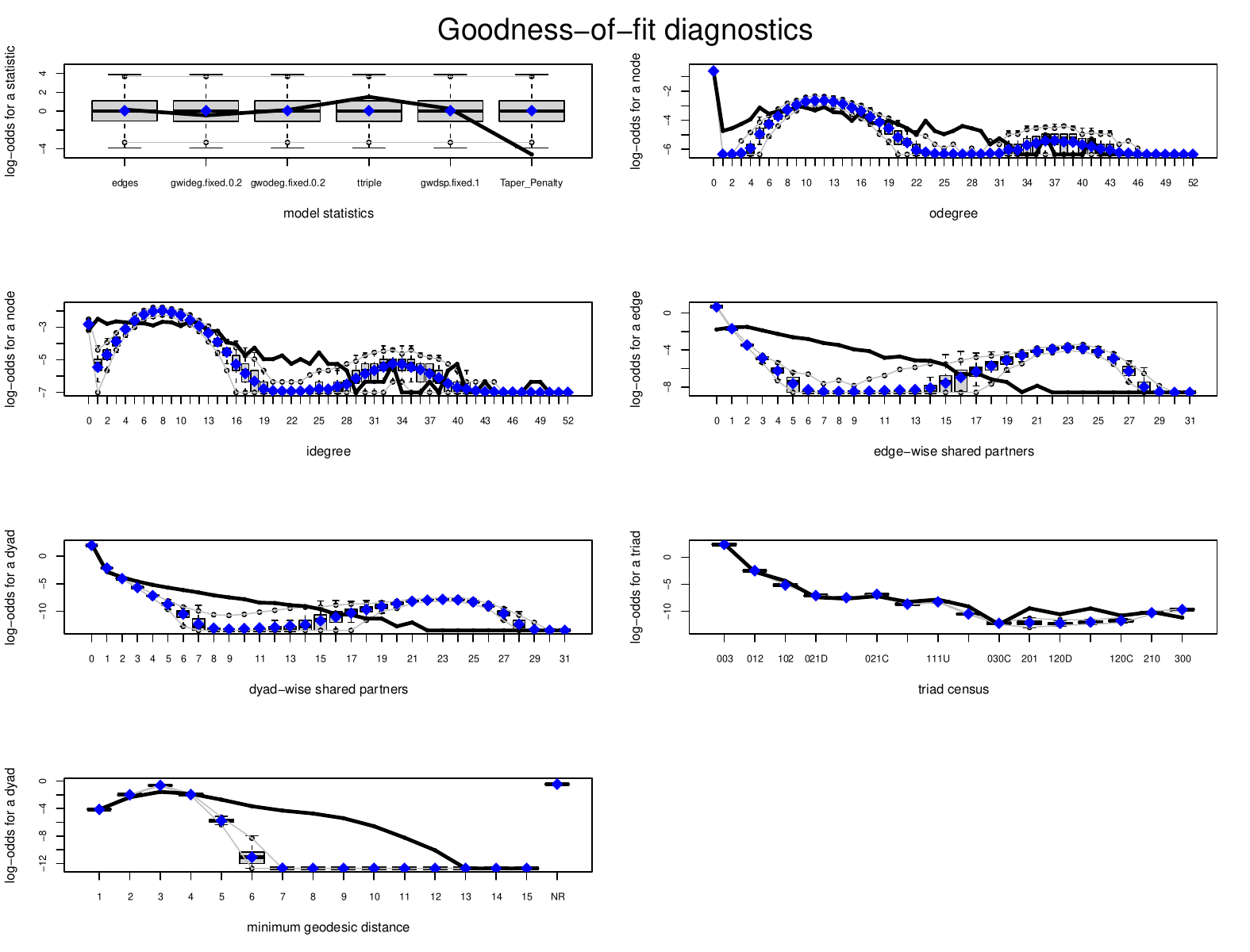}}
  \caption{Goodness-of-fit plots for the tapered ERGM models of the
    Cook \textit{C.~elegans} connectome
    (Table~\ref{tab:celegans_tapered_estimations}).}
  \label{fig:celegans_tapered_gof}
\end{figure}

\begin{figure}
  \centering
  \subfigure[Model diagnostic plots]{\includegraphics[scale=.5]{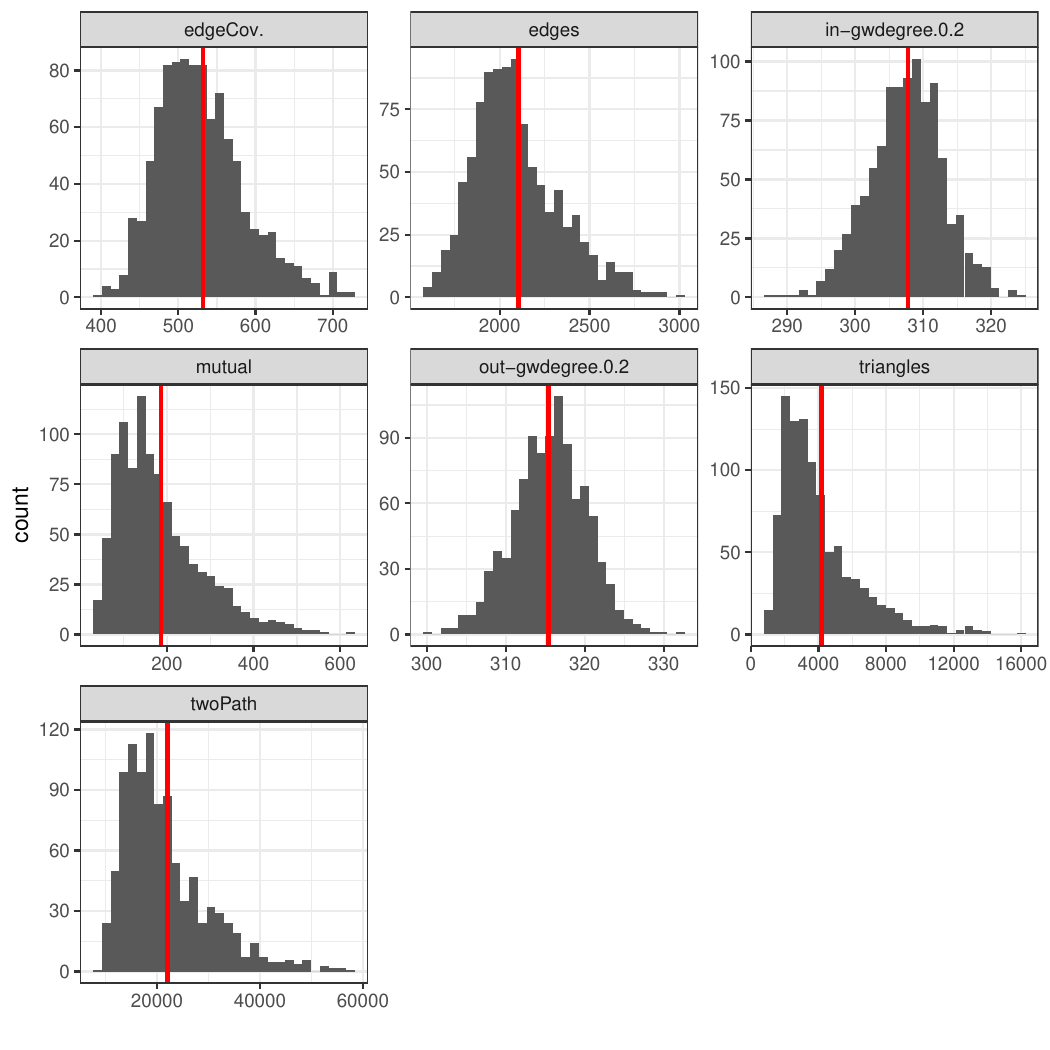}}
  \subfigure[Edges]{\includegraphics[scale=0.4]{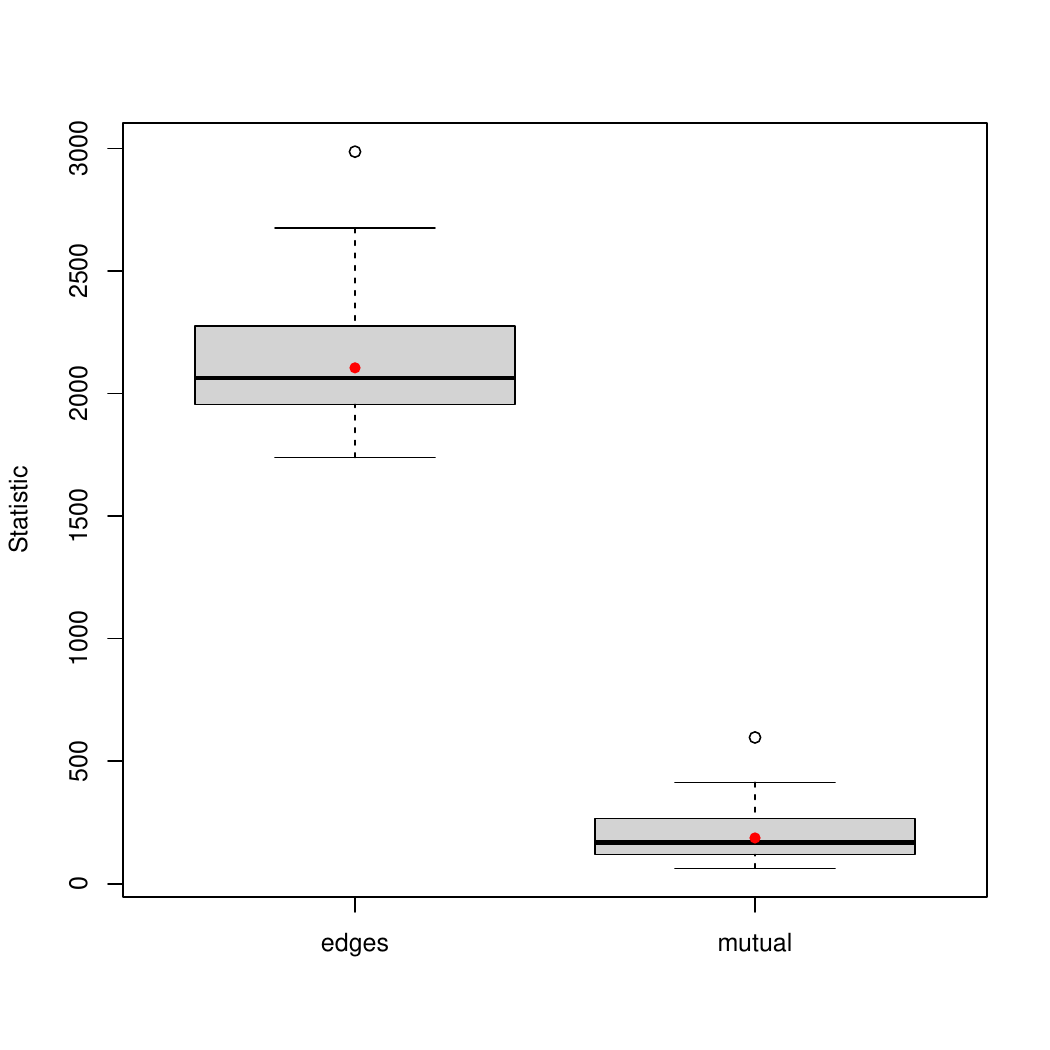}}
  \subfigure[In-degree distribution]{\includegraphics[scale=0.3]{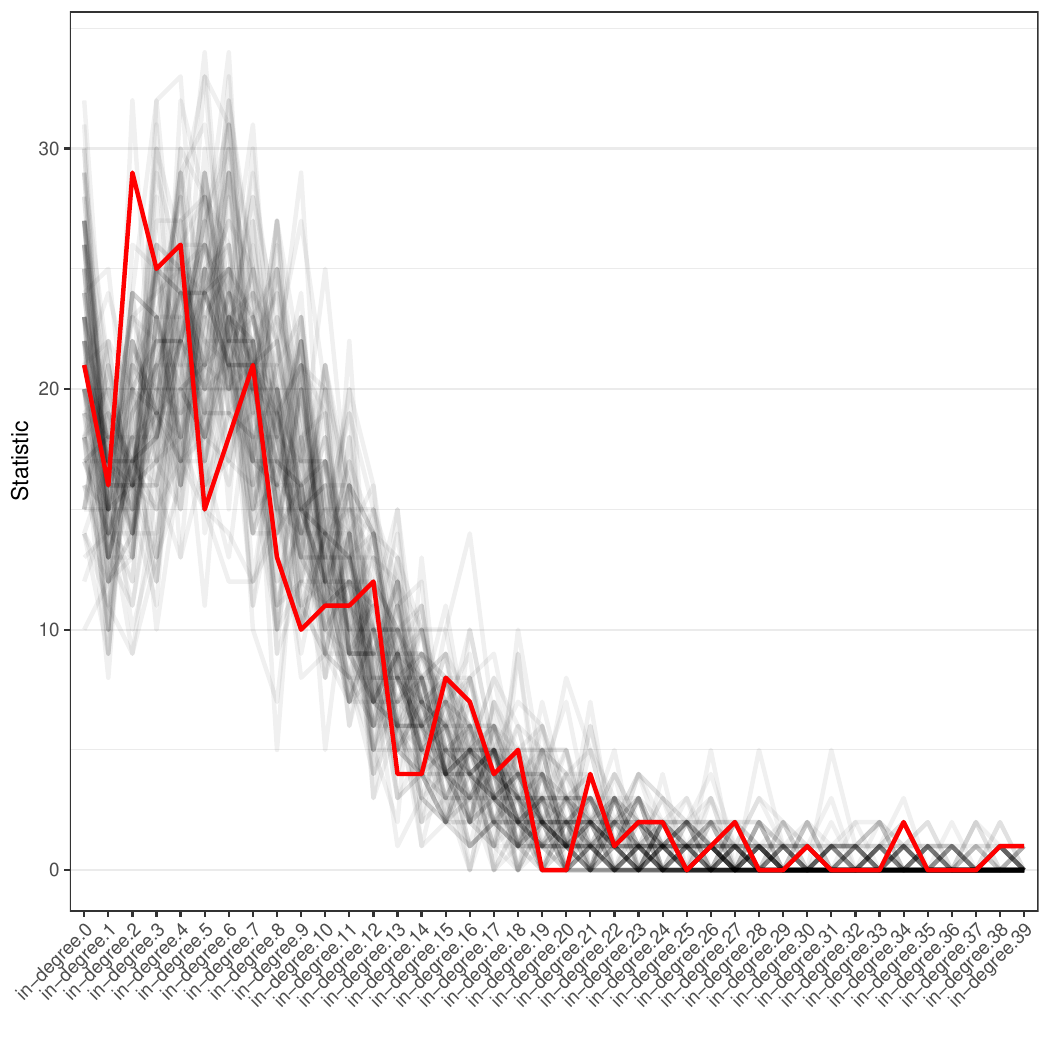}}
  \subfigure[Out-degree distribution]{\includegraphics[scale=0.3]{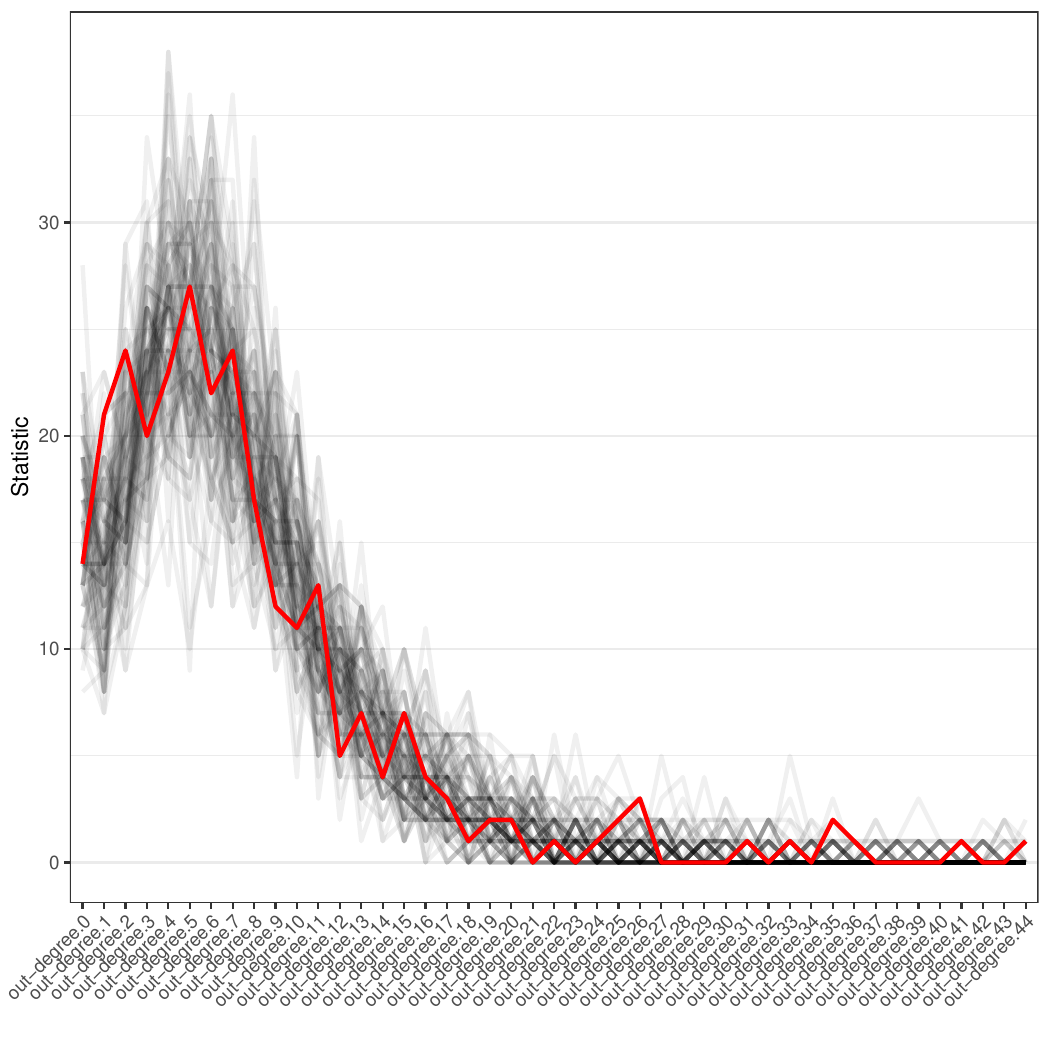}}
  \subfigure[Edgewise shared partners]{\includegraphics[scale=0.3]{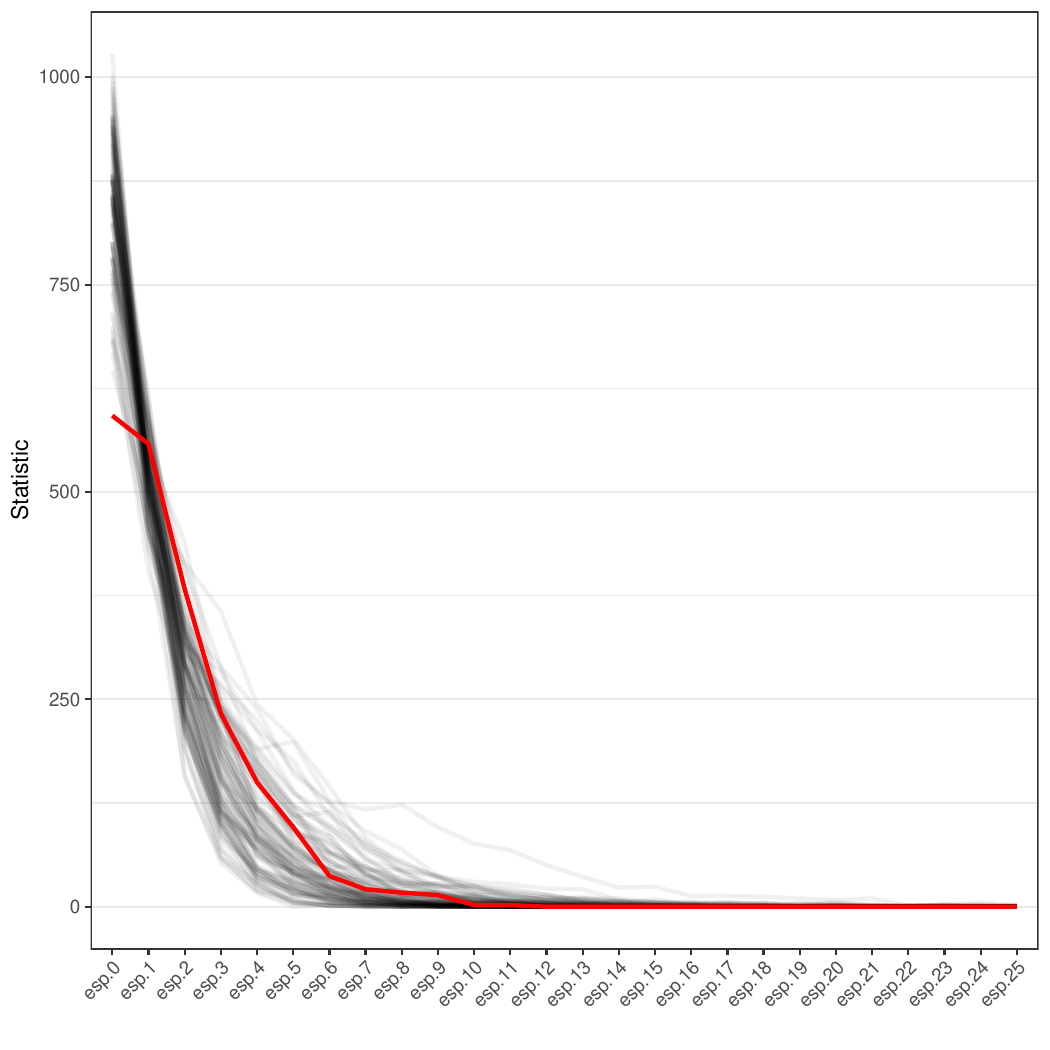}}
  \caption{Model diagnostic and goodness-of-fit plots for the Kaiser
    \textit{C.~elegans} neural network LOLOG Model 1
    (Table~\ref{tab:celegans277_neural_lolog}).}
  \label{fig:celegans277_neural_lolog_gof1}
\end{figure}

\begin{figure}
  \centering
  \subfigure[Model diagnostic plots]{\includegraphics[scale=.5]{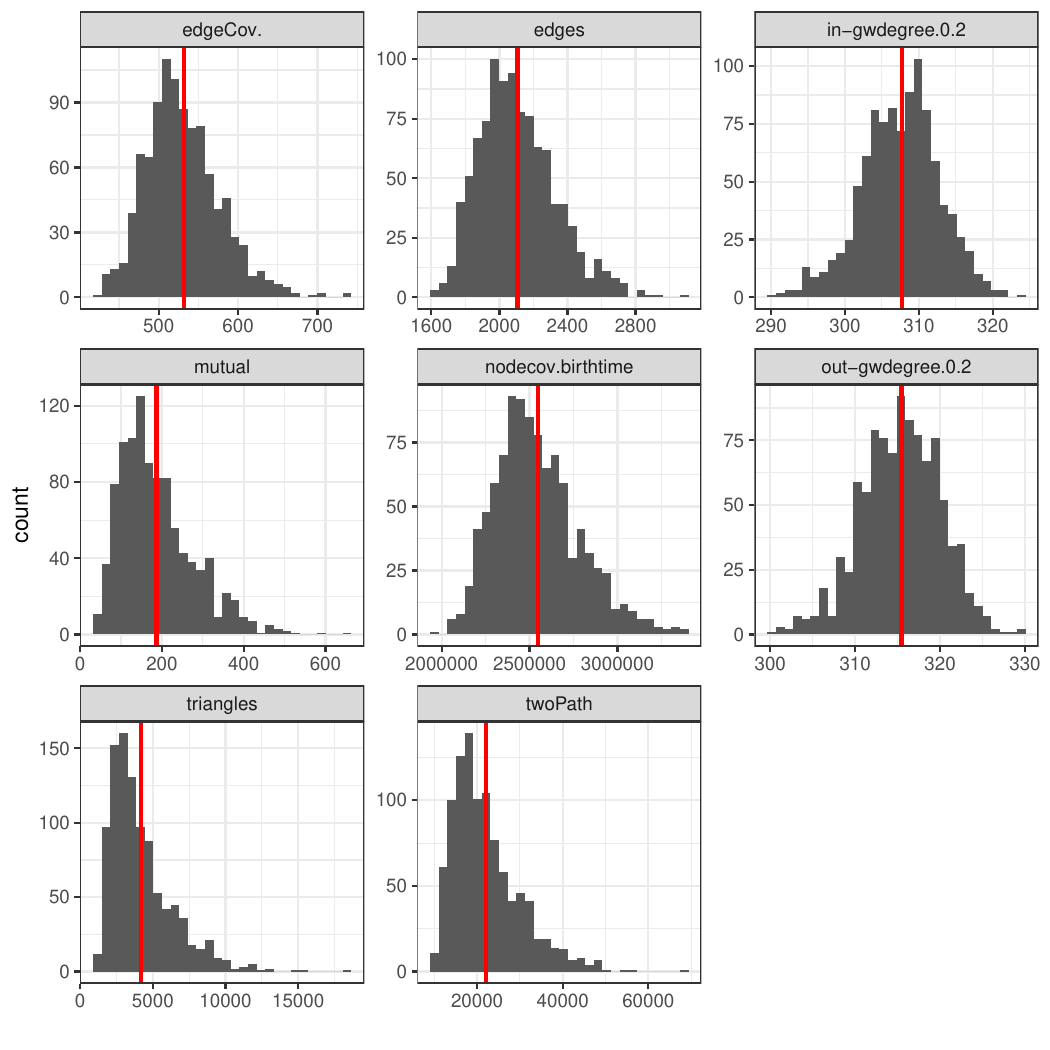}}
  \subfigure[Edges]{\includegraphics[scale=0.4]{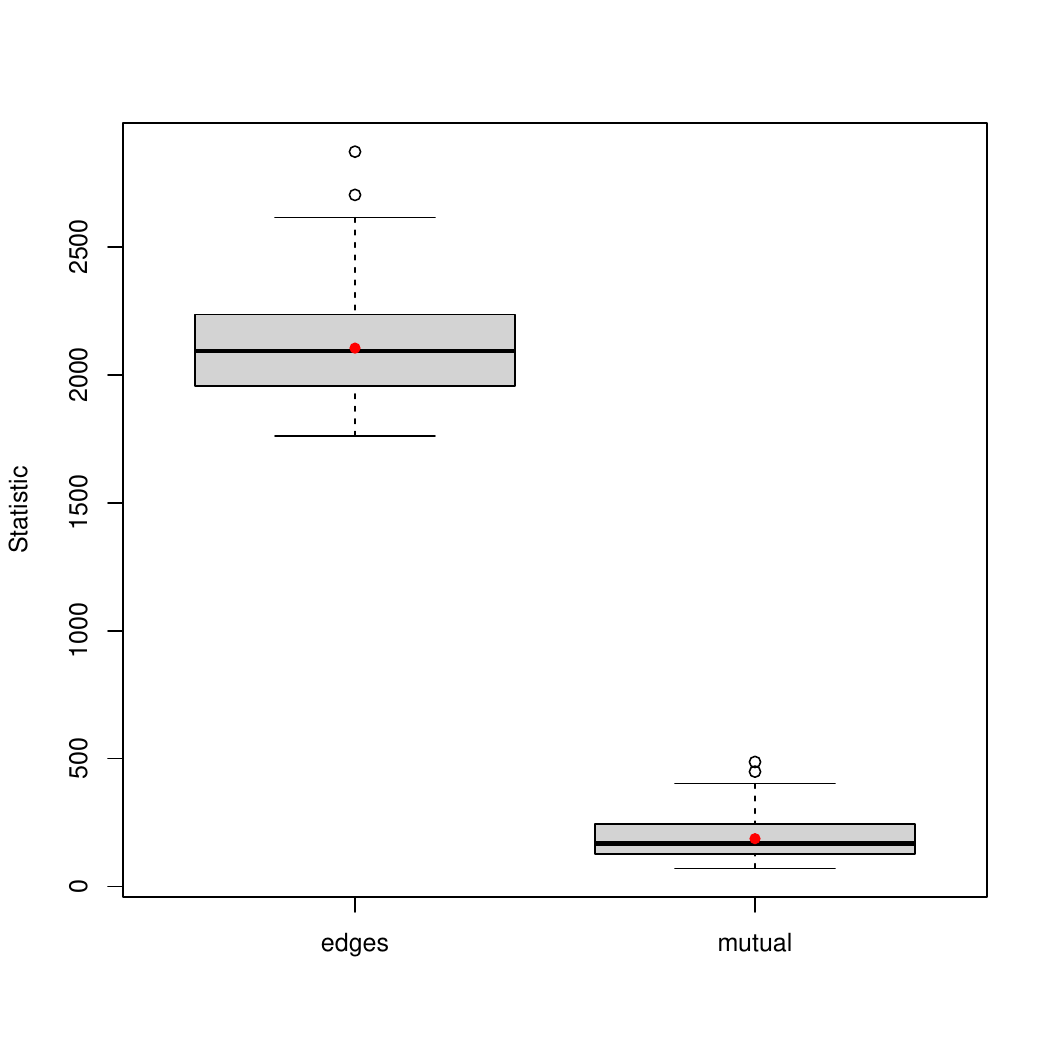}}
  \subfigure[In-degree distribution]{\includegraphics[scale=0.3]{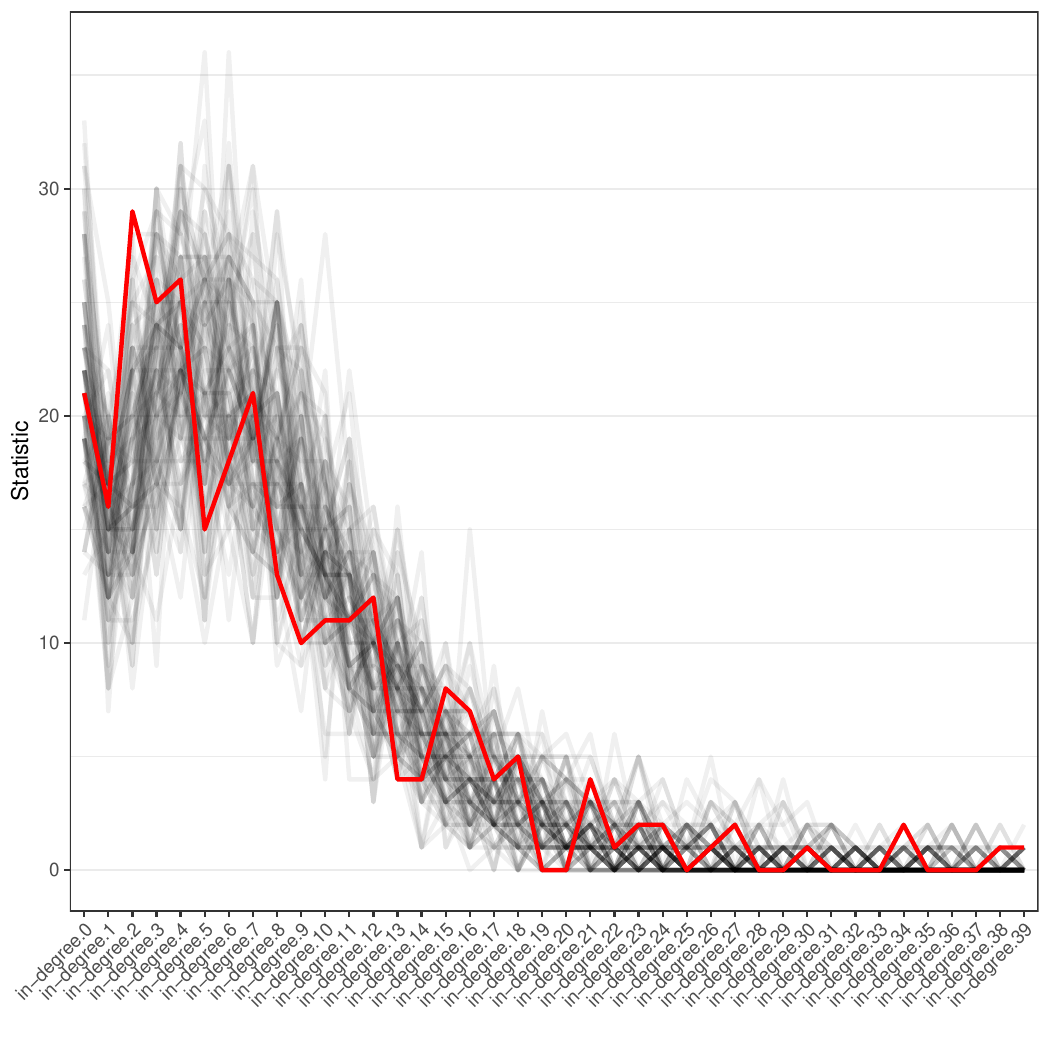}}
  \subfigure[Out-degree distribution]{\includegraphics[scale=0.3]{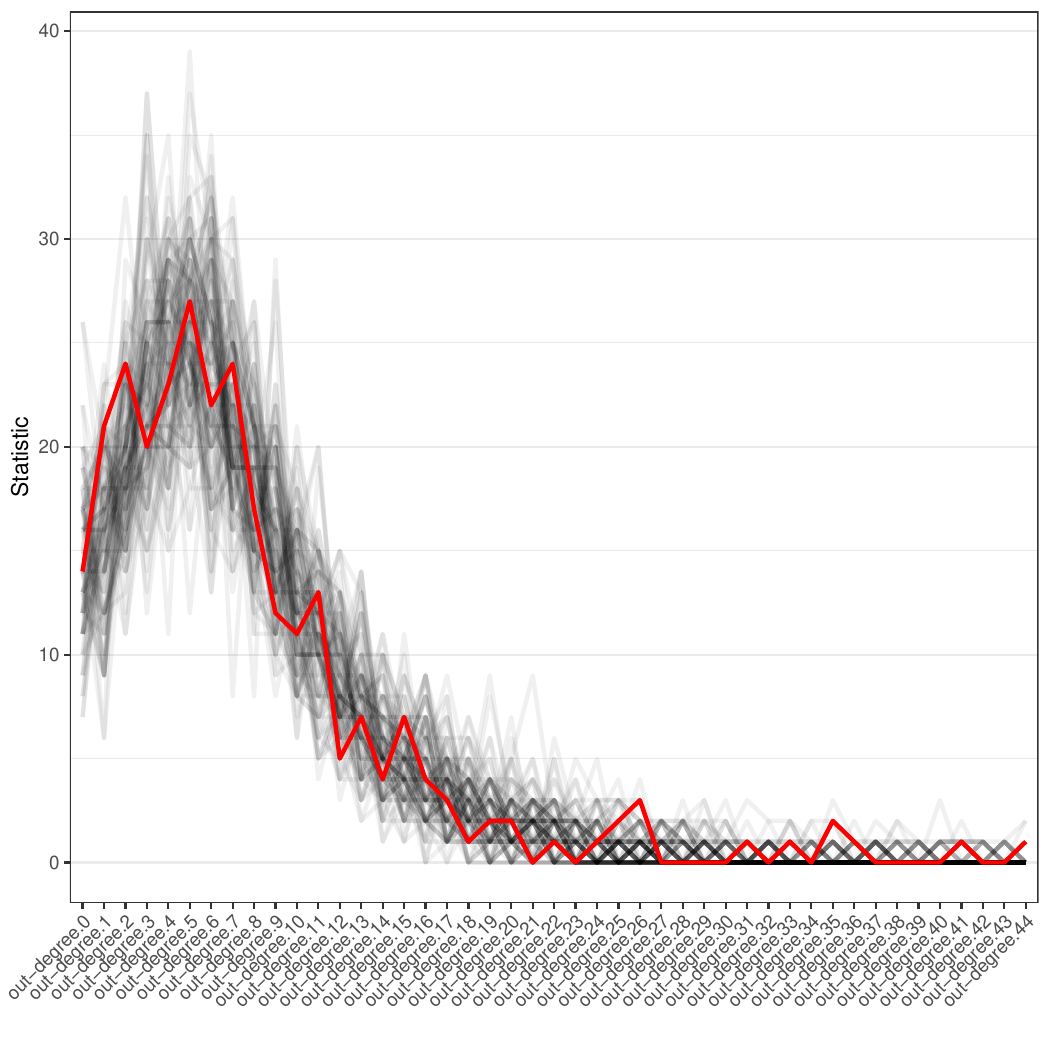}}
  \subfigure[Edgewise shared partners]{\includegraphics[scale=0.3]{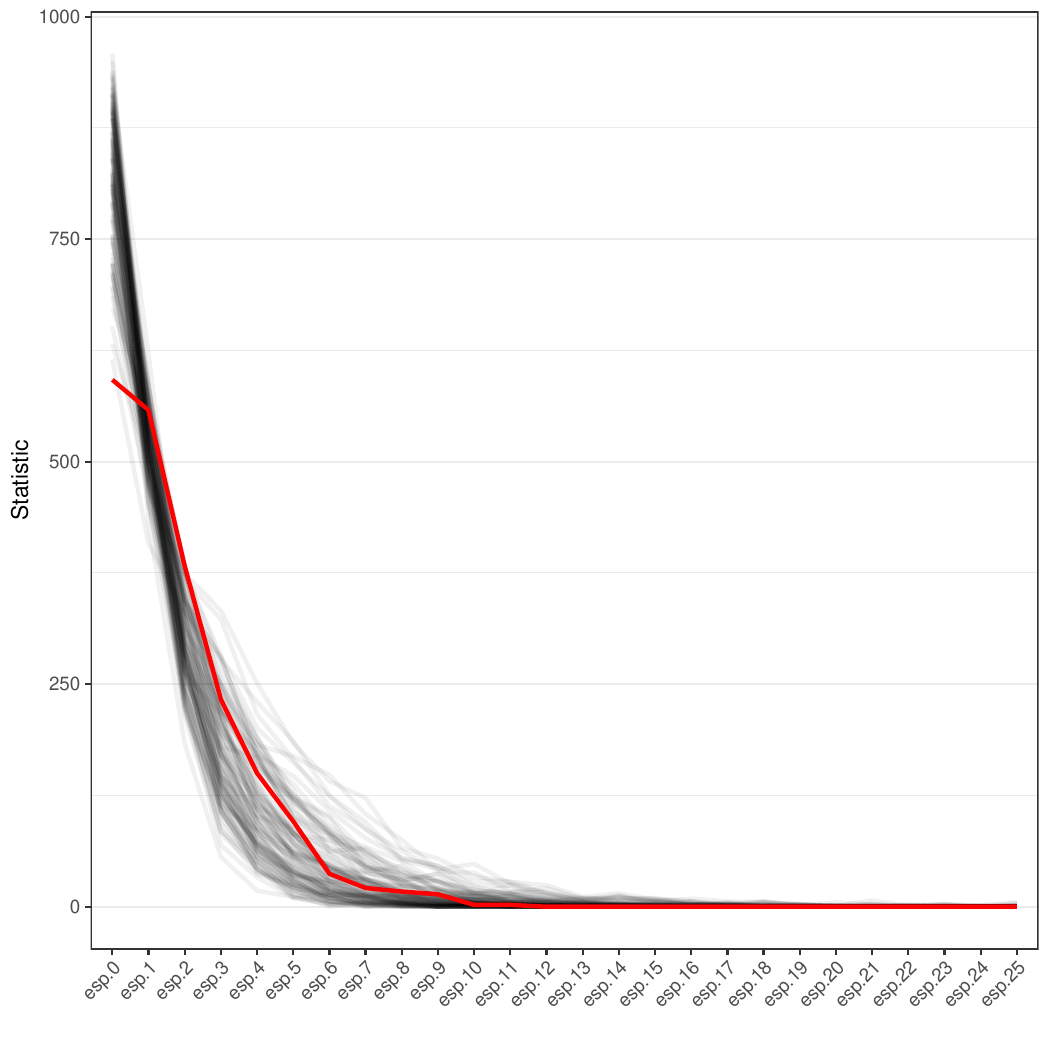}}
  \caption{Model diagnostic and goodness-of-fit plots for the Kaiser
    \textit{C.~elegans} neural network LOLOG Model 2
    (Table~\ref{tab:celegans277_neural_lolog}).}
  \label{fig:celegans277_neural_lolog_gof2}
\end{figure}

\begin{figure}
  \centering
  \subfigure[Model diagnostic plots]{\includegraphics[scale=.5]{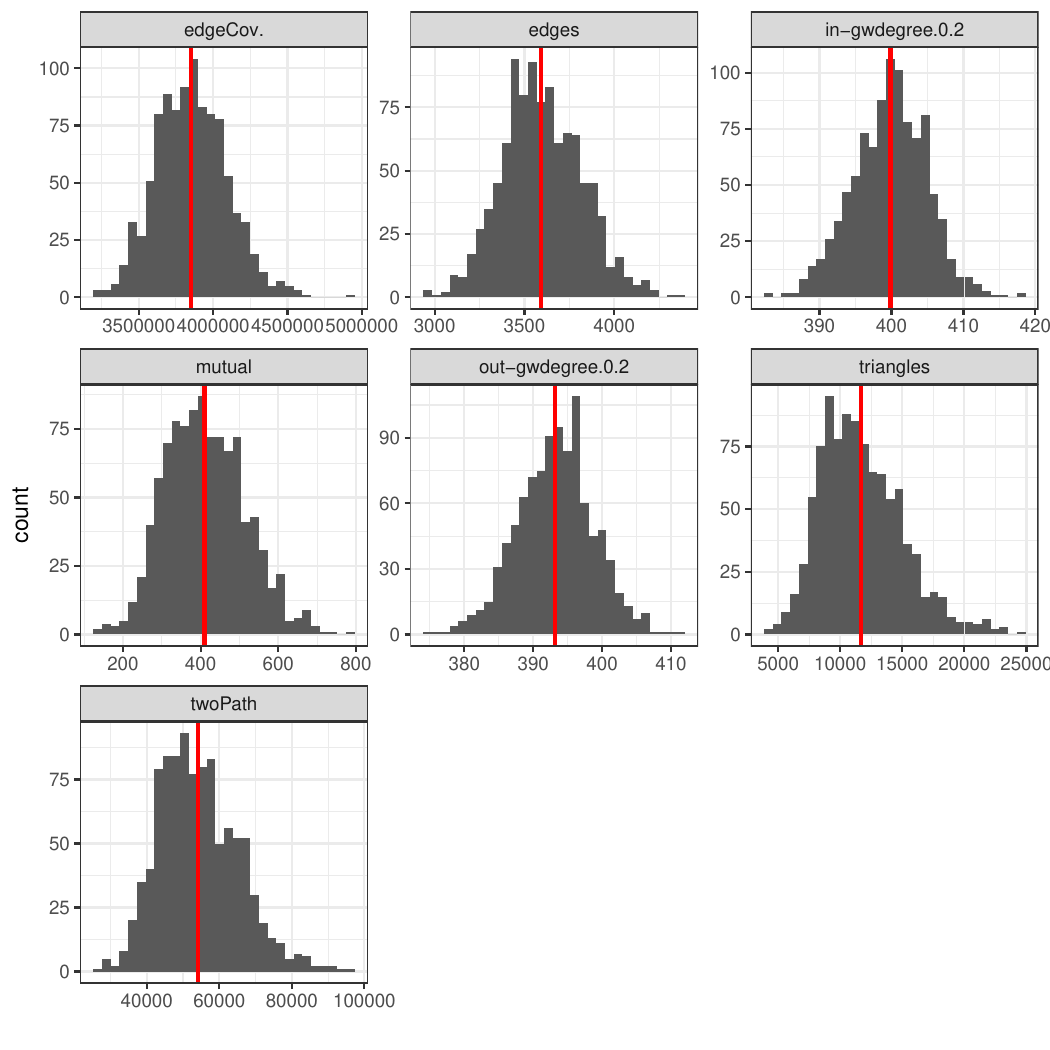}}
  \subfigure[Edges]{\includegraphics[scale=0.4]{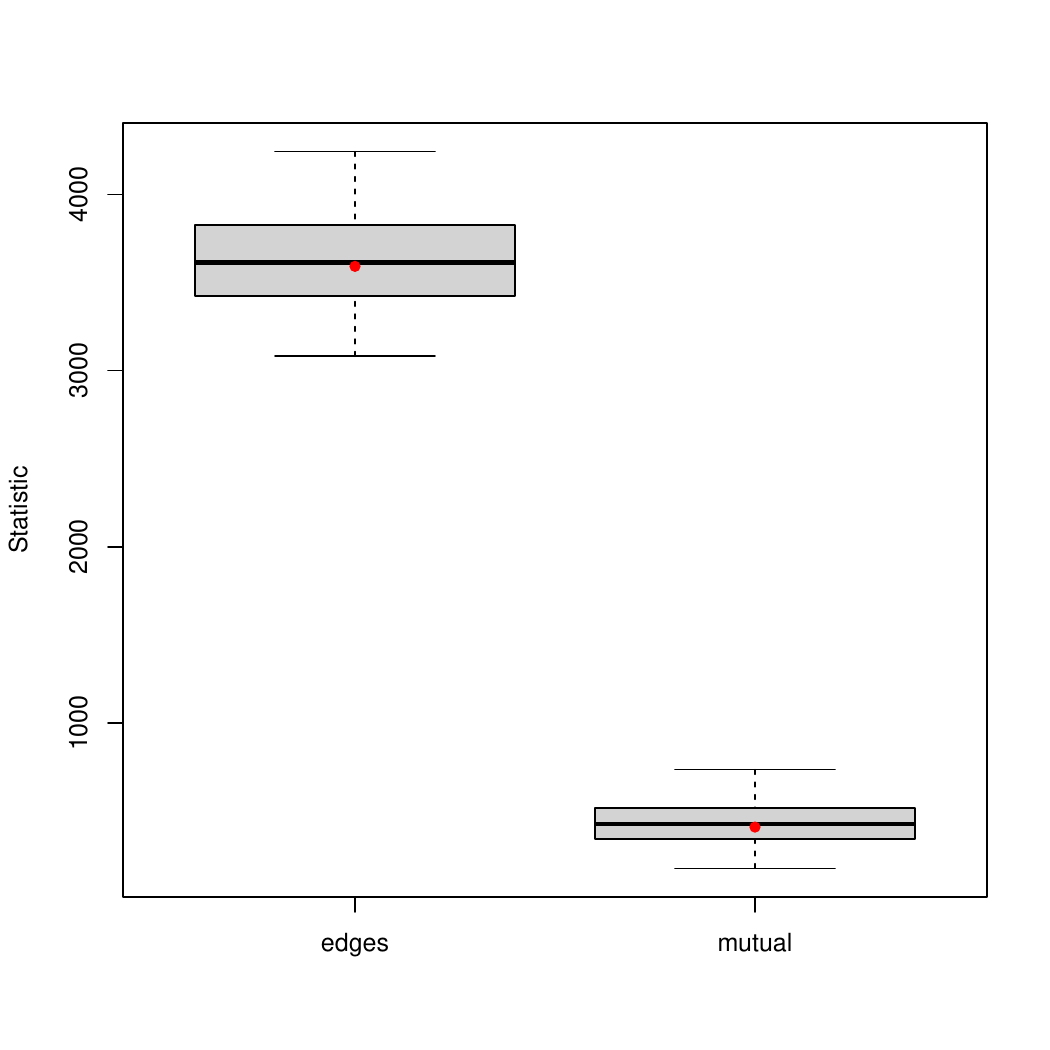}}
  \subfigure[In-degree distribution]{\includegraphics[scale=0.3]{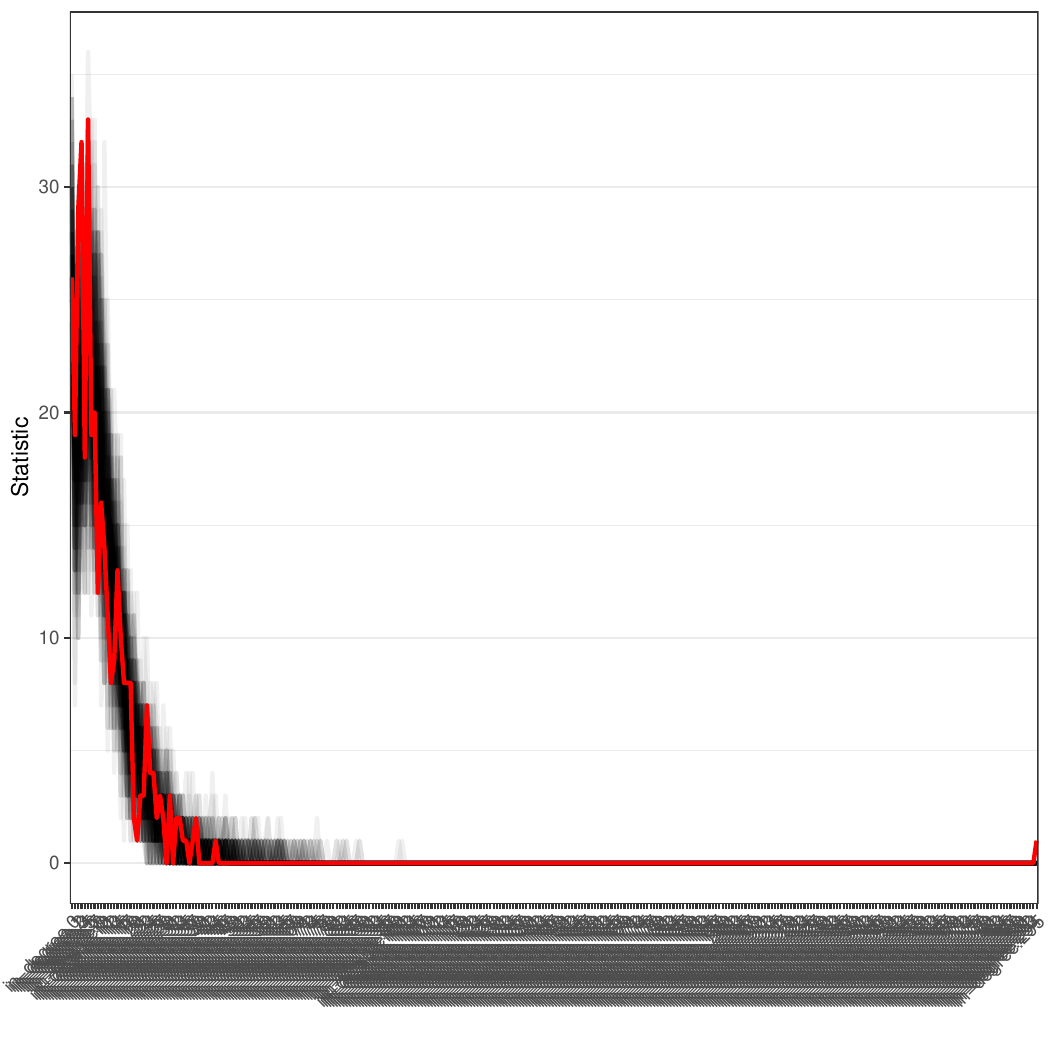}}
  \subfigure[Out-degree distribution]{\includegraphics[scale=0.3]{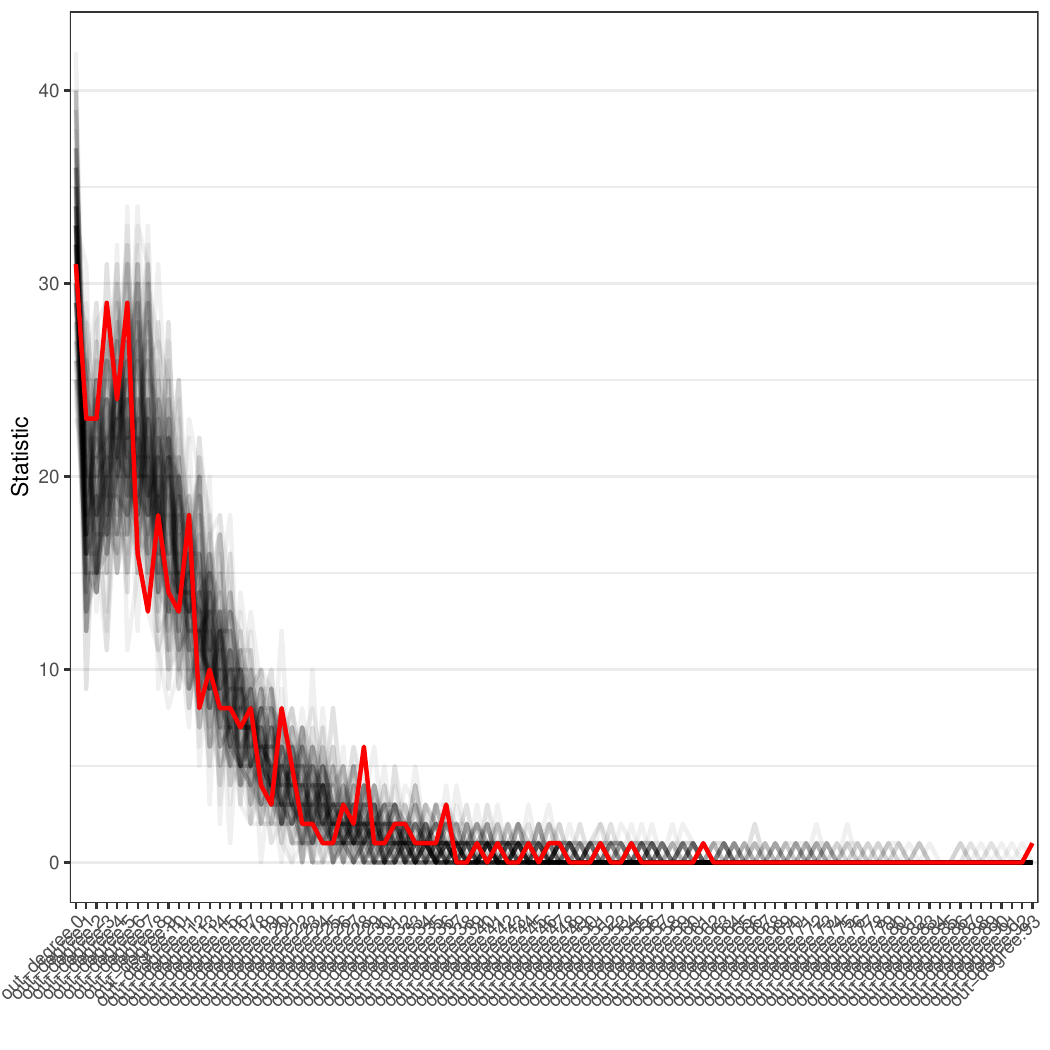}}
  \subfigure[Edgewise shared partners]{\includegraphics[scale=0.3]{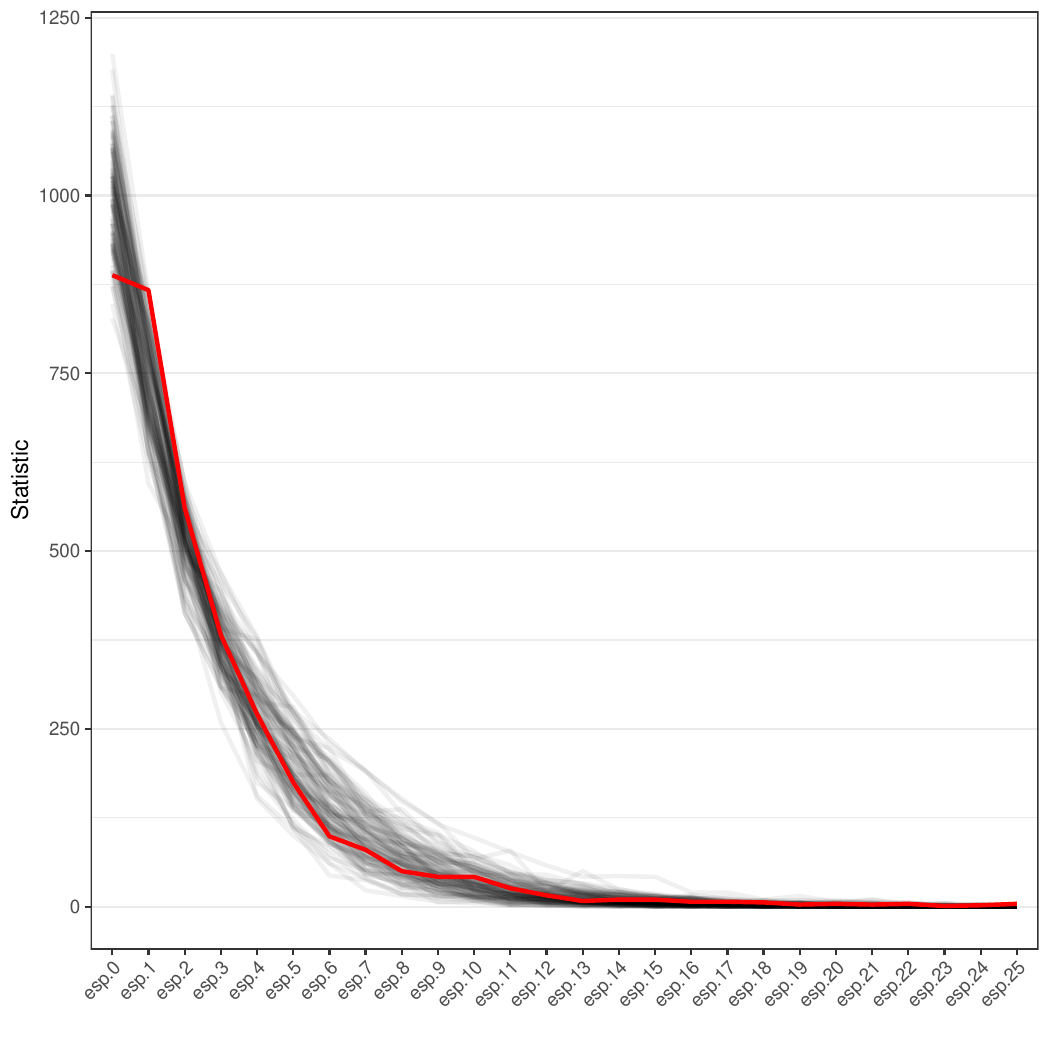}}
  \caption{Model diagnostic and goodness-of-fit plots for the
    \textit{Drosophila} medulla network LOLOG model,
    Table~\ref{tab:drosophila_medulla_lolog}.}
  \label{fig:drosophila_medulla_lolog_gof}
\end{figure}

\begin{figure}
  \centering
  \includegraphics[angle=270,width=\textwidth]{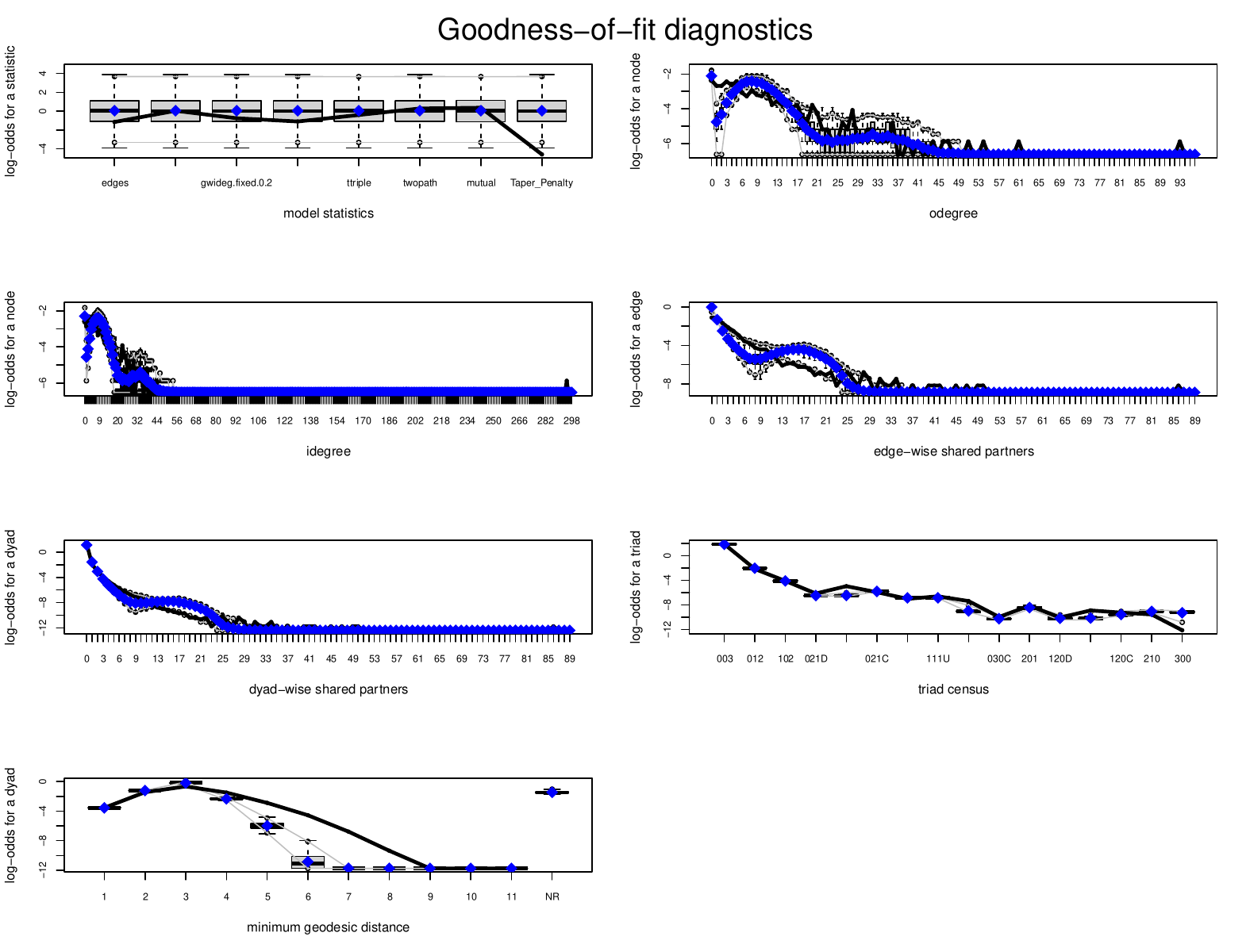}
  \caption{Goodness-of-fit plots for the tapered ERGM model of the
    \textit{Drosophila} medulla network
    (Table~\ref{tab:drosophila_medulla_tapered_estimations}).}
  \label{fig:drosophila_medulla_tapered_gof}
\end{figure}

\end{document}